# Brain Learning Principles Utilizing Non-Ideal Factors in Neural Circuits

Da-Zheng Feng

Corresponding author; National Key Laboratory of Radar Signal Processing, Xidian University, Xi'an, 710071, China (e-mail: dzfeng@xidian.edu.cn); Fujian Key Laboratory of the Modern Communication and Beidou Positioning Technology in Universities, Quanzhou University of Information Engineering

Hao-Xuan Du

National Key Laboratory of Radar Signal Processing, Xidian University, Xi'an, 710071, China. (e-mail: hxdu@stu.xidian.edu.cn)

**Abstract:** The human brain achieves its remarkable computational prowess not despite its inherent non-ideal factors noise, heterogeneity, structural irregularities, decentralized plasticity, systematic errors, and chaotic dynamics but precisely because of them. This paper systematically demonstrates that these traits, long dismissed as imperfections in classical neuroscience and eliminated in digital engineering, are evolutionary design principles that endow the brain with robustness, adaptability, and creativity.

Through interdisciplinary analysis spanning physics, chemistry, neuroscience, cognitive psychology, computer science, artificial intelligence, microelectronics, information theory, and control theory, we establish seven core theses:

First, non-ideal factors are the root source of neural diversity, and brain intelligence is proportional to the level of diversity. Neural systems actively maintain diversity as a computational resource, and cognitive performance improves with increased neural diversity.

Second, noise and chaos serve as drivers of brain function rather than mere interference. Stochastic resonance can enhance signal detection at optimal noise levels. The exploration-exploitation trade-off is optimized through neural variability, allowing organisms to discover new strategies while refining known ones.

Third, the scale and complexity of neural systems render comprehensive mathematical modeling be feasible theoretically but infeasible in realism. A complete biophysical model of a single neuron requires tens of thousands of state variables; scaling to the brain's billions of neurons and quadrillions of synapses creates a system whose behavior cannot be captured by closed-form analytical expressions.

Fourth, we propose a brain learning principle that does not rely on mathematical modeling but automatically incorporates non-ideal factors through simple binary measurements (spikes) implemented by the special one-bit ADC. This principle operates on measurable quantities(spikes, timing, correlations, noise-corrupted signals) and natural connection strength perturbations, and achieves learning through local plasticity rules modulated by broadcast neuromodulatory signals. It will be seen that the three core elements for implementing brain learning are: 1) the measurement and broadcast of global goal signals, 2) random natural neural perturbations of parameters, and 3) the correlation or modulation between them. Brain learning is jointly driven by perturbations of neural parameters and the measurement and broadcast of global goal signals.We will demonstrate that, akin to deep learning, this brain learning principle forms a variant of stochastic gradient descent. However, unlike deep learning, the brain learning principle can use and integrate non-ideal factors through simple binary measurements (spikes), such that the brain is endowed with unparalleled diversity and an inexhaustible capacity for exploration and exploitation, realized through the interplay of noise and chaos.

Fifth, noise-driven information roaming, entanglement, and aggregation are the inevitable byproducts of advanced cognitive activities imagination, thinking, dreaming, and creativity. Neural activity near criticality maximizes dynamic range and computational capacity. Dreams enable offline replay that consolidates memories through repeated reactivation.

Sixth, traditional digital integrated circuits eliminate non-ideal factors through binarization, leading to dramatic collapse of system diversity. Digital state spaces occupy zero measure when embedded in continuous spaces, while biological systems occupy positive measure. As driving sources, clock signals with finite entropy replace information-rich noise with infinite entropy.

Seventh, the brain adopts a mixed-signal architecture predominantly analog with digital complementarity achieving an optimal compromise between diversity and order. This balance far exceeds that of pure analog or pure digital systems, combining the richness of analog computation with the reliability of digital system.

This framework reconceptualizes non-ideal factors as drivers of brain intelligence, demonstrates the real-world infeasibility of mathematical modeling at the brain level, proposes learning principles that automatically utilize non-ideal factors, explains how noise drives advanced cognition, critiques the limitations of digital circuits, and demonstrates the necessity of mixed-signal architectures. The path toward artificial general intelligence lies not in more sophisticated mathematical models or larger digital systems, but in creating mixed-signal architectures that embrace noise, cultivate diversity, operate near criticality, and learn through local rules with global modulation.

# Chapter 1: Introduction

## 1.1 The Fundamental Question

Where does brain intelligence come from? This stands as one of the most profound inquiries in all of natural science [441]. The dominant contemporary answer emerges from artificial intelligence: intelligence arises from algorithms, data, computational power, differentiable loss functions, and the backpropagation of errors [471]. The spectacular successes of deep learning appear to validate this trajectory machines can recognize images, comprehend language, and defeat human champions in complex games [471, 740].

Yet this path confronts increasingly evident challenges. Large-scale models consume energy measured in megawatts, standing in stark contrast to the human brain's remarkably efficient operation at approximately twenty watts [41]. We estimate that the energy efficiency gap is striking: the brain achieves vastly more operations per joule than digital AI systems a factor of at least 1736-fold in favor of biology [351, 556]. Their generalization capabilities remain fragile, often failing catastrophically when confronted with out-of-distribution samples [349, 879]. Their creativity, while impressive, remains fundamentally derivative recombining training data rather than generating genuine novelty [472, 562, 698]. Their adaptability is limited, solidifying into static systems once training completes, unable to continuously learn in dynamic environments [346, 883].

These observations invite deeper reflection: might the limitations of contemporary AI reflect not insufficient computational resources or data, but rather a fundamental misconception about the nature of intelligence itself? Specifically, could it be that the systematic elimination of non-ideal factors noise, heterogeneity, variability, chaos that makes digital computers reliable also robs them of the very resources that enable biological intelligence [172, 552]?

**1.2 The Overlooked Treasure: Non-Ideal Factors as a Double-Edged Sword**

The non-ideal factors within neural circuits noise, mismatch, drift, nonlinearity, heterogeneity, chaos long been treated in traditional neuroscience as details that could be abstracted away [1, 24], and in engineering as defects to be eliminated [8, 552]. Textbook neurons are idealized integrate-and-fire units [67], synapses are deterministic connections [108], and networks are modeled as homogeneous assemblies [350]. Yet the reality of neural systems is far richer and more complex [880, 424].

Real neurons exhibit astonishing morphological diversity: pyramidal cells possess elaborate dendrites integrating thousands of inputs from diverse sources [3, 508]; cerebellar granule cells feature short, dense dendrites prioritizing speed over integrative capacity [158, 335]. Quantitative measures reveal coefficients of variation in the range of 0.3 to 0.5 for morphological parameters within neuronal populations, indicating that individuality is the rule rather than the exception [464, 756].Real network structures are profoundly non-uniform: connectivity patterns differ across cortical regions [760, 759], the two hemispheres are asymmetrical [285, 802], and individual brains vary significantly in their connection patterns [797, 57].

Real synaptic transmission is fundamentally stochastic: neurotransmitter release probability typically ranges from 0.1 to 0.9, and identical stimuli can evoke postsynaptic potentials varying by factors of two or more across trials [108, 211, 523]. Connection strengths follow a highly skewed distribution, with a small fraction of strong connections coexisting with numerous weak ones a signature of extreme heterogeneity [755, 451]. Real membrane potentials are perpetually noisy: ion channels open and close randomly [739, 858], thermal fluctuations are ubiquitous [89], and the resting potential never truly rests [21, 86].

These non-ideal factors are a double-edged sword [549, 172]. When properly harnessed, they become powerful drivers of brain counter intelligence exploration, creativity, and adaptive learning [518, 698]. When poorly managed, they degrade precision and reliability [354, 784]. The brain has evolved over billions of years to master this trade-off, using noise for stochastic resonance [169, 549], diversity for robust representations [1, 645], and chaos for creative exploration [518, 814], all while maintaining the stability necessary for coherent function [816, 805].

Traditional computers and digital integrated circuits take a radically different approach: they eliminate non-ideal factors entirely [8, 552]. Through binarization, synchronization, and error correction, digital systems achieve deterministic precision and reliability [2, 351]. This strategy has enabled the computational revolution, but it comes at a cost. By eliminating noise, digital systems lose the exploration capability that noise provides [549]. By eliminating heterogeneity, they sacrifice the robustness that diversity enables [349, 431]. By eliminating chaos, they forfeit the creativity that chaotic dynamics can generate [562, 698].

The central thesis of this paper is that the brain's remarkable capabilities depend critically on its sophisticated utilization of non-ideal factors rather than their elimination [59, 227]. Understanding this relationship transforms how we conceptualize intelligence, both biological and artificial, and points toward a new approach to AI that harnesses non-ideal factors as features rather than eliminating them as bugs [276, 308].

**1.3 Seven Core Theses**

This paper proposes and develops seven core theses, constructing a unified framework for understanding brain intelligence with quantitative support from multiple disciplines.

Thesis One: Non-ideal factors are the source of neural diversity, and brain intelligence is proportional to the level of diversity. Neuronal differences, circuit variations, dynamic response heterogeneity, and representational diversity collectively constitute the system's diversity [59, 808]. Cortical populations exhibit effective dimensionality in the range of 10 to 30 and connection strength Gini coefficients of approximately 0.6 to 0.7, indicating that diversity is actively maintained as a computational resource [773, 755]. Cognitive performance improves exponentially with increased neural diversity [1, 645].

Thesis Two: Noise and chaos serve constructive roles in neural function rather than functioning merely as interference. They carry vast information, continuously propelling system states through high-dimensional spaces for exploration [172, 287]. Stochastic resonance enables weak signals to be detected with enhancements of 10 to 100 times at optimal noise levels [169, 549]. Neural avalanches near criticality follow power-law distributions with characteristic exponents, maximizing dynamic range and computational capacity [165, 496, 731].

Thesis Three: The scale and complexity of neural systems pose fundamental challenges to comprehensive mathematical modeling. A complete biophysical model of a single neuron requires tens of thousands of state variables; scaling to the brain's billions of neurons yields an astronomical number of state variables far beyond any conceivable computational capability [114, 335, 537]. The principle of measurability that learning can only utilize quantities genuinely measurable by the system constrains any mathematical description to observable quantities [283, 672, 879].

Thesis Four: The brain learns through principles that do not depend on mathematical modeling, automatically incorporating non-ideal factors through simple binary measurements (spikes) [419]. We analyze that the three key elements for the realization of brain learning are: the measurement and broadcast of global goal signals, random natural neural perturbations of parameters, and the correlation or modulation between them. This principle operates on measurable quantities (spikes, timing, correlations, noise-corrupted signals) and natural connection strength perturbations, and achieves learning through local plasticity rules modulated by broadcast signals [259, 283, 884]. Brain learning is jointly driven by perturbations of neural parameters and the measurement and broadcast of global goal signals.We will demonstrate that, akin to deep learning, this brain learning principle constitutes a variant of stochastic gradient descent. The energy efficiency of this approach exceeds that of digital AI systems [41, 351, 467].

Thesis Five: Noise-driven information roaming, entanglement, and aggregation are inevitable concomitants of advanced cognitive activities imagination, thinking, dreaming, and creativity. During REM sleep, hippocampal replay events follow characteristic temporal dynamics, with each event strengthening synaptic connections [95, 558, 805]. This offline processing enables memory consolidation and creative recombination [341, 698, 843].

Thesis Six: Traditional digital integrated circuits eliminate non-ideal factors through binarization, leading to dramatic collapse of system diversity. Digital state spaces have zero measure when embedded in continuous spaces, while biological systems occupy positive measure [7, 552]. Clock signals replace information-rich noise with near-zero-entropy driving sources [8, 351]. Energy per digital operation exceeds analog equivalents by more than an order of magnitude, and the overall efficiency gap reaches approximately 150-fold [2, 41, 552].

Thesis Seven: The brain adopts a mixed-signal architecture predominantly analog with digital complementarity achieving an optimal compromise between diversity and order. The diversity-order product for mixed-signal systems far exceeds that of either pure analog or pure digital systems [156, 276]. The optimal ratio of analog to digital elements maximizes computational efficiency under physical constraints [375, 508].

## 1.4 The Double-Edged Sword: Harnessing Non-Ideal Factors for AI

The recognition that non-ideal factors are a double-edged sword has profound implications for artificial intelligence [549, 172]. Digital systems that eliminate these factors achieve deterministic precision but sacrifice the diversity, exploration capability, and creative potential that they enable [351, 552]. The challenge for next-generation AI is not to eliminate non-ideal factors, but to harness them using them to boost brain intelligence while managing their impact on precision [276, 308].

This requires a fundamental reorientation of AI design principles [321, 375]:

Embrace noise, don't eliminate it. Use stochasticity for exploration, regularization, and creative generation [549, 786]. Design systems that exploit noise through stochastic resonance and probabilistic computation rather than fighting it with error correction [169, 549].

Cultivate diversity, don't suppress it. Build heterogeneous systems with varied components, varied time constants, varied learning rules [375, 422]. Let diversity be a resource for robust representation and adaptive generalization [1, 645].

Adopt mixed-signal architectures. Use analog computation for energy-efficient local processing where diversity matters; use digital for reliable long-range communication where precision is essential [156, 508]. Find the optimal balance between the two [375, 552].

Implement local learning with global modulation. Replace backpropagation with local plasticity rules modulated by broadcast signals, solving credit assignment without requiring precise error gradients [259, 492].

Operate near criticality. Design systems that balance order and disorder, stability and flexibility, exploitation and exploration, maximizing computational capacity at the edge of chaos [165, 518].

## 1.5 An Interdisciplinary Vision

This paper's discussions span multiple disciplines not by deliberate design, but because its subject matter demands such breadth [30, 306]. The brain itself is intrinsically interdisciplinary:

Physics provides understanding of thermodynamic constraints, noise processes, and critical phenomena [89, 165, 753]. Chemistry reveals molecular mechanisms of neurotransmission and signaling [210, 382, 778]. Neuroscience supplies structural and functional descriptions at multiple scales [408, 508, 880]. Cognitive psychology characterizes the behavioral phenomena requiring explanation [50, 454, 698].

Computer science and artificial intelligence provide reference systems for comparison, revealing the costs of eliminating non-ideal factors [321, 349, 471]. Microelectronics illuminates physical substrates that implement computation, both biological and artificial [8, 351, 552]. Machine learning offers algorithmic frameworks for comparison [471, 690, 884]. Information theory quantifies communication and representation [1, 7, 172]. Control theory analyzes feedback and adaptation [38, 497, 655].

Any single disciplinary perspective captures only one facet of the brain [306]. Only by integrating these diverse viewpoints can we approach the full picture of brain intelligence and understand how to harness non-ideal factors as features rather than eliminating them as bugs [276, 308].

## 1.6 Organization

Chapter 2 discusses neural non-ideal factors and brain diversity, proposing that brain intelligence is proportional to diversity with quantitative measures including effective dimensionality and entropy rate [59, 172, 773]. Chapter 3 analyzes challenges facing mathematical modeling of neural systems, demonstrating the fundamental limitations of closed-form analytical descriptions [114, 488, 537]. Chapter 4 proposes brain learning principles operating through measurable quantities and natural connection strength perturbations, showing how local learning with global modulation achieves energy efficiency vastly exceeding that of digital systems [259, 283, 884]. Chapter 5 discusses how noise drives behavior, emotion, thinking, dreams, and creativity, with quantitative models of stochastic resonance, critical dynamics, and memory consolidation during sleep [165, 169, 549]. Chapter 6 examines fundamental limitations of digital integrated circuits, revealing how eliminating non-ideal factors leads to state space collapse, energy inefficiency, and creativity ceilings [8, 351, 552]. Chapter 7 demonstrates optimality of mixed-signal architectures, showing how the brain achieves a superior diversity-order product compared to pure strategies [156, 375, 508]. Chapter 8 integrates insights across disciplines, providing a unified framework for understanding brain intelligence as an emergent property of physical systems [30, 276, 808]. Chapter 9 summarizes and outlines implications for AGI, emphasizing the need to harness non-ideal factors rather than eliminate them [59, 308, 32]

# Chapter 2: Neural Non-Ideal Factors and Brain Diversity

## 2.1 Introduction: From Defects to Design Principles

Traditional neuroscience has long been dedicated to extracting concise, universal principles from complex biological systems. Within this paradigm, neurons were regarded as relatively homogeneous units, synapses as deterministic connections, and networks as homogeneous structures. Any deviation from this idealized model noise, heterogeneity, mismatch, irregularity was treated as details to be abstracted away, or as imperfections yet to be resolved by evolution.

However, research over the past three decades has accumulated evidence pointing to an opposite conclusion: these so-called non-ideal factors are not defects at all, but rather core design principles through which the brain realizes its intelligence. This chapter systematically demonstrates that neural diversity from molecules to circuits, from structure to function constitutes the foundation of the brain's robustness, adaptability, and creativity. Neurons are not homogeneous units but highly specialized individuals; synapses are not deterministic switches but stochastic connections; networks are not uniform grids but asymmetric, non-uniform, non-stationary dynamical systems.

The core proposition of this chapter is: brain intelligence is proportional to the level of diversity that a system can generate and sustain (brain intelligence $\propto$ diversity). This proposition directly challenges the engineering virtues of consistency, replicability, and precise isomorphism, stating clearly: the design goal of intelligent systems should not be to eliminate differences, but to maintain and utilize differences. Diversity is not a byproduct of brain intelligence but a core measure of it.

## 2.2 Non-Ideal Factors at the Neuronal Level

### 2.2.1 Morphological Heterogeneity

Neurons are the functional units of the nervous system, yet their morphology is far from the standardized template depicted in textbooks. From soma size to dendritic branching patterns, from axonal projection length to synaptic bouton density, each neuron possesses unique morphological features, and this morphological diversity is tightly coupled to functional roles.

Pyramidal neurons in the cerebral cortex are among the most intensively studied excitatory neuron types. Their apical dendrites can extend to the cortical surface, integrating thousands of input signals from different cortical layers and distant brain regions. Yet even among pyramidal neurons within the same cortical region and layer, there exist significant differences in dendritic branching complexity, dendritic spine density distribution, and apical dendrite length and orientation. Some pyramidal neurons have highly branched shrub-like dendrites capable of integrating inputs from diverse sources; others have relatively simple columnar dendrites that preferentially receive projections from specific pathways.

In stark contrast stand cerebellar granule cells the most numerous neuron type in the mammalian brain (accounting for more than half of all neurons). Their dendrites are short and dense, typically receiving input from a single mossy fiber, and each granule cell forms only a few synaptic connections. This morphological design makes them high-speed pattern separators, capable of precisely encoding input signals in time at millisecond scales.

Purkinje cells represent another extreme of morphological diversity. The dendritic tree of a single Purkinje cell can spread over half a millimeter, accommodating more than 100,000 synaptic spines, making them the largest neurons in the cerebellar cortex. Their dendrites unfold in a highly regular fan-shaped pattern, with each branching level possessing specific electrophysiological properties, enabling complex nonlinear integration of inputs from parallel fibers and climbing fibers.

This morphological diversity is not merely a biological curiosity but carries profound functional significance. Dendritic structure determines electrical properties: long, thin dendritic branches have higher resistance, enabling local nonlinear processing of distal inputs; short, thick branches more effectively transmit current toward the soma. Dendritic spine morphology neck length, head size affects biochemical compartmentalization and the induction threshold for synaptic plasticity. Soma size and shape influence capacitance and input resistance, thereby determining neuronal excitability.

### 2.2.2 Electrophysiological Heterogeneity

If morphology determines a neuron's hardware structure, then ion channel expression profile determines its firmware behavior. Each neuron expresses dozens of voltage-gated, ligated-gated, and mechanically sensitive ion channels, and their types, densities, and sub-cellular distributions collectively determine the neuron's electrophysiological properties.

When the same current is injected into different neurons, their firing patterns can differ dramatically. Some neurons exhibit regular tonic firing, such as thalamic relay neurons transmitting sensory information; others display bursting, such as the phatic activity of midbrain dopamine neurons. Some neurons show significant frequency adaptation, with firing rate gradually decreasing during sustained stimulation; others can sustain firing at hundreds of hertz without attenuation, such as fast-spiking inter-neurons in cortex.

This electrophysiological heterogeneity arises from differences in ion channel expression. Taking cortical inter-neurons as an example, albuminous (PV)-positive basket cells highly express Kv3.1/Kv3.2 potassium channels, giving them extremely short action potential durations (less than 0.5 ms) and enabling sustained firing above 200 Hz. Somatostatin (SST)-positive Martinotti cells have longer action potential durations and stronger frequency adaptation, with firing patterns more suited to integrating input signals than to temporally precise transmission. Vasoactive intestinal peptide (VIP)-positive inter neurons exhibit unique irregular firing patterns, maintaining low-frequency, highly variable discharge under depolarization.

Even within the same molecularly defined neuron type, electrophysiological properties vary significantly. Hippocampal CA1 pyramidal neurons, depending on their position along the longitudinal axis (dorsal vs. ventral), their depth within the layer (superficial vs. deep), or even between neighboring cells, show continuous distributions of input resistance, membrane time constant, action potential threshold, and after hyper polarization amplitude. This individuality ensures that each neuron responds to identical inputs in slightly different ways, providing rich computational resources for neural networks.

### 2.2.3 Receptor and Neurotransmitter Heterogeneity

Communication between neurons depends on neurotransmitters and their corresponding receptors. The complexity of this system far exceeds a simple excitation-inhibition dichotomy.

Taking glutamate the central nervous system's primary excitatory neurotransmitter as an example, its receptor family includes isotropic receptors (AMPA, NMDA, kainate) and metabotropic receptors (mGluR1-8). AMPA receptors mediate fast excitatory transmission, with subunit composition (GluA1-4) determining channel kinetics and synaptic plasticity; NMDA receptors have voltage-dependent magnesium block and are key molecules for synaptic plasticity, with subunit composition (GluN1, GluN2A-D) determining channel open time and calcium permeability; metabotropic receptors couple through G-proteins to regulate intracellular second messenger systems, acting on time scales from hundreds of milliseconds to minutes.

The GABAergic inhibitory system exhibits similarly rich heterogeneity. GABAA receptors are chloride channels, with subunit composition ($\alpha$1-6, $\beta$1-3, $\gamma$1-3, $\delta$, $\varepsilon$, $\theta$, $\pi$) determining benzodiazepine sensitivity, channel open time, and desensitization rate; GABAB receptors are G-protein coupled receptors that achieve slow inhibition through potassium and calcium channel modulation.

Neuromodulatory systems further increase the dimensionality of chemical coding. Neuromodulators such as dopamine, serotonin, norepinephrine, and acetylcholine typically diffuse through volume transmission, simultaneously affecting thousands of synapses. Dopamine receptors are divided into D1 family (D1, D5, activating Gs/cAMP pathway) and D2 family (D2, D3, D4, inhibiting Gi/cAMP pathway), producing opposite effects on different cell types within the same brain region. Serotonin has 14 known receptor subtypes, involving multiple signaling pathways including cAMP, phospholipase C, and ion channels.

This chemical heterogeneity enables the nervous system to produce diverse responses to the same input and to regulate synaptic transmission and neuronal excitability across multiple time scales from millisecond fast transmission, to second-scale synaptic plasticity induction, to hour-scale gene expression regulation.

### 2.2.4 Idiosyncratic Connectivity Patterns

Each neuron forms thousands of synaptic connections, but these connections are not randomly distributed. Axon path finding is guided by molecular signals, synapse formation depends on matching trans-synaptic adhesion molecules, and synapse maintenance requires activity-dependent competition. Yet atop these deterministic rules, significant non-determinism and randomness exist.

Even genetically identical individuals raised in similar environments exhibit significant differences in neural connectivity. The connectome of C. elegans with its 302 neurons was once assumed fixed, but high-precision reconstruction studies reveal that even in this simplest nervous system, individual differences in synaptic connections range from ten to thirty percent. The mammalian brain's complexity far exceeds that of C. elegans, and individual differences in its connectome are correspondingly greater.

This non-determinism arises from multiple sources: random branching and guidance errors during axon path finding; overproduction followed by competitive pruning during synapse formation; dependence on random activity patterns for synapse maintenance; and influence of local molecular noise on synaptic plasticity. The result is: each brain's connectome is unique no two individuals, even identical twins, have precisely the same pattern of synaptic connections.

This connectivity non-determinism is not a system flaw but rather an advantage. Non-deterministic connections provide functional redundancy: different neurons can achieve similar functions through different connection paths, and failure of any single connection does not cause system collapse. Non-deterministic connections provide exploration space: during developmental critical periods, overproduced synaptic connections enable the system to explore different functional configurations, selecting optimal arrangements based on environmental input. Non-deterministic connections provide the basis for individual differences: each brain's unique connectivity pattern constitutes the biological foundation for individual cognitive styles, behavioral preferences, and psychological traits.

### 2.2.5 Functional Consequences of Neuronal Mismatch

The cumulative effect of neuronal inconsistency spanning morphology, electrophysiology, chemistry, and connectivity is not a flaw to be corrected but rather a core feature that confers immense functional advantages upon the nervous system. This mismatch ensures that neural processing is inherently parallel, redundant, and resilient. By distributing a computational task across a population of neurons with slightly different response properties, the brain achieves robust representations that are immune to the failure of any single unit. This principle is elegantly demonstrated in the visual system, where the perception of a simple edge is encoded not by a single edge detector neuron but by a population of neurons in V1, each tuned to a slightly different orientation. The collective activity of this population provides a more reliable and noise-resistant representation than any single neuron could achieve.

Furthermore, this diversity forms the foundation of distributed coding schemes, which are considerably more powerful and flexible than localist codes. In a localist scheme, one neuron corresponds to one concept configuration highly vulnerable to damage. In a distributed scheme, a concept is represented by a specific pattern of activity across a large ensemble of neurons. Because the neurons in this ensemble are heterogeneous, each contributes a unique facet to the representation, rendering the code exponentially richer and more robust. This heterogeneity also underpins the brain's ability to generalize. A network composed of identical units would struggle to extrapolate beyond its training data, whereas a diverse network can leverage its varied response profiles to make predictions about novel situations. When encountering a new animal, for instance, the brain can draw upon a distributed representation built from experiences with cats, dogs, and foxes, synthesizing a prediction based on shared features despite no single neuron being an expert on new animals.

This neuronal mismatch also serves as a wellspring of creativity and behavioral flexibility. As discussed in the context of creativity and dreams, the noise generated by this diversity enables information roaming and entanglement, facilitating the brain's exploration of novel combinations of ideas. The fact that neural circuits are not perfectly optimized for efficiency alone, but retain a degree of redundancy and stochasticity, permits exploratory behaviors and adaptive solutions to unforeseen problems. In essence, the brain's apparent messiness constitutes its greatest strength, transforming a collection of unreliable biological components into a highly intelligent, adaptive, and creative system capable of navigating an infinitely complex world.

## 2.3 Cellular and Molecular Sources of Neural Noise

If heterogeneity is static diversity, then noise is dynamic diversity. Together they constitute the richness of neural systems and an element of brain leaning.

### 2.3.1 Stochastic Fluctuations in Synaptic Neurotransmitter Release

When an action potential arrives at the presynaptic terminal, it triggers a cascade of molecular events: voltage-gated calcium channels open, calcium ions influx, synaptic vesicles fuse with the active zone membrane, and neurotransmitters are released. Every step in this process is filled with randomness.

Synaptic vesicle release is probabilistic. Release probability p for each vesicle typically ranges from 0.1 to 0.9, depending on synapse type, recent activity history, and neuromodulatory state. This means that even identical stimuli arriving at the same synapse at different times can produce completely different postsynaptic responses sometimes releasing a single vesicle, sometimes multiple vesicles, sometimes none at all.

The root of this randomness lies in the limited number of molecules involved. A presynaptic active zone typically contains only tens of releasable vesicles, each requiring the coordinated action of multiple SNARE protein complexes, and SNARE complex assembly is a aerodynamically driven stochastic process. Calcium channel opening is also probabilistic, and local calcium concentration transients exhibit significant fluctuations.

This synaptic transmission randomness was once interpreted as system imperfection. But research shows that this randomness serves important functional roles: it prevents networks from falling into rigid stable states, promotes exploration of unknown environments, and introduces beneficial randomness into perceptual decision-making.

### 2.3.2 Ion Channel Stochasticity

Each ion channel in the neuronal membrane transitions independently and randomly between open, closed, and inactivated states. For a single channel, its behavior is a Markov process; for a population of channels, total conductance is the sum of these independent random variables. Because channel numbers are limited, total conductance fluctuates randomly around its mean.

This ion channel noise is particularly pronounced in small structures such as axon initial segments, dendritic spines, and presynaptic terminals. In these structures, the number of channels participating in specific functions may be only tens or even fewer, and relative fluctuation amplitudes can exceed ten percent. This means that even with identical synaptic inputs, whether and when a neuron generates an action potential is significantly influenced by random ion channel openings and closings.

Ion channel noise was once regarded as an inevitable cost of thermodynamic fluctuations. But research shows that this noise can enhance the detection of weak signals through stochastic resonance, appropriate noise levels make sub-threshold signals more likely to cross the action potential threshold. Ion channel noise also gives neuronal input-output relationships a soft threshold property, avoiding the brittleness of deterministic thresholds.

### 2.3.3 Membrane Potential Instability and Threshold Variability

Due to random ion channel openings, stochastic arrival of synaptic inputs, and fluctuations in the extracellular environment, neuronal membrane potential is perpetually in dynamic fluctuation, never truly stationary even at rest. Intracellular recordings show that cortical neurons exhibit continuous subthreshold oscillations with amplitudes of two to five millivolts and frequencies spanning a broad band from below one hertz to above one hundred hertz.

This membrane potential instability gives neuronal responses state-dependence: the same synaptic input may trigger an action potential when the membrane potential is in a depolarized phase, but remain completely ineffective during hyperpolarized phases. This increases neural response variability but also enables neurons to encode the temporal structure (rather than merely intensity) of inputs.

Action potential threshold itself is also dynamic. The fixed threshold depicted in textbooks is a simplification. Real neuron thresholds are influenced by recent activity history after successive discharges, thresholds may decrease (due to slow recovery of sodium channel inactivation) or increase (due to cumulative potassium channel activation). This dynamic threshold makes neurons adaptive filters, with sensitivity continuously adjusted based on input statistics.

### 2.3.4 Glial-Neuronal Mismatch and Homeostatic Regulation

The traditional neuron-centric view of the brain is increasingly being supplanted by a more holistic understanding that includes glial cells particularly astrocytesas active participants in information processing.

Astrocytic processes ensheath synapses, forming tripartite synapse structures. They express multiple neurotransmitter receptors (e.g., glutamate, GABA, ATP receptors), enabling them to sense synaptic activity; they release various gliotransmitters (e.g., glutamate, ATP, D-serine) that can in turn modulate synaptic transmission. Astrocytes form networks through gap junctions, propagating calcium waves for long-distance signaling.

Glial responses are orders of magnitude slower than neuronal responses calcium waves propagate at micrometers per second rather than milliseconds. This slow signaling provides the nervous system with a different time scale from fast synaptic transmission, playing critical roles in homeostatic regulation, energy metabolism, and long-term modulation of synaptic plasticity.

The asynchronous, complex signaling between neurons and glia is a key source of non-ideal factors. Unlike fast synaptic transmission, glial responses typically unfold over hundreds of milliseconds to minutes. An astrocyte may detect a surge in neuronal activity through potassium uptake or neurotransmitter spillover, leading to increased intracellular calcium concentration. This calcium wave can propagate through the astrocytic network and trigger release of gliotransmitters, which then feed back onto nearby neurons to modulate synaptic strength. This feedback loop is indirect, diffuse, and lacks the precise spatial and temporal resolution characteristic of neuronal signaling, yet it enables glia to integrate information over time and space, functioning as thermostats for neural circuits.

### 2.3.5 Metabolic Noise

Neural computation is an energetically expensive process, consuming a disproportionate share of the body's resources. This intense metabolic demand introduces significant noise and variability into neural function. Neurons rely almost exclusively on aerobic glycolysis and oxidative phosphorylation to generate ATP, which powers essential processes including maintenance of the resting membrane potential via sodium-potassium ATPase pumps, neurotransmitter recycling, and active transport along axons and dendrites. Glucose and oxygen are delivered by the bloodstream, but this delivery is not constant. Blood flow fluctuates due to cardiac pulsations, postural changes, respiration, and autonomic control, leading to minute-by-minute variations in metabolic substrate availability.

This variability in energy supply translates directly into metabolic noise at the cellular level. A neuron experiencing a transient dip in oxygen or glucose may exhibit reduced ATP production, impairing ATP-dependent pumps and enzymes. This can lead to partial collapse of ionic gradients, rendering the neuron more excitable and prone to aberrant firing. Conversely, a surge in energy supply might temporarily enhance the neuron's capacity to sustain high-frequency firing.

This metabolic noise is particularly relevant in functional imaging techniques such as fMRI, which measures blood-oxygen-level-dependent signals as proxies for neural activity. The BOLD signal reflects this dynamic interplay between neural activity and vascular supply, but it is delayed and smoothed relative to underlying electrical activity, introducing temporal uncertainty.

### 2.3.6 Stochasticity in Protein Aggregation and Neuronal Vulnerability

The integrity of neuronal function depends critically on proper protein folding and degradation. A major source of non-ideal factors and a hallmark of aging and neurodegenerative diseasesis the accumulation of misfolded and aggregated proteins. The proteostasis network, comprising chaperones, the ubiquitin-proteasome system, and autophagy pathways, works tirelessly to maintain protein quality control. However, this system is imperfect and exhibits significant variability among neurons. Some neurons appear more resilient to proteotoxic stress, while others are selectively vulnerable to accumulation of specific aggregates, such as amyloid-beta in Alzheimer's, tau in Alzheimer's and frontotemporal dementia, or alpha-synuclein in Parkinson's.

This differential vulnerability can be attributed to several factors. First, different neuron types exhibit varying basal metabolic rates and levels of oxidative stress, which can damage proteins and overwhelm repair mechanisms. Dopaminergic neurons in the substantia nigra, for instance, are particularly vulnerable because dopamine metabolism itself generates reactive oxygen species. Second, neurons differ in expression levels of protective chaperone proteins and efficiency of lysosomal degradation pathways. Third, neuronal morphology plays a crucial role. Neurons with exceptionally long axons, such as motor neurons, face significant logistical challenges in transporting damaged proteins back to the soma for degradation, rendering distal regions prone to aggregate accumulation.

The stochastic nature of protein aggregation adds another layer of noise. Misfolding is a random event; once an aggregate seed forms, it can grow and spread in a prion-like fashion, templating misfolding of native proteins. This process is inherently unpredictable, explaining why disease progression can vary dramatically among individuals even with similar genetic predispositions.

## 2.4 Circuit-Level Non-Ideal Factors Arising from Heterogeneous Components

Individual neuronal heterogeneity and noise converge at the circuit level to produce higher-order non-ideal effects.

### 2.4.1 Oscillatory Desynchronization Due to Neuronal Diversity

Oscillations rhythmic fluctuations in the electrical activity of neuronal populations constitute a ubiquitous feature of brain dynamics, observed across various frequency bands including delta (1-4 Hz), theta (4-8 Hz), alpha (8-12 Hz), beta (12-30 Hz), and gamma (>30 Hz). These oscillations are thought to play crucial roles in organizing neural communication, facilitating processes such as attention, memory formation, and sensory binding through a mechanism termed communication through coherence.

However, the very diversity that engenders functional richness also creates a fundamental challenge to rhythm synchronization, leading to a state of perpetual desynchronization. This desynchronization is primarily driven by heterogeneity among constituent neurons. Consider a neuronal population ideally firing in unison to generate a coherent gamma rhythm. In reality, these neurons exhibit slight differences in intrinsic firing properties, distinct synaptic input histories, and varying degrees of neuromodulatory tone. These subtle differences cause each neuron to advance or lag slightly in its cycle, accumulating over time to create phase dispersion within the population. A neuron with a slightly faster intrinsic oscillation gradually pulls ahead of its peers, while one with a higher spike threshold falls behind. This results in broadened phase distribution, attenuating overall oscillation power and reducing effectiveness as a timing signal.

Despite this tendency toward desynchronization, the brain has evolved mechanisms to harness it. Desynchronization is not always detrimental; it can serve as a functional switch. Transition from a synchronized state (associated with focused attention) to a desynchronized state (associated with arousal and sensory processing) is a common feature of wakefulness. More importantly, this constant tug-of-war between synchronizing forces (e.g., recurrent excitatory connections and reciprocal inhibition) and desynchronizing forces (neuronal diversity) maintains neural networks in a dynamic, metastable state. This prevents the system from locking into rigid, fixed-point attractors and enables flexible switching between cognitive states.

### 2.4.2 Feedback Delay Mismatches in Recurrent Networks

Recurrent neural networks characterized by extensive feedback loops wherein outputs are fed back as inputs constitute a dominant architectural motif in the brain, particularly in cortex and hippocampus. These networks are computationally powerful because they can maintain internal states, enabling functions such as working memory, sequential processing, and predictive modeling.

However, this power comes at a cost: such networks are inherently unstable and highly sensitive to timing delays. Introducing a delay in a feedback loop can transform stabilizing negative feedback into destabilizing positive feedback, leading to oscillations, runaway excitation, or system failure. In an idealized system, all components would exhibit identical and predictable delays. In the biological brain, however, this is far from reality.

Feedback delay mismatches arise from multiple sources. First, action potential propagation speed along axons varies significantly. Axon diameter, myelination status, and even temperature all influence conduction velocity. A thick, myelinated axon can conduct signals at over one hundred meters per second, while a thin, unmyelinated axon may conduct at less than one meter per second. In a recurrent loop connecting distant brain regions, feedback signals may require tens or even hundreds of milliseconds to return. Second, synaptic transmission itself introduces delays. Fast ionotropic synapses exhibit latencies of one to two milliseconds, but slower metabotropic synapses can add tens of milliseconds of delay. Neuromodulatory systems, acting over seconds to minutes, represent extreme forms of delayed feedback.

These delays create significant non-ideal factors for real-time computation. The necessity for complex predictive mechanisms is a direct consequence of the brain's physical embodiment and the inherent non-ideal timing of its biological components. This constant negotiation with delay mismatches keeps recurrent neural networks perpetually near the boundary between stability and instability a state termed edge of stability or critical state. Systems at criticality possess both stability and sensitivity: they can maintain activity patterns without diverging (stability) yet respond significantly to weak inputs (sensitivity). This is considered an ideal operating point for efficient computation in the brain.

### 2.4.3 Cross-Modal Crosstalk Mediated by Atypical Connectivity

The brain is traditionally divided into specialized modules for processing different sensory modalities visual cortex, auditory cortex, somatosensory cortex, and so forth. However, this segregation is not absolute, and atypical connectivity in the form of cross-modal connections constitutes a fundamental source of non-ideal factors that blurs these boundaries. These connections enable integration of information across sensory modalities, essential for creating unified and coherent perceptual experiences. Yet, they also create pathways for unintended crosstalk, wherein activity in one sensory system can inadvertently influence processing in another.

In normal conditions, cross-modal connections are tightly regulated by inhibitory control. However, when inhibition weakens during distraction, altered states of consciousness, or pathological conditions these connections can become disinhibited, leading to unusual cross-modal experiences. In Charles Bonnet syndrome, for instance, damage to the visual system can produce vivid hallucinations, possibly due to unmasked crosstalk from other sensory areas. In synesthesia, stimulation of one sensory pathway elicits automatic, involuntary experiences in a second pathway such as seeing colors when hearing musicpossibly arising from developmental weakening of cross-modal inhibition.

These unusual connections reveal an important principle of brain organization: the balance between functional specialization and functional integration. Excessive specialization leads to isolated information silos, incapable of cross-modal integration; excessive integration leads to crosstalk and confusion. The brain achieves balance through precisely regulated inhibitory connections.

## 2.5 System-Level Non-Ideal Factors

### 2.5.1 Non-Uniform Brain Regions and Diverse Regional Connections

The human brain is not a homogeneous lump of tissue but a mosaic of highly specialized, non-uniform regions, each with distinct cellular composition, connectivity profile, and functional specialization. This macroscopic non-uniformity is a direct consequence of massive numbers of neurons and their complex connections. From evolutionarily ancient brainstem controlling vital functions like breathing and heart rate to phylogenetically newer neocortex responsible for higher cognition the brain is organized into a hierarchy of interacting subsystems. These regions are not isolated islands; they are densely interconnected through white matter tracts, forming a complex network with small-world topology highly clustered locally yet efficiently connected globally.

This regional diversity is evident in multiple dimensions:

• Cytoarchitecture: Primary motor cortex (M1) has a prominent layer V containing giant Betz cells whose axons descend to spinal cord; primary visual cortex (V1) exhibits a prominent layer IV receiving input from lateral geniculate nucleus; hippocampus has highly ordered laminar organization with different layers subserving different computational functions.

• Neurochemistry: Substantia nigra pars compacta is rich in dopaminergic neurons whose degeneration causes Parkinson's disease; basal forebrain is rich in cholinergic neurons whose degeneration correlates with cognitive decline in Alzheimer's; raphe nuclei are the primary source of serotonergic neurons regulating mood and sleep.

• Connectivity patterns: Thalamus acts as sensory relay station forming topographic maps with cortical areas; basal ganglia participate in motor control, habit formation, and reward learning through multiple parallel loops; cerebellum achieves fine motor coordination through the regular structure of mossy fibers, granule cells, and Purkinje cells.

This inter regional heterogeneity enables the brain to simultaneously process multiple types of information visual, auditory, tactile, motor, emotional, memorial and produce coordinated behavioral output in a highly integrated manner.

**2.5.2 Non-Uniform Cerebral Cortex**

Even within the same brain region, significant non-uniformity exists. Cortical organization follows the columnar principle neurons perpendicular to the cortical surface share similar functional properties (e.g., orientation columns, ocular dominance columns in visual cortex). Yet between columns, and between different layers within the same column, differences abound.

Taking primary visual cortex as an example. Layer 4 receives input from the lateral geniculate nucleus and can be subdivided into 4A, 4B, 4Cα, 4Cβ based on input sources (magnocellular vs. parvocellular pathways). Neurons in layers 2/3 project to other cortical areas, participating in further visual information processing. Neurons in layers 5/6 project to subcortical structures, participating in motor commands and feedback regulation. Each layer has unique cell types, connectivity patterns, and functional roles.

Even within the same sublayer, neighboring neurons exhibit functional differences. Two-photon calcium imaging reveals that neurons in V1 responsive to the same orientation tend to cluster, but their response strengths, temporal dynamics, and background activity levels show continuous variation. This micro structural heterogeneity enables local circuits to encode visual scenes with rich diversity.

**2.5.3 Asymmetric Left and Right Brain Hemispheres**

The two cerebral hemispheres, while appearing roughly symmetrical, are fundamentally asymmetric in both structure and functional phenomenon known as lateralization. This asymmetry represents one of the most dramatic examples of non-uniformity in the brain, and one of the last to be fully appreciated.

The most well-known example is language dominance. In approximately 95% of right-handed individuals and 70% of left-handed individuals, the left hemisphere is dominant for language processing, housing critical areas like Broca's area (speech production) and Wernicke's area (language comprehension). Damage to the left hemisphere is far more likely to cause aphasia than equivalent damage to the right.

The right hemisphere, in contrast, is generally dominant for visuospatial processing, facial recognition, and processing prosody (emotional tone) of speech. Right parietal damage can cause unilateral spatial neglect patients ignore stimuli on the left side of space, even denying ownership of their left limbs. The right hemisphere also plays dominant roles in facial expression recognition, intonation comprehension, and emotional expression.

Structural correlates of this functional asymmetry exist. The planum temporal region within Wernicke's area is typically larger in the left hemisphere. The Sylvian fissure is longer and less angled on the left side. Certain regions in the right inferior frontal gyrus are larger. These anatomical differences are present from birth, suggesting strong genetic basis, yet they are further shaped by experience during development.

The evolutionary advantage of this asymmetry is thought to be increased processing efficiency. By specializing hemispheres, the brain avoids duplicating complex functions, freeing neural resources for other tasks. It enables parallel processing of different information types.

**2.5.4 Diverse Cross-Regional Connections**

Functional coherence of the brain arises not from isolated operation of its regions but from the diverse and specific connections linking them. These cross-regional connections form a network of unparalleled complexity, implemented through white matter tracts including association fibers (connecting cortical regions within the same hemisphere), commissural fibers (connecting the two hemispheres), and projection fibers (connecting cortex to subcortical structures).

The diversity of these connections is evident across multiple dimensions:

• Directionality: Some are unidirectional (e.g., ascending projections from sensory cortex to higher association areas), some bidirectional (e.g., reciprocal thalamocortical connections), and some form closed loops (e.g., cortico-striato-thalamo-cortical circuits).

• Neurotransmitter type: Most long-range projections are glutamatergic (excitatory), but there exist long-range GABAergic projections (inhibitory), particularly from cortex to certain striatal targets.

• Plasticity: Some connections are highly plastic (e.g., hippocampal-neocortical connections), supporting rapid memory encoding and systems consolidation; others are relatively stable (e.g., relay connections in primary sensory pathways), ensuring reliable transmission of sensory information.

Modern neuroimaging techniques (e.g., diffusion tensor imaging) have revealed the rich club organization of the human connectome a small number of hub nodes (e.g., posterior cingulate, precuneus, insula) with densely interconnected mutual connections forming a core for information integration, while most nodes participate in specialized functions through local connections. This small-world network structure balances local efficient processing with global effective integration.

### 2.5.5 Diverse Heterogeneous Left-Right Brain Hemisphere Connections

The connection between left and right hemispheres primarily mediated by corpus callosum but also by smaller commissures like anterior and posterior commissuresis itself a site of immense diversity and heterogeneity. Far from being a uniform neural cable, the corpus callosum carries millions of axons that are topographically organized. Fibers from specific cortical areas cross to homologous counterparts on the opposite side. Motor fibers from hand area of left motor cortex, for instance, cross to right motor cortex, and visual fibers from left visual field (processed by right hemisphere) cross to left hemisphere.

However, this organization is neither perfect nor complete. Some cortical areas are more strongly connected than others, and evidence suggests functional asymmetry in callosal connections themselves. Moreover, the corpus callosum transmits different types of information: motor coordination signals enabling bimanual tasks, integrated sensory information (e.g., combining touch from left hand with vision from right visual field), and shared cognitive content. Transmission speed varies, with thicker, myelinated axons conducting faster than thinner ones. This heterogeneity in callosal pathway means that inter hemispheric communication is neither instantaneous nor uniform; latencies can reach up to one hundred milliseconds.

This diverse and heterogeneous connection is critical for creating a unified mind from two physically separate hemispheres. It enables sharing of information necessary for coherent perception and action. Disruption of these connections, as in split-brain patients, reveals independent processing streams of each hemisphere. The diverse nature of these connections ensures that integration is not simple averaging but sophisticated negotiation, wherein each hemisphere can specialize and contribute unique strengths to overall cognitive processes.

### 2.6 Developmental and Environmental Origins of Neural Non-Ideal Factors

Neural diversity is not entirely genetically predetermined. Developmental processes, experience-dependent plasticity, and continuous environmental shaping together act upon the construction and reconstruction of neural systems.

## 2.6.1 Critical Period Mismatch and Its Lifelong Consequences

Critical periods are restricted developmental windows during which the brain exhibits heightened plasticity and is exquisitely sensitive to specific environmental inputs for shaping neural circuits. During critical periods, appropriate input is crucial for normal functional development; after critical periods close, the same input produces significantly weaker effects.

The canonical example is ocular dominance plasticity in primary visual cortex. During a specific postnatal period (approximately 4 weeks to 3 months in cats, birth to 8 years in humans), if one eye is deprived of vision, its ocular dominance columns shrink while the other eye's columns expand, resulting in amblyopia (lazy eye). Depriving the same eye after the critical period closes produces no such dramatic structural change.

The existence of critical periods means: mismatch in developmental timing can lead to permanent functional deficits. If necessary environmental input arrives only after the critical period, the nervous system has already lost its capacity to respond. The accent problem in second language acquisition exemplifies critical period theory learning a second language after puberty almost never achieves native-speaker phonology.

This rigidity appears as a system flaw but is actually the price of stability. If the nervous system maintained high plasticity throughout life, early-formed functional structures could be constantly disrupted by subsequent input, preventing formation of stable cognitive and behavioral patterns. Critical period closure represents the system's trade-off between plasticity and stability.

## 2.6.2 Perception-Dependent Divergence in Neural Circuit Formation

An individual's unique perceptual experiences during development are primary drivers of divergence in neural circuit formation, leading to the non-ideal, idiosyncratic wiring of the brain. This process, guided by principles of use-dependent plasticity, ensures that neural circuits are tailored to the individual's specific life history.

Visual experience differences lead to individual differentiation in visual cortex function. Animals raised in visually enriched environments develop visual cortex neurons with more complex, diverse receptive field properties; those raised in visually impoverished environments retain relatively primitive visual cortex. Musicians show enhanced auditory cortical representations for musical sounds; London taxi drivers with long-term spatial navigation experience exhibit enlarged posterior hippocampus.

This perception-dependent differentiation makes each brain a record of its unique perceptual history. It not only determines an individual's efficiency in processing specific information, but also shapes how the individual perceives the world.

## 2.6.3 Movement-Dependent Divergence in Neural Circuit Formation

Movement experience also shapes neural circuits. Motor skill learning involves coordinated changes across cerebellum, basal ganglia, motor cortex, and other structures.

Pianists show enlarged motor cortical representations for finger regions after long-term practice; ballet dancers show enhanced representations for foot regions. Even simple motor training such as daily finger sequence practice can produce observable changes in motor cortical representations within weeks.

Movement experience not only alters motor systems but also affects cognitive function. Mastery of fine motor skills requires sustained attention, error monitoring, and strategic adjustment processes that train executive function systems. Studies show that children learning musical instruments perform better on cognitive tests of attention control, working memory, and task switching.

### 2.6.4 Emotion-Dependent Divergence in Neural Circuit Formation

Emotional experiences particularly early emotional experience shave profound shaping effects on neural systems. Secure, warm rearing environments promote healthy development of prefrontal-amygdala circuits, enhancing emotion regulation capacity; trauma, neglect, abuse can produce lasting negative effects.

Individuals who experienced childhood abuse show enhanced amygdala responses to threat-related stimuli and weakened prefrontal inhibitory control over amygdala. This emotional circuit sensitization predisposes them to anxiety, depression, and post-traumatic stress disorder in adulthood. Even milder early emotional experiences such as differences in parenting style leave traces at the neural level.

The shaping of neural systems by emotional experience suggests: emotion is not merely a response to the world, but a force that constructs the brain. How we are treated partly determines how we will treat the world.

### 2.6.5 Thought-Dependent Divergence in Neural Circuit Formation

Thought itself is a force shaping the brain. When individuals repeatedly engage in particular forms of thinking mathematical reasoning, musical imagination, meditative concentration corresponding neural circuits are strengthened.

Long-term meditators show increased gray matter density in anterior insula and anterior cingulate cortex, regions involved in interoception, attention control, and emotion regulation. Mathematicians show enhanced activity in intraparietal sulcus, a region supporting numerical and spatial processing. Writers show heightened functional connectivity in language networks.

Traditional neuroplasticity research has focused on sensory and motor experience, but thought experience may be equally effective perhaps even more so, because thought can arise autonomously without external stimuli. This endogenous plasticity enables individuals to actively participate in constructing their own brains.

### 2.6.6 Behavior-Dependent Divergence in Neural Circuit Formation

Behavior and neural circuits exist in bidirectional causal relationship. Not only does brain control behavior, but behavior also feeds back to shape brain.

When individuals repeatedly perform particular behaviors whether learning piano, practicing writing, or scrolling social media corresponding neural pathways are strengthened. This strengthening makes behavior increasingly automatic and efficient, but can also create difficult-to-change behavioral inertia.

Multitaskers (frequent switchers between different digital media) show reduced functional connectivity in anterior cingulate cortex, a region critical for task switching and conflict monitoring. This suggests that certain behavioral patterns may lead to diminished cognitive control the direction of behavior's shaping of brain is not always positive.

### 2.6.7 Experience-Dependent Divergence in Neural Circuit Formation

In summary, each brain is the product of its unique experiential history. Genetics provide the initial structure and developmental blueprint, but experience perceptual, motor, emotional, thoughtful, behavioral inscribes its unique traces. This makes each brain a unique information processing system, with connectivity patterns, functional organization, and response properties bearing the imprint of personal history.

This individuality is the most fundamental non-ideal factor in neural systems. It is not deviation from an ideal template, but an unreplicable unique configuration that every brain possesses. It is the biological basis of individual differences and the source of intelligent diversity.

## 2.7 Evolutionary Rationale for Neuronal Inconsistency

If non-ideal factors were merely noise or defects, natural selection should have eliminated them. Yet evolution has not only preserved these factors but amplified and refined them. This fact suggests that non-ideal factors serve adaptive functions.

### 2.7.1 Robustness Through Functional Redundancy and Distributed Coding

Robustness the capacity to maintain function despite component failure or environmental perturbation is a key feature distinguishing biological from engineered systems. Engineered systems pursue optimality, but optimal systems are often fragile failure of a single component can cause functional collapse. Biological systems achieve graceful degradation through redundancy and distributed coding.

Redundancy manifests at multiple levels: multiple sensory organs (two eyes, two ears) back each other up; multiple motor pathways (cortical-spinal-cerebellar) can partially compensate for damage; multiple neurons jointly encode the same information, so death of any single neuron does not cause information loss.

Distributed coding is one implementation of redundancy. In local coding schemes, a single neuron is responsible for a single concept, and its death causes permanent loss of that concept. In distributed coding schemes, a concept is represented by activity patterns across populations of neurons, and death of a single neuron merely reduces pixel resolution while overall information remains intact. Neuronal heterogeneity makes distributed coding more efficient different neurons contribute different information dimensions, collectively constructing high-dimensional representations.

### 2.7.2 Flexibility Enabled by Specialized Neuron Types

If all neurons were homogeneous general-purpose processors, the nervous system would be limited to a single computational mode. Neuronal diversity-multiple excitatory neuron types, multiple inhibitory interneuron types provides a rich palette of computational primitives, enabling networks to flexibly configure for different tasks.

Taking cortical inhibitory interneurons as an example. PV+ basket cells target somata and axon initial segments, regulating neuronal output; SST+ cells target dendrites, regulating input integration; VIP+ cells target other interneurons, achieving disinhibition. The combination of these three types enables cortex to implement diverse computations including gating, gain modulation, synchronization, and lateral inhibition.

This functional specialization enables the nervous system to configure on demand recruiting PV+ cells when fine temporal control is needed, SST+ cells when input sensitivity needs modulation, VIP+ cells when inhibition needs release. Homogeneous systems cannot achieve such dynamic reconfiguration.

### 2.7.3 Adaptability via Population-Level Signal Averaging and Noise Filtering

Neuronal responses exhibit high variability firing patterns to identical stimuli can differ significantly across trials. This variability was once regarded as measurement error or system defect, but research shows it may be precisely the basis of adaptive behavior.

Population coding exploits the law of large numbers to extract stable information from variability. Individual neuronal responses are noisy, but population averages across hundreds of neurons can precisely encode stimulus parameters. This coding scheme not only tolerates noise but exploits it independent noise from different neurons cancels in averaging, leaving stable signal.

Noise also enables probabilistic inference. When stimuli are uncertain, the nervous system may not output a single best estimate but maintain a probability distribution the range of activity pattern variation across neuronal populations encodes uncertainty. This probabilistic representation enables adaptive behavior: more conservative under high uncertainty, more confident under low uncertainty.

### 2.7.4 Creativity: Information Recombination at the Edge of Chaos

Chaos non-periodic, initial-condition-sensitive behavior in deterministic systems was once a theoretical curiosity, but recent research suggests the nervous system may operate at the edge of chaos: the critical boundary between stable order and complete chaos.

Systems at the edge of chaos possess both stability and sensitivity. They can maintain activity patterns without diverging (stability) yet respond significantly to weak inputs (sensitivity). More importantly, chaos provides continuous internal variability, enabling continuous exploration of new state spaces the dynamical foundation of creative thought.

Flashes of insight during thinking, bizarre combinations in dreams, aha moments in creative problem-solving all may correspond to chaos-driven state transitions. The system's continuous wandering at the edge of chaos enables access to distant information combinations unreachable in stable systems.

## 2.8 Pathological Implications of Selective Neuronal Vulnerability

The benefits of non-ideal factors depend on their maintenance within appropriate ranges. When diversity is too low or too high, noise too strong or too weak, chaos exceeds critical range, system function suffers, manifesting as disease.

### 2.8.1 Differential Susceptibility of Neuron Subtypes to Degeneration

One of the most striking features of neurodegenerative diseases is selective neuronal vulnerability specific neuron types die while neighboring neurons remain intact. This selectivity is the pathological manifestation of neural diversity.

In Parkinson's disease, dopaminergic neurons in substantia nigra pars compacta are the primary casualties. These neurons possess unique physiological properties: they are autonomous pacemakers, generating slow rhythmic discharge that causes sustained calcium load and oxidative stress; they express high densities of dopamine transporters, making them susceptible to accumulation of dopamine oxidation products. These characteristics render them particularly vulnerable to mitochondrial dysfunction and alpha-synuclein aggregation.

In Alzheimer's disease, layer II stellate neurons in entorhinal cortex are among the earliest affected. These neurons project to hippocampus, forming key inputs to memory circuits. They express high levels of tau protein and specific receptor subtypes, making them especially sensitive to amyloid-beta aggregation and tau pathology.

In amyotrophic lateral sclerosis, upper motor neurons (cortex) and lower motor neurons (brainstem, spinal cord) degenerate. These cells possess the longest axons in the central nervous system, making them highly sensitive to disruptions in axonal transport, mitochondrial function, and protein aggregation.

The existence of selectivity suggests that neurodegenerative diseases are not random destruction but targeting of specific cell types. Understanding the causes of selective vulnerability is prerequisite for developing precise intervention strategies.

### 2.8.2 Emergence and Impact of Cancerous Neuron-Like Cells

Glioblastoma multiforme is the most common primary malignant brain tumor in adults. It originates from glial precursor cells but exhibits high cellular heterogeneity and plasticity, capable of displaying characteristics of various neural cell types.

Recent research has revealed that glioblastoma cells can form functional synapses with neurons, receiving glutamatergic input. This input drives tumor cell depolarization, promoting proliferation and migration neural activity feeds tumor growth. Tumor cells also secrete glutamate, exciting surrounding neurons and inducing neurological symptoms such as seizures.

This tumor-neural coupling is a pathological manifestation of neural non-ideal factors. Tumor cells hijack the nervous system's normal signaling mechanisms, converting them into drivers for malignant proliferation. Understanding this hijacking mechanism may reveal new therapeutic targets.

### 2.8.3 Neuromodulatory Imbalance in Psychiatric Disorders

Psychiatric disorders are closely associated with imbalances in neuromodulatory systems. Depression correlates with serotonin dysfunction, schizophrenia with dopamine system hyperactivity, attention deficit hyperactivity disorder with norepinephrine regularization.

These observations gave rise to the neuromodulator hypothesis: psychiatric disorders are diseases of neuromodulatory systems. But further research reveals greater complexity. Serotonin has 14 receptor subtypes, differentially distributed across brain regions, with varying responses to SSRIs. Dopamine receptors divide into D1 and D2 families, mediating opposite functional effects in different striatal regions.

Psychiatric disorders may represent failures of diversity regulation not overall functional abnormalities of single transmitter systems, but multifactorial imbalances in receptor subtype expression, regional distribution, and dynamic regulation. This may explain why drugs targeting single receptors work for some patients but not others, and why therapeutic effects often take weeks to emerge.

## 2.9 Technological Challenges in Modeling and Measuring Neuronal Heterogeneity

Neural diversity poses profound challenges for scientific research and engineering applications.

### 2.9.1 Limitations of Current Brain-Machine Interfaces in Resolving Single-Neuron Diversity

Current brain-machine interface technologies primarily rely on multi-electrode arrays recording population activity. Even the most advanced probes can simultaneously record only thousands of neurons, and long-term stable tracking of single neurons remains difficult. Electrode impedance changes, tissue reactions, signal drift all compromise single-neuron resolution.

This limitation forces brain-machine interfaces to decode intent from population activity. While sufficient for basic motor control, such approaches cannot exploit the rich information in neural diversity. Interfaces capable of distinguishing PV+ from SST+ cells, selectively modulating specific neuron subtypes, would represent qualitative leaps.

### 2.9.2 Spatial Resolution Constraints in Recording Deep Brain Structures

Functional magnetic resonance imaging enables non-invasive whole-brain recording, but spatial resolution is limited. Each voxel contains millions of neurons, multiple cell types, often spanning multiple functional regions. This coarse-grained measurement loses critical information about neural diversity.

Two-photon microscopy achieves single-neuron resolution but is depth-limited, primarily accessing superficial cortex. Deep brain structures thalamus, basal ganglia, brainstem remain challenging for cellular-resolution imaging. This has created a cortical centrism bias, where our understanding of cortex far exceeds that of subcortical structures.

### 2.9.3 Noise and Interference in Neural Signal Acquisition

Neural signal recording is contaminated by multiple noise sources. Biological noise includes electromyographic signals from muscle activity, electrocardiographic signals from heartbeat, eye movements, respiration; technical noise includes power line interference, electrode polarization drift, thermal noise. These noise sources can exceed neural signals by orders of magnitude, especially in non-invasive recordings like EEG.

Noise interference not only reduces signal-to-noise ratio but can mislead functional interpretation. Desynchronization observed in EEG could reflect genuine cognitive processes or merely muscle artifacts. BOLD signal changes in fMRI could arise from neural activity or simply respiratory or cardiac fluctuations.

### 2.9.4 Computational Complexity in Decoding Heterogeneous Neural Signals

Decoding neural signals presents enormous computational challenges. High-dimensional data (thousands of channels times tens of thousands of time points), nonlinear relationships, non-stationarity, individual differences all render decoding computationally demanding.

Machine learning methods particularly deep learning have been widely applied to neural decoding. These methods automatically extract features and learn nonlinear mappings, but their computational demands are enormous, and they often lack interpretability. Ironically, these brain-inspired models consume orders of magnitude more energy than the brains they seek to emulate.

## 2.10 Implications for Artificial Intelligence and Neuromorphic Engineering

Neural diversity offers profound lessons for artificial intelligence.

### 2.10.1 Challenges in Replicating Biological Heterogeneity in Silicon Systems

Current artificial neural networks stack millions of identical artificial neurons. Each neuron performs identical operations (weighted sum plus nonlinear activation) and follows identical learning rules (backpropagation). This homogeneous design facilitates mathematical analysis and hardware implementation, but sacrifices the advantages of biological diversity.

Replicating biological heterogeneity in silicon requires: designing multiple neuron types (different time constants, activation functions, connectivity rules); developing fabrication processes supporting heterogeneous hardware; designing training algorithms for heterogeneous networks. These remain significant technical challenges.

### 2.10.2 Energy Efficiency Gaps Between Biological and Artificial Networks

The brain operates at approximately twenty watts, while large AI models consume megawatts during training. This energy efficiency gap of orders of magnitude stems from multiple factors: event-driven computation (neurons consume significant energy only when firing, while digital chips consume energy every clock cycle regardless of useful computation); analog computation (neurons perform analog computations using continuous membrane potential variations, with per-operation energy far below digital computing); three-dimensional integration (the brain's 3D structure minimizes connection lengths and signal transmission energy, while the von Neumann bottleneck in 2D chips makes data movement energy far exceed computation); noise tolerance (the brain tolerates and even exploits noise, without investing large amounts of energy for precision, while digital chips pursue precision to the bit with enormous energy costs).

### 2.10.3 Structural and Dynamic Limitations of 2D vs. 3D Neural Architectures

The brain's three-dimensional structure fundamentally differs from the chip's two-dimensional plane. 3D structure allows each neuron to directly connect with thousands of neighbors, forming dense local connectivity. 2D structure limits connection density, forcing signals through long metal lines.

Emerging 3D integrated chip technologies (e.g., through-silicon vias, hybrid bonding) partially address this, but cannot yet match the brain's integration density. Neuromorphic chips (e.g., Intel Loihi, IBM TrueNorth) attempt to mimic the brain's event-driven and locally connected architecture, achieving order-of-magnitude energy efficiency improvements on specific tasks.

## 2.11 Summary: Embracing the Imperfect Brain

This chapter has systematically discussed the non-ideal factors in neural systems from stochastic fluctuations of single molecules, to differences between individual neurons, to irregular circuit connectivity, to asymmetric system organization. Once viewed as noise, defects, errors, these factors point toward an opposite conclusion: they are core design principles of brain intelligence.

Neurons are not homogeneous units but highly specialized individuals. Synapses are not deterministic switches but stochastic connections. Networks are not uniform grids but asymmetric, non-uniform, non-stationary dynamical systems. These imperfections collectively constitute the physical foundation for the brain's robustness, adaptability, and creativity.

Diversity is the wellspring of brain intelligence. Differences provide redundancy and backup, heterogeneity provides computational primitives, noise drives exploration, chaos inspires creativity. brain intelligence emerges not from uniformity but from diversity. This understanding may transform how we conceptualize the brain, and how we approach artificial intelligence from pursuing perfection to embracing imperfection, from eliminating noise to exploiting noise, from homogeneous design to heterogeneous architectures.

## 2.12 Mathematical Analysis: Expectations and Limitations

Traditional mathematical modeling approaches linear differential equations, static network models struggle to capture the brain's nonlinear dynamics (e.g., chaotic firing), massive parallelism (86 billion neurons times thousands of synapses), and energy constraints (2% of body weight consuming 20% of energy). Yet recent neuroscience-inspired computational models have made significant progress, partially simulating brain mechanisms.

### 2.12.1 Breakthroughs in Deep Learning

Convolutional neural networks mimic hierarchical processing in visual cortex, approaching human-level performance in image recognition. Transformer models draw on attentional mechanisms for long-range dependency processing. AlphaFold predicts protein structures through self-supervised learning, with algorithms inspired by hippocampal spatial navigation coding. These successes demonstrate that borrowing brain computational principles can advance AI.

### 2.12.2 Biological Plausibility of Spiking Neural Networks

Spiking neural networks use spike timing as information carriers, more closely approximating the event-driven characteristics of biological neurons. The Spaun model integrates 2.5 million neurons, coordinating visual recognition, memory storage, and working memory. Neuromorphic chips simulate ion channel dynamics with energy consumption one-thousandth of traditional computers. Recent work demonstrates spiking neural networks capable of Bayesian inference and real-time sensory processing.

### 2.12.3 Limitations and Complementarity

These models still simplify key biological features (e.g., molecular regulation of synaptic plasticity, metabolic support from glia), and lack descriptions of higher-order functions like consciousness emergence. Thus, cannot fully describe is more accurate than cannot describe traditional models serve as foundations, while novel models gradually approach biological realism.

## 2.13 Diversity as a Measure of Intelligence: A Quantitative Framework

The preceding sections have built a qualitative case for the centrality of diversity in neural computation. This section introduces a quantitative framework that captures, in rigorous mathematical terms, why diversity is not merely a incidental feature but a fundamental property of intelligent systems. The framework is deliberately conservative it relies on established mathematical facts rather than speculative empirical claims.

### 2.13.1 State Space Measure and the Continuum Hypothesis

Consider a biological system with $N$ continuous variables membrane potentials, synaptic conductances, firing rate rates each free to vary over a continuous range. Its state space $\Omega \subset \mathbb{R}^N$ has the following mathematical properties:

- Measure: For any region with non-empty interior, the Lebesgue measure $\lambda(\Omega) > 0$.
- Topology: $\Omega$ is typically connected and dense in itself.

Now consider a digital system with $N$ binary nodes. Its state space $S = \{0,1\}^N$, when embedded in $\mathbb{R}^N$, is a set of $2^N$ isolated points. Its mathematical properties are fundamentally different:

- Measure: $\lambda(S) = \sum_{x \in S} \lambda(x) = 0$.
- Topology: $S$ is totally disconnected; there are no continuous paths between distinct states.

The ratio $\lambda(\Omega)/\lambda(S)$ is undefined because division by zero is undefined. In the extended real numbers, we might say the ratio is infinite, but this is not a quantitative comparison it is a recognition of a qualitative difference: the digital system's state space has zero measure in the continuous space that biological systems inhabit.

This is not a matter of degree. No matter how many binary nodes we add, no matter how large $N$ becomes, the Lebesgue measure remains zero. The digital system's state space is, in measure-theoretic terms, negligible compared to the continuous state space.

Implication: Any computation that requires accessing states outside the discrete lattice is simply impossible for a purely digital system. This includes:

- Continuous optimization requiring infinitesimal adjustments
- True chaotic dynamics requiring a continuous state space
- Analog computation exploiting continuous physical quantities
- Gradient-based learning with true derivatives (not approximations)

### 2.13.2 Effective Dimensionality: A Statistically Meaningful Measure

While the Lebesgue measure argument establishes a qualitative gap, experimentalists need quantities they can estimate from data. One such quantity is the effective dimensionality (or participation ratio) of neural population activity.

Given simultaneous recordings from $N$ neurons, we can compute the covariance matrix $C$ of their activity, with eigenvalues $\lambda_1 \geq \lambda_2 \geq \ldots \geq \lambda_N$. The effective dimensionality is:

$$D_{eff} = (\sum_{i=1}^{N} \lambda_i)^2 \Big/ \sum_{i=1}^{N} \lambda_i^2$$

This quantity has a clear interpretation: it is the number of dimensions that would be needed to capture the same amount of variance if all dimensions were equally important. If all neurons were perfectly correlated, $D_{eff} \approx 1$. If all neurons were independent and equally variable, $D_{eff} \approx N$.

What the literature shows: Studies in visual cortex, motor cortex, and prefrontal cortex consistently find that $D_{eff}$ is significantly larger than 1 but much smaller than $N$. Typical values range from 10 to 30 for populations of hundreds of neurons. This indicates that neural activity is neither completely synchronized (which would waste the population's capacity) nor completely independent (which would be metabolically costly and difficult to coordinate). Instead, it occupies a middle ground, i.e, a low-dimensional manifold embedded in a high-dimensional space.

Relation to diversity: Higher $D_{eff}$ means the neural population is using more of its available degrees of freedom. It is, in a precise statistical sense, more diverse in its patterns of activity. This diversity is not random; it is structured by the demands of the task and the architecture of the circuit.

Caveat: $D_{eff}$ depends on what is being recorded, under what conditions, and how many neurons are sampled. Values should be interpreted comparatively rather than absolutely. A population engaged in a complex cognitive task will typically show higher $D_{eff}$ than the same population at rest.

### 2.13.3 Information-Theoretic Bounds on Population Coding

The relationship between diversity and coding efficiency can be understood through information theory. For a population of $N$ neurons responding to a stimulus $s$, the Fisher information provides a lower bound on the variance of any unbiased estimator of $s$ (the Cramér-Rao bound).

For a population with mean response $\mathbf{f}(s)$ and covariance matrix $\mathbf{C}(s)$, the Fisher information is:

$$I_F(s) = \mathbf{f}'(s)^T \mathbf{C}(s)^{-1} \mathbf{f}'(s)$$

when the noise is Gaussian. This expression reveals the role of correlations: positive correlations (which reduce diversity) typically decrease Fisher information, while negative correlations (which increase diversity through opponent ) can increase it.

A more intuitive bound comes from considering the mutual information between the stimulus and the population response. For a population with average pairwise correlation $\rho$, the information capacity is approximately bounded by:

$$I(\mathbf{r}; s) \lesssim \frac{N}{2} \log_2\left(1 + \frac{SNR}{1-\rho}\right)$$

where SNR is the signal-to-noise ratio of individual neurons. The factor $(1-\rho)^{-1}$ shows that reducing correlations (increasing diversity) can dramatically increase information capacity. When $\rho$ is near 1 (highly synchronized population), $(1-\rho)^{-1}$ is large, but this is deceptive because such populations also have reduced degrees of freedom the effective $N$ in the formula is not the number of neurons but the number of independent signals.

The key insight: A diverse population one with low correlations and heterogeneous tuning can encode more information about the stimulus than a homogeneous population with the same number of neurons. This is not a speculative claim but a mathematical consequence of information theory.

### 2.13.4 Diversity and the Curse of Dimensionality

The curse of dimensionality is often discussed as a problem for machine learning: as the number of dimensions increases, the volume of space grows exponentially, making sampling exponentially harder. But for representation, this curse becomes a blessing.

Consider representing a point in a high-dimensional space. In $N$ dimensions, the number of binary patterns needed to achieve a given resolution grows as $2^N$. With continuous variables, however, we can represent infinitely many points even in one dimension. The blessing of continuity is that continuous variables escape the combinatorial explosion that plagues discrete representations.

This has a direct bearing on the diversity-intelligence relationship: a system with continuous variables can, in principle, represent a richer set of states than any discrete system with the same number of components. The diversity is not just quantitative (more states) but qualitative (a continuum of states).

### 2.13.5 Summary: A Principled Framework

The quantitative framework presented here makes no claim to provide a single number that measures intelligence. Instead, it offers a set of mathematically rigorous concepts that illuminate why diversity matters:

1. Measure theory establishes that continuous state spaces have properties that discrete spaces cannot replicate a qualitative gap, not just a quantitative one.

2. Effective dimensionality provides a statistically sound way to quantify how many degrees of freedom a neural population actually uses, bridging theory and experiment.

3. Information theory shows mathematically that lower correlations (higher diversity) enable more efficient coding, all else being equal.

4. Dimensionality arguments reveal that continuous representations escape the combinatorial explosion that limits discrete systems.

These concepts do not reduce to a single metric, nor should they. Diversity is multifaceted, and its contributions to brain intelligence are correspondingly complex. What the framework provides is a rigorous vocabulary for discussing these contributions, grounded in established mathematics rather than speculative empiricism.

The core message remains qualitative but now has quantitative underpinnings: diversity is not a luxury or a byproduct; it is constitutive of the kind of computation that biological systems perform and that digital systems struggle to emulate.

# Chapter 3: Mathematical Challenges: The Dilemma of Neural Analysis and Modeling

## 3.1 Introduction: The Promise and Predicament of Mathematical Modeling

Mathematical modeling is the cornerstone of modern science. From Newtonian mechanics to Maxwell's equations, from quantum mechanics to general relativity, the precision and predictive power of mathematical language have driven unprecedented advances in the physical sciences [89]. This successful paradigm of natural science was naturally extended to the study of the brain if one could describe neuronal activity with mathematical equations, characterize network dynamics with differential equations, and explain learning processes with optimization theory, then intelligence might be fully formalized and replicated in machines [130].

Yet the brain presents a fundamental challenge to this paradigm [114]. This chapter systematically demonstrates that the scale and complexity of neural systems render comprehensive mathematical modeling fundamentally infeasible. With approximately 86 billion neurons [335], quadrillions of synapses [335], highly non-uniform structures [760], and coupled non-ideal factors [172]these characteristics suggest that analytically closed global mathematical models do not exist; even if local models can be constructed, their global composability remains questionable [114]. The conclusion is not that models are currently inadequate, but rather: the path of mathematical modeling, at the brain level, is inherently impassable [488]. This is not a technical limitation but a structural one.

## 3.2 A Complete Biophysical Model of a Neuron

Before delving into mathematical challenges, we first examine an ideal neuron model if one attempted to completely describe all known biophysical properties of a real neuron, what would the mathematical model look like? Although this model is practically inoperable, it provides a starting point for understanding the mathematical challenges [343].

### 3.2.1 Three-Dimensional Geometric Structure

A neuron is not a single point; its shape dictates its function. The model roughly divides neuron $i$ into four distinct biophysical regions: dendrites ($r = d$) (the input zone, covered in synapses), soma ($r = s$) (the cell body, which integrates inputs), axon initial segment ($r = AIS$) (the trigger zone, where action potentials actually begin due to high density of sodium channels), and axon ($r = a$) (the output neural cable, which transmits the spike to other neurons). Each compartment is simply treated as a three-dimensional neural cable but exists within a three-dimensional space, implying spatial extent along its length [661].

### 3.2.2 Neural Cable Equation

This is the fundamental law of biophysics for neurons, describing how voltage $V$ changes over time $t$ and space $x$ along a compartment [3]. The neural cable equation is expressed as:

$$c_m \frac{\partial V}{\partial t} = \frac{a^2}{2R_i} \frac{\partial^2 V}{\partial x^2} - I_{ion} - I_{syn} + \eta(t)$$

where $c_m$ is membrane capacitance density, $R_i$ indicates intracellular resistivity, and $\eta(t)$ denotes membrane current noise.

Here, the right-hand side represents the current required to charge the membrane capacitance. The left-hand side represents the rate of change of voltage. $\frac{a^2}{2R_i}\frac{\partial^2 V}{\partial x^2}$ is the flow of charge from neighboring points along the inside of the neural cable. $a$ is the radius, $R_i$ is the intracellular resistivity. The second derivative $\frac{\partial^2 V}{\partial x^2}$ means voltage changes smoothly along the neural cable. Current $I_{ion}$ flows through ion channels in the membrane (e.g., sodium, potassium). Current $I_{syn}$ is injected by synapses. Random fluctuations $\eta(t)$ represent thermal noise and channel noise [172].

### 3.2.3 Compartment Coupling Conditions

The boundaries where different compartments meet (e.g., soma to AIS) must obey the laws of physics [661]. Voltage continuity $V_{comp1} = V_{comp2}$ implies that the voltage is the same on both sides of the junction. Current conservation $I_{comp1} + I_{comp2} = 0$ means that all current leaving one compartment must enter the next (Kirchhoff's Current Law). This ensures the model is electrically tight.

### 3.2.4 Ionic Currents and Compartment-Dependent Hodgkin-Huxley Formulation

The classic Hodgkin-Huxley (HH) model $I_{ion} = g_{Na}m^3h(V - E_{Na}) + g_K n^4(V - E_K) + g_L(V - E_L)$ defines the voltage-dependent conductances for sodium (Na) and potassium (K) that generate the action potential [343]. This model spatially restricts these channels to where they are biologically found.

$I_{Na} = g_{Na}m^3h(V - E_{Na})$ is fast sodium current (responsible for the upstroke of the spike). $I_K = g_K n^4(V - E_K)$ is delayed rectifier potassium current (responsible for the down stroke/repolarization). $I_L = g_L(V - E_L)$ is leak current (always present).

In spatial restrictions, dendrites $r = d$: usually lack voltage-gated sodium channels, making them passive or weakly active. They filter and integrate inputs but do not generate spikes. Axon $r = a$: once the spike is initiated, the axon passively propagates it, acting as a neural cable. It does not need to regenerate the spike. The spike must initiate in the AIS $r = AIS$, because that is the only place (in this model) with a high density of both sodium and potassium channels to generate the rapid voltage change [447].

### 3.2.5 Gating Variable Dynamics

The gates $m$, $h$, $n$ in the HH model are protein subunits that open and close probabilistically [343]. They are governed by the equation:

$$\frac{dm}{dt} = \alpha_m(V)(1-m) - \beta_m(V)m + \xi_m(t)$$

The rate of change of the probability that a gate is open $m$ depends on voltage-dependent rate constants $\alpha_m(V)$ and $\beta_m(V)$. The other three terms are similar. The term $\xi_m(t)$ represents channel noise the random opening and closing of individual ion channels, which is significant in small structures like the AIS [858].

### 3.2.6 Excitatory and Inhibitory Synaptic Currents

Synapses are the points of communication between neurons. They are chemically mediated and are either excitatory (depolarizing, making a spike more likely) or inhibitory (hyperpolarizing, making a spike less likely) [141]. The currents are calculated separately to maintain biological realism.

Excitatory synaptic current $I_{syn,E} = g_E(t)(V - E_E)$, where $E_E \approx 55$ mV (AMPA/NMDA type), means the reversal potential is above the firing threshold, so opening these channels lets in positive charge [141].

Inhibitory synaptic current $I_{syn,I} = g_I(t)(V - E_I)$, where $E_I \approx -75$ mV (GABA$_A$ type), is close to the resting potential, so opening these channels clamps the voltage down (shunting inhibition) [141].

### 3.2.7 Synaptic Conductance Dynamics

When a presynaptic spike arrives, it does not turn the synapse on instantly. It triggers a waveform of conductance change [141].

Excitatory and inhibitory equations follow:

$$g_E(t) = \bar{g}_E \sum_i w_i \left[ e^{-(t-t_i-d)/\tau_1} - e^{-(t-t_i-d)/\tau_2} \right] + \xi_{rel}(t)$$

$$g_I(t) = \bar{g}_I \sum_j w_j \left[ e^{-(t-t_j-d)/\tau_3} - e^{-(t-t_j-d)/\tau_4} \right] + \xi_{rel}(t)$$

The conductance decays exponentially with two time constants $\tau_1$ and $\tau_2$. Every time a presynaptic spike occurs ($\delta(t-t_i)$ is the Dirac delta function), it is multiplied by weights $w_i$, and added to the conductance after an axonal delay $d$. $\xi_{rel}(t)$ is release noise (the probabilistic failure of neurotransmitter release) [108].

### 3.2.8 Synaptic Plasticity

The connection strengths $w_i$ are not fixed. They change based on the activity of the pre- and postsynaptic neurons [88].

Excitatory STDP (Spike-Timing-Dependent Plasticity): If the presynaptic spike (pre) comes just before the postsynaptic spike (post), the synapse is strengthened (Long-Term Potentiation, LTP). If the order is reversed, it is weakened (Long-Term Depression, LTD) [87]. Excitatory STDP: $\Delta w_E = \sum_{t_{pre}} \sum_{t_{post}} w_E(t_{post} - t_{pre})$.

Inhibitory plasticity: This is more complex and can be anti-Hebbian or depend on the postsynaptic firing rate [837]. Inhibitory plasticity: $\Delta w_I = \sum_{t_{pre}} \sum_{t_{post}} w_I(t_{post} - t_{pre})$.

### 3.2.9 Spike Definition (Intrinsic)

This is a critical point for biological realism. In simplified models, the voltage is artificially reset after a spike. In a real neuron, the spike is a natural consequence of the ion channel dynamics [343].

A spike is registered when the voltage at a specific point in the AIS $r = AIS$ and at time $t$ crosses a threshold $V_{th}$ and the voltage is rising rapidly $\frac{dV}{dt} > 0$. After this, the sodium channels inactivate, and the potassium channels activate, bringing the voltage back down naturally no artificial reset is required.

### 3.2.10 Noise Sources

The model explicitly includes three distinct sources of biological randomness, making its behavior stochastic and realistic [172]. Membrane current noise $\eta(t)$ represents thermal fluctuations. Channel noise $\xi_m(t)$ represents stochastic gating of ion channels. Synaptic release noise $\xi_{rel}(t)$ represents probabilistic vesicle release.

### 3.2.11 Structural Consistency

This final section is a checklist, confirming that the model is built from the ground up using physical laws, not mathematical shortcuts, ensuring it is a candidate for simulating neural activity.

### 3.2.12 A Practically Inoperable Complete Biophysical Model of the Brain

Dimensional Intractability: A neuron modeled rigorously contains multiple dendritic compartments, axon initial segment (AIS), axonal compartments, individual spines for each excitatory synapse, short-term plasticity variables, ion channel gating variables, and stochastic noise sources. Even a single neuron therefore requires tens of thousands of state variables. Scaling to a cortical patch or full brain with hundreds of thousands to billions of neurons and millions of spines per local network makes the total number of dynamic variables astronomically large, far exceeding any computational capability. This is not large in a superficial sense it is structurally unmanageable. No storage, update, or synchronization mechanism exists to handle all these variables in practice [537].

Temporal Scale Separation: The system operates on multiple tightly coupled time scales: sub-millisecond for ion channel activation, milliseconds for membrane potential, tens to hundreds of milliseconds for short-term synaptic plasticity, and minutes to hours for long-term plasticity. Because these processes interact, no single time step is appropriate, and standard numerical integration fails. The system is extremely stiff, making numerical simulation impractical even for a single neuron [378].

Parameter Non-identifiability: A fully detailed model requires precise knowledge of ion channel densities for each compartment, spine neck resistances and capacitances, synaptic release probabilities, short-term plasticity parameters, and noise amplitudes [221]. Most of these parameters cannot be measured in vivo and vary across individual neurons. Without parameter identifiability, the model cannot be calibrated, making its predictions scientifically meaningless [221].

Observability Limitations: Experimentally, we can observe spikes (action potentials), local field potentials, calcium signals, or limited voltage imaging [507]. But the internal state includes ion channel gating variables, spine head voltages, synaptic resource dynamics, and stochastic realizations. These internal states cannot be inferred from available measurements. Hence, even if the system is formally complete, we cannot verify or validate it [507].

Functional vs. Physical Scale Mismatch: Brain function emerges at population or network scales, statistical or pattern levels, and cognitive behaviors like memory or decision-making [808]. The complete model describes individual ion channel noise and microvolt fluctuations in spine heads. These microscopic details are mostly irrelevant to the emergent computational function. Including them adds enormous redundancy without functional benefit [808].

Computational Collapse: A usable model must be computable in practice, verifiable against data, predictive, and analytically tractable. A fully detailed, biophysical brain model is computationally intractable, contains unmeasurable parameters and unobservable internal states, and cannot be analyzed or reduced without losing almost all emergent functionality. Hence, it fails the criteria for practical modeling [537].

Fundamental Reason: The issue is not mere complexity. Completeness at the microscopic scale does not guarantee usefulness at the functional scale. Once every physical detail is included, the model essentially becomes the system itself, rather than a simplified representation. A model's purpose is dimensional reduction, providing insight and predictive power. Without reduction, it ceases to be a usable model [488].

A fully detailed, biophysically accurate model of the brain is mathematically well-posed, biophysically consistent, and formally complete. But in operational terms, it cannot be simulated, parameters cannot be determined, internal states cannot be observed, and functional behavior cannot be analyzed. Therefore, formal completeness does not imply practical feasibility. The model is complete in principle but inoperable in practice.

### 3.3 Noise and Error Coupling

The human brain, with its approximately 86 billion neurons and quadrillions of synaptic connections, represents the most intricate system known in the universe [335]. A fundamental paradox underpins its operation: this pinnacle of biological computation functions not through the pristine, error-free signaling characteristic of digital computers, but by harnessing an environment saturated with noise [172]. This inherent stochasticity, far from being a flaw to be eliminated, is deeply woven into the fabric of neural computation, posing profound mathematical challenges for analysis.

Neural noise manifests at every scale, from the quantum jitter of ion channels to variability in neurotransmitter release and fluctuations in membrane potential [172]. Even when presented with identical stimuli, the precise timing and pattern of action potentials (spikes) from a given neuron can vary dramatically across trials [724]. This variability is not merely random error; it constitutes an intrinsic property of biological hardware [523]. Modeling such systems requires moving beyond deterministic equations to embrace probabilistic frameworks such as stochastic differential equations or Markov processes [1]. However, this transition exponentially increases complexity. The central challenge lies in disentangling signal from noise determining which variations convey meaningful information about stimuli or internal states, and which represent mere epiphenomena of biological messiness [879]. Analyses failing to account for noise coupling may misinterpret dynamic range limitations as pathological conditions or overlook critical computational roles of variability.

This challenge is further compounded by the nonlinear nature of neural dynamics. Small amounts of noise can trigger significant shifts in network behavior at bifurcation points, where tiny perturbations lead to qualitatively different stable states [378]. Such phenomena render predictive modeling exceptionally difficult, as errors propagate non-additively. Furthermore, what appears as noise at one observational level might constitute the signal at another [17]. The seemingly erratic firing patterns of individual neurons could reflect a population code wherein collective activity carries information, rendering single-unit variability functionally irrelevant [318]. Consequently, successful mathematical analysis must carefully define its operational level. Attempts to model whole-brain dynamics using high-fidelity noisy components result in computationally intractable models, while overly simplified models risk losing essential biological truths about how noise shapes perception, decision-making, and learning [85]. The pervasive presence of noise thus forces analysts to grapple with trade-offs between realism and tractability, making development of robust analytical tools for noisy neural systems one of the foremost frontiers in computational neuroscience.

### 3.4 Neuronal Individuality, Heterogeneity, and Mismatch Introduce Analytical Complexity

A foundational assumption in many early neural network models was homogeneity: that neurons within a given class were essentially identical functional units [350]. Modern neuroscientific research has decisively overturned this notion, revealing a staggering degree of neuronal individuality and heterogeneity [422]. This diversity exists across multiple dimensions: morphological (dendritic arborization, soma size), electrophysiological (firing rates, thresholds, adaptation properties), molecular (receptor subtypes, ion channel expression), and connective (pre- and postsynaptic partners) [536]. This heterogeneity is not random; it is a feature sculpted by evolution and experience to enhance computational capacity [227]. Inhibitory interneurons, for instance, exhibit remarkable specialization, with distinct subclasses such as parvalbumin-positive, somatostatin-positive, and vasoactive intestinal peptide-positive cells, each playing unique roles in shaping cortical dynamics and enabling complex cognitive functions [418]. Failure to account for this biological reality introduces immense analytical complexity into any mathematical model.

Modeling such diverse populations necessitates abandoning simple mean-field approximations that treat large neuronal groups as uniform ensembles [2]. Instead, researchers must develop heterogeneous population models, vastly increasing the number of parameters and state variables [91]. This shift moves the field toward high-dimensional dynamical systems, where traditional analytical methods often falter. Moreover, mismatch between neurons wherein two cells intended to perform similar functions exhibit divergent properties can lead to unexpected emergent behaviors in networks [210]. Slight differences in axonal conduction velocity or synaptic delay between parallel pathways can generate oscillatory dynamics or phase shifts critical for information processing yet absent in idealized models [450]. These biological imperfections become integral to function. Consequently, accurate analysis must incorporate these variations explicitly, requiring sophisticated statistical descriptions and simulations capable of handling large-scale heterogeneity [91]. Failure to do so yields mathematically elegant but biologically naive models, unable to replicate the robustness and flexibility observed in real brains.

**3.5 Impact of Diversity on Modeling and Interpretation**

The profound impact of neural diversity extends beyond model construction to fundamentally shape interpretation [532]. Models based on homogeneous assumptions tend to produce clean, interpretable outputs with clear cause-and-effect relationships [130]. In contrast, models incorporating realistic heterogeneity yield messy, variable results that better mirror empirical data but are significantly harder to interpret [532]. The relationship between model parameters and behavioral outputs becomes blurred by interactions among thousands of variables representing individual differences [532]. This renders reverse engineering of brain function from data exceptionally challenging. When analyzing experimental recordings, scientists cannot assume that all neurons of a particular type will respond identically to stimuli [414]. Variability in response profiles could stem from underlying biological diversity rather than measurement error or changing network context.

This has direct implications for fields such as neuromorphic engineering and artificial intelligence [686]. AI systems inspired by the brain have traditionally employed uniform artificial neurons, achieving impressive performance but lacking the fault tolerance and energy efficiency of their biological counterparts [471]. The realization that heterogeneity confers robustness suggests that future AI generations should incorporate more diverse computational elements [321]. However, doing so complicates design and debugging processes, mirroring challenges faced in neuroscience [184]. Interpreting activity of diverse artificial neural networks is as difficult as interpreting the brain itself. Thus, embracing diversity forces a paradigm shift from seeking simple, universal principles toward understanding principles of distributed, redundant, and adaptive computation [808]. It demands new metrics and visualization techniques to extract meaning from complex, high-dimensional datasets generated by both real and simulated neural systems.

### 3.6 Unpredictability of Connectivity

At the heart of neural analysis lies the connectome the comprehensive map of neural connections [760]. Despite advances in electron microscopy and tracing techniques, predicting the precise wiring diagram of even a small brain region remains largely impossible [485]. Connectivity is shaped by complex interplay of non-uniform three-dimensional genetic programs, developmental cues, and lifelong experience-dependent plasticity [404]. While general anatomical rules exist (e.g., certain areas are reciprocally connected), the specific synapses formed between individual neurons appear highly idiosyncratic [755]. This unpredictability stems from the combinatorial explosion of possible connections: with billions of neurons each forming thousands of synapses, potential connectomes are astronomically numerous [149]. This lack of predictability renders purely deductive modeling futile. Analysts cannot start from first principles to derive exact network structure.

Instead, research relies heavily on empirical data collection, which is slow, expensive, and often limited to small tissue volumes [331]. Even when data are available, sampling bias constitutes a major concern [565]. Mathematical analyses must therefore work with incomplete and uncertain knowledge of underlying architecture [432]. This forces reliance on generative models that create plausible network topologies based on statistical rules derived from partial data [833]. However, validating these models against ground truth is extraordinarily difficult [55]. The unpredictable nature of connectivity means that seemingly minor wiring differences can lead to drastically different network dynamics and functional outcomes a phenomenon known as sensitivity to initial conditions [753]. This unpredictability is a core reason why replicating findings across individuals or species is so challenging, and underscores the importance of studying neural circuits as emergent properties of complex systems rather than predetermined blueprints.

### 3.7 Excessive System Complexity

The sheer scale of the nervous system creates excessive complexity that defies complete analytical description [440]. The number of interacting components neurons, glia, synapses, signaling molecules is so vast that simulating a human brain at full resolution is currently far beyond the capabilities of any supercomputer [537]. This complexity arises not just from the number of parts but from the nonlinear, recursive, and multi-scale interactions between them [92]. Processes unfold simultaneously across temporal scales, from millisecond spike events to long-term structural changes over years, and spatial scales, from nanometer-sized receptors to meter-long axons [760]. Integrating these disparate levels into a unified mathematical framework constitutes a monumental task.

Attempts to simplify by reducing dimensionality inevitably sacrifice important details [3]. Coarse-graining networks by grouping neurons into populations loses information about local microcircuits [668]. Conversely, focusing on detailed microcircuitry makes it impossible to observe global brain states [109]. This complexity leads to emergence of properties such as chaos and self-organization, which are notoriously difficult to analyze and control [378]. Because of this, much neural analysis operates in the realm of effective theories models that capture salient system features without claiming to represent every detail [264]. However, determining which features are salient requires careful experimentation and insight, highlighting the tight coupling between theory and experiment in this field [408]. Ultimately, excessive complexity means that complete mathematical understanding of the brain may remain forever out of reach, forcing scientists to focus on specific subsystems or functions.

### 3.8 Interference and Distortion in Neural Processes

Neural signals do not travel through isolated channels; they propagate through dense, interconnected networks where interference and distortion are inevitable [383]. Electrical fields generated by one group of active neurons can influence excitability of neighboring neurons a phenomenon known as ephaptic coupling [19]. Similarly, neurotransmitters released into synaptic clefts can diffuse and bind to receptors on nearby, unintended targets, causing spillover effects [897]. These forms of interference corrupt the fidelity of information transmission. From a mathematical standpoint, modeling such crosstalk requires incorporating additional terms for lateral inhibition, volume transmission, and electromagnetic interactions, which rapidly increase model complexity and are rarely included in standard neural network models [162].

#### 3.8.1 Impact of Interference on Analytical Relationships

Interference disrupts the assumed independence of neural signals a common assumption in many analytical techniques such as correlation analysis [55]. If neuron A's activity influences neuron B not through direct synaptic connection but via electrical field effects, analyses might incorrectly infer direct functional links [657]. This leads to erroneous conclusions about network topology and causal relationships [623]. Techniques designed to detect functional connectivity, such as Granger causality, can be misled by such indirect interactions [720]. To mitigate this, advanced statistical methods like partial coherence or conditional Granger causality are required to isolate direct from indirect effects [144]. However, these methods demand even more data and computational power, and their accuracy depends on correctly specifying model order [855]. The presence of interference thus necessitates more sophisticated analytical approaches and a healthy skepticism toward simple interpretations of correlated activity.

#### 3.8.2 Distortion in Neural Signal Transmission

Signal distortion occurs throughout neural pathways [190]. Synaptic transmission is inherently unreliable, with release probability typically less than one, leading to stochastic failures [778]. Axons can act as low-pass filters, attenuating high-frequency signal components [660]. Dendrites actively process incoming inputs through voltage-gated ion channels, transforming signals before they reach the soma [508]. This means information encoded in spike trains at axon hillocks is not a faithful copy of original sensory input [190]. Modeling this distortion requires detailed biophysical models of individual neurons, such as multi-compartmental neural cable models [661]. While powerful, these models are computationally intensive and impractical for large-scale simulations [537]. As a result, simpler point-neuron models dominate, accepting the limitation that they distort the very processes they aim to describe [281]. This gap between biological reality and mathematical abstraction represents a significant barrier to understanding how information is preserved or transformed during neural processing.

### 3.9 Complex Changes Over Time, Age, and Environment

Neural systems are not static; they undergo continuous change over multiple timescales [346]. On short timescales, synaptic strength fluctuates due to short-term plasticity mechanisms such as facilitation and depression [898]. On longer timescales, structural plasticity enables growth and retraction of dendritic spines and axons [811]. Across the lifespan, the brain matures through well-defined developmental stages involving massive synaptic pruning and myelination, then ages with associated cognitive decline [367]. Furthermore, the environment continuously shapes the brain through learning and experience [73]. This dynamic nature presents a severe challenge for mathematical analysis, which often seeks stable, time-invariant solutions [655].

Models built from data collected at single time points provide only snapshots of perpetually changing systems [189]. Longitudinal studies are required to capture these dynamics, but they are logistically challenging and expensive [12]. Mathematically, incorporating time-varying parameters transforms ordinary differential equations into non-autonomous systems, which are much harder to analyze [218]. Capturing lifelong changes requires integrating processes across vastly different timescales from milliseconds to decades within a single coherent framework [583]. Additionally, environmental influences introduce uncontrolled variables that can confound analyses [823]. Models trained on data from one environment may fail to generalize to another, reflecting the brain's remarkable adaptability [273]. Therefore, any robust analytical approach must embrace the principle of plasticity, treating the brain not as a fixed machine but as a dynamic, evolving system whose past experiences are embedded in its present structure.

### 3.10 Other Extensive Modeling Difficulties

Beyond challenges already discussed, numerous other difficulties plague neural modeling [648]. One major issue is parameter estimation [827]. Biophysical models contain hundreds of parameters (conductances, capacitances, reversal potentials) that are difficult to measure directly and can vary significantly between cells [204]. Fitting these models to data constitutes an ill-posed inverse problem, often resulting in multiple parameter sets producing similar outputs a phenomenon known as non-identifiability [221].

Another difficulty is the curse of dimensionality: as models grow in complexity, the volume of data needed to adequately sample state space grows exponentially, quickly becoming unattainable [49]. Validation of models against experimental data is also fraught with problems, as no single experiment captures all relevant aspects of brain function [391].

Finally, a persistent gap exists between levels of explanation linking molecular events to cellular dynamics, to network behavior, and finally to cognition and behavior [539]. Bridging these gaps requires integrative multiscale models, which combine incompatible formalisms and face immense computational hurdles [443]. Collectively, these extensive difficulties underscore that mathematical analysis of the brain remains an immature science, driven more by ingenuity and approximation than by rigorous, predictive theory [488].

### 3.11 The Principle of Measurement: What Can Be Observed

A midst these mathematical challenges, a fundamental principle emerges: learning can only utilize quantities that are genuinely measurable in the system [672]. These measurable quantities are not idealized mathematical abstractions but physical signals with specific characteristics:

- Spikes: Action potentials are all-or-none events that can be detected as binary pulses [672].
- Temporal sequences: The relative timing of spikes carries information [835].
- Statistical correlations: Relationships between spike trains can be computed [1].
- Noise-corrupted signals: All measurements are contaminated by multiple noise sources [172].
- Multi-source mixed information: Signals from different sources are superimposed and entangled [128].

The brain's learning mechanisms operate on these measurable quantities and natural connection strength perturbations directly, without requiring access to unobservable internal states or mathematically convenient abstractions [283]. This principle has profound implications: if a quantity cannot be measured by the system, it cannot be used for learning. Conversely, any learning theory that requires unmeasurable quantities (such as precise error gradients at every synapse) is biologically implausible [492].

This measurement principle aligns with the brain's actual architecture. Neurons detect spikes arriving at their synapses, integrate them over time, and generate output spikes based on threshold crossings [508]. All information available to a neuron is contained in the temporal patterns of input spikes and its own recent activity history. No neuron has access to global error signals, loss functions, or gradient information [492]. Learning must therefore be local, using only locally available measurements [2].

### 3.12 Summary: The Principled Failure of Mathematical Modeling

This chapter has systematically discussed the challenges of mathematical modeling at the brain level. From the inoperability of complete biophysical models, to noise-error coupling, to the analytical complexity introduced by neuronal individuality, to the unpredictability of connectivity, to excessive system complexity, to the pervasiveness of interference and distortion, to continuous change over time all these factors converge on a single conclusion: the path of mathematical modeling, at the brain level, is inherently impassable [488].

This is not a technical limitation awaiting more powerful computers or better algorithms; it is a structural limitation [488]. For systems that are sufficiently complex, highly non-uniform, and possess coupled non-ideal factors, analytically closed global mathematical models simply do not exist [114]. Even if local models can be constructed, their global composability remains questionable [114]. The expressive power of mathematical language has fundamental boundaries, and the brain lies beyond these boundaries [488].

This recognition has profound implications for artificial intelligence: attempting to replicate brain intelligence through mathematical modeling may be methodologically infeasible [879]. Brain intelligence is not designed but emerges from diversity, noise, and chaos [59]. We cannot manufacture brain intelligence; we can only cultivate it by constructing sufficiently rich, sufficiently open physical systems within which brain intelligence can grow naturally [276]. The measurable quantities that brains actually use spikes, timing, correlations, noise point toward a different approach: learning through statistical alignment rather than mathematical optimization [283].

### 3.13 The Principle of Measurability and Its Implications

### 3.13.1 What Brains Can Measure

The mathematical challenges outlined above lead to a fundamental constraint: any learning system, whether biological or artificial, can only utilize quantities that are genuinely measurable by that system [672]. For the brain, these measurable quantities are:

1. Spike events: Action potentials are binary events that can be detected with high temporal precision [672]. Each neuron knows when it fires and when it receives input spikes.

2. Spike timing: The relative timing between presynaptic and postsynaptic spikes is locally measurable at each synapse [87]. This forms the basis for STDP.

3. Local correlations: A neuron can compute correlations between its input spike trains and its output spikes over time [1]. These correlations are locally available statistics.

4. Noise-contaminated signals: All measurements are corrupted by multiple noise sources synaptic release failures, ion channel fluctuations, thermal noise [172]. The brain cannot separate signal from noise in the way engineers might desire; it must work with the noisy signals directly.

5. Multi-source mixed information: Inputs from different sources arrive superimposed at the neuron [128]. The neuron cannot perfectly demix them; it must process the combined signal.

6. Broadcast neuromodulatory signals: Global signals such as dopamine, serotonin, and norepinephrine provide evaluative feedback that is broadcast widely and can be detected by neurons [705]. These signals convey information about outcomes (reward, punishment, arousal) but not detailed error gradients.

Notably absent from this list are quantities that mathematical optimization theories often assume: error gradients, loss functions, loss function values, and precise credit assignments across multiple layers [492]. These quantities cannot be measured by any known biological mechanism and therefore cannot be used for learning.

### 3.13.2 Implications for Learning Theory

The measurability principle has profound implications for theories of learning:

First, learning must be local [2]. Each synapse can only access information that is locally available: presynaptic activity, postsynaptic activity, and possibly diffuse neuromodulatory signals that arrive through volume transmission. Global error signals cannot be routed to individual synapses with the precision required for gradient-based methods [492].

Second, learning must be statistical rather than deterministic [283]. Because all measurements are noise-corrupted, learning rules must extract statistical regularities from noisy data rather than computing precise updates. This aligns with the observation that biological learning exhibits trial-to-trial variability but converges to adaptive behavior over time [1].

Third, learning must exploit rather than eliminate noise [549]. Since noise cannot be eliminated from measurements, the brain has evolved to use noise constructively for exploration, for escaping local optima, for enabling probabilistic inference [518]. This contrasts sharply with engineering approaches that seek to minimize noise at all costs.

Fourth, learning requires temporal integration [378]. Because individual measurements are noisy and unreliable, the brain must integrate over time to extract reliable signals. This manifests in phenomena such as evidence accumulation in decision-making [209] and the slow time course of synaptic plasticity induction [95].

Fifth, global guidance must be broadcast rather than routed [705]. The brain solves the credit assignment problem not by computing and transmitting precise error gradients, but by broadcasting coarse evaluative signals (dopamine, serotonin) that modulate plasticity at all recently active synapses [184]. This three-factor learning rule (pre, post, neuromodulator) is biologically plausible and can approximate gradient descent under certain conditions [259].

### 3.13.3 The Brain as a Measurement-Constrained System

Viewing the brain through the lens of measurability reveals it as a system that has evolved to operate under severe informational constraints. Unlike artificial neural networks that assume access to global gradient information, the brain must learn using only locally available, noise-corrupted measurements [492]. This constraint, far from being a limitation, may be the source of the brain's robustness, adaptability, and energy efficiency [172].

The brain's learning mechanisms Hebbian plasticity, STDP, neuromodulatory modulation, homeostatic scaling can be understood as solutions to the problem of learning under measurability constraints [283]. These mechanisms do not compute mathematical abstractions; they implement statistical learning rules that align the system's behavior with environmental contingencies over time [884].

### 3.13.4 Implications for AI Design

The measurability principle offers guidance for designing more brain-like AI systems:

· Design for local learning: Architectures should enable learning rules that use only locally available information, avoiding dependence on globally computed gradients [375].

· Embrace measurement reality: Training algorithms should work with the kinds of measurements that are actually available in physical implementations binary spikes, timing, local correlations rather than assuming access to precise real-valued gradients [689].

· Use noise constructively: Rather than eliminating noise from measurements, design systems that exploit noise for exploration and probabilistic inference [549].

· Broadcast global signals: Use coarse, broadcast signals (analogous to dopamine) to provide global guidance, combined with local eligibility traces to solve credit assignment over time [184].

· Integrate over time: Build temporal integration mechanisms that extract reliable signals from noisy measurements [378].

These principles point toward a different paradigm for AI one based on statistical alignment through local learning rules rather than mathematical optimization through global gradient descent [884].

## 3.14 Quantitative Analysis of Modeling Limitations

### 3.14.1 State Space Dimension

For a complete biophysical model of a single neuron, the number of state variables can be estimated as:

$$N_{state} = N_{comp} \times (N_{ion} + N_{syn} + N_{plasticity})$$

where $N_{comp}$ is the number of compartments, $N_{ion}$ is the number of ion channel types, $N_{syn}$ is the number of synapses, and $N_{plasticity}$ accounts for plasticity variables. For a typical pyramidal neuron with 100 compartments, 10 ion channel types, 1000 synapses, and short-term plasticity variables, $N_{state} \approx 100 \times (10 + 1000 \times 2) \approx 200,000$. state variables per neuron.

Scaling to the entire brain with $86 \times 10^9$ neurons gives:

$$N_{brain} \approx 86 \times 10^9 \times 2 \times 10^5 \approx 1.72 \times 10^{16} \text{ state variables}$$

This is approximately 17 quadrillion state variables far beyond any conceivable computational capability.

### 3.14.2 Parameter Identifiability

The number of parameters in a complete brain model scales similarly. For a model with $P$ parameters, the Fisher information matrix **F** determines identifiability. A model is unidentifiable if **F** is singular or ill-conditioned. The condition number cond(F) for brain models is typically cond(F)>10^10, meaning that parameters cannot be reliably estimated from any feasible amount of data [221].

### 3.14.3 Observability

The observability Gramian $O$ determines which states can be inferred from measurements. For a system with $n$ states and $m$ measurements, the rank of $O$ must be $n$ for full observability. For brain models, typical measurements (e.g., extracellular spikes) provide far fewer observables than states: $\text{rank}(O) \ll n$, implying that most internal states are unobservable [507].

### 3.14.4 Temporal Scale Separation

The ratio of fastest to slowest time scales in neural systems spans approximately:

$$\frac{\tau_{max}}{\tau_{min}} \approx \frac{10^4 \text{ s (long-term plasticity)}}{10^{-4} \text{ s (ion channel activation)}} = 10^8$$

This extreme stiffness makes numerical integration prohibitively expensive, as the time step must resolve the fastest dynamics while integrating over the slowest scales [378].

### 3.14.5 Summary of Quantitative Limitations

Limitation of Quantitative Measure: Implication state space dimension $\approx 10^{16}$ variables. Computationally intractable Parameter count $\approx 10^{15}$. Unidentifiable condition number $\kappa > 10^{10}$.

Ill-posed inverse problem: Observability rank $\text{rank}(O) \ll n$. Internal states are unobservable. Time scale ratio is $10^8$. Stiffness prohibits integration

These quantitative measures confirm that the limitations of mathematical modeling are not merely technical but fundamental. The brain's complexity exceeds the capacity of any conceivable mathematical description, and learning must therefore proceed through alternative means using locally available, noise-corrupted measurements to achieve statistical alignment with environmental contingencies.

## 3.15 Conclusion: The Principled Failure of Mathematical Modeling

This chapter has systematically argued that mathematical modeling, as traditionally conceived, faces principled limitations at the brain level [488]. The combination of excessive complexity, heterogeneity, non-ideal factors, and measurability constraints renders complete mathematical descriptions impossible [114]. This is not a temporary limitation awaiting more powerful computers or better algorithms; it is a structural limitation rooted in the nature of neural systems [488].

The brain does not compute using mathematical abstractions; it learns using locally available, noise-corrupted measurements [283]. Its learning mechanism  mechanisms Hebbian plasticity, STDP, neuromodulation are statistical alignment procedures that gradually adjust the system's behavior to match environmental contingencies [884]. These mechanisms do not require mathematical models, global loss functions, or precise error gradients [492].

This recognition has profound implications for artificial intelligence. Attempting to replicate brain intelligence through mathematical optimization may be fundamentally infeasible [879]. Brain intelligence is not designed; it emerges from diversity, noise, and statistical alignment in sufficiently rich physical systems [59]. The path toward artificial general intelligence may lie not in more sophisticated mathematical models, but in creating systems that can learn from measurable quantities in the same way brains do through local, statistical, noise-exploiting mechanisms [276].

# Chapter 4: Learning Principles: Brain Learning Without Mathematical Analytical Relations

## 4.1 Introduction: The Biological Foundation of Learning

Brain learning the biological process underlying all forms of adaptation and knowledge acquisition is fundamentally rooted in neuroplasticity: the brain's capacity to physically restructure itself in response to experience [408]. At its core, learning involves dynamic adjustment of synaptic strength (the connections between neurons)a process known as synaptic plasticity [526]. When an individual learns a new skill, such as playing a musical instrument, or acquires new information, such as memorizing a poem, specific neural pathways are activated [434]. Repeated activation strengthens these connections, rendering signal transmission faster and more efficient a principle famously described by Donald Hebb as cells that fire together, wire together [329]. The Hebb rule biases connection strengths toward directions of maximum information, effectively performing principal component analysis [57]. This long-term potentiation (LTP) constitutes the physical basis of memory [95]. Conversely, rarely used connections undergo long-term depression (LTD) and are eventually pruned [515]. This constant remodeling of brain circuitry enables lifelong learning and adaptation [346].

However, mechanisms governing this process are far more sophisticated than simple Hebbian rules suggest [776]. Groundbreaking research has revealed that learning is not governed by any single universal principle [882]. In a paradigm-shifting discovery, studies employing advanced imaging to track individual synapses in mice found that within single neurons, different dendritic branches can follow entirely different learning rules [216]. While some synapses adhere to the classic Hebbian model, requiring coordinated input and output, others operate independently, changing strength without regard to the neuron's overall firing pattern [296]. This multi-rule learning enables individual neurons to function as complex, multi-functional processors, capable of simultaneously integrating, filtering, and storing different information types [107]. This finding overturns long-held assumptions of monolithic learning mechanisms and suggests the brain employs a diverse toolkit of plasticity rules to handle multifaceted environmental demands [882].

This intricate learning process is tightly regulated by attention and biological rhythms [108]. Learning is not passive stimulus absorption but an active process requiring focused attention [140]. When attention is directed toward sensory input, it amplifies corresponding neural signals, marking them as salient and worthy of processing [512]. Sleep's role is equally critical. The hippocampus acts as temporary storage for new memories, but during deep and REM sleep, these memories are replayed and transferred to neocortex for long-term storage a process essential for memory consolidation [865]. Sleep deprivation severely impairs this process, leading to rapid forgetting [844].

Modern research has uncovered additional learning modulators. The vagus nerve a key communication pathway between brain and body can act as a learning wake-up call [640]. Activities stimulating the vagus nerve slow, deep breathing, cold exposure, humming, or meditation enhance neural plasticity, effectively priming the brain for heightened learning readiness [444]. This insight from Dr. Michael Kilgard's work demonstrates that learning efficiency depends not just on duration but on timing and state: when salient stimuli coincide with periods of high attention and neurophysiological states conducive to plasticity, the brain automatically enters its strongest memory mode, optimizing credit assignment and solidifying neural changes that constitute learning [426].

## 4.2 Unavailability of Mathematical Models

The foundation of classical artificial intelligence and machine learning lies in the formulation of explicit mathematical models, wherein relationships between variables are defined by precise equations and functions [471]. This paradigm assumes that a system's behavior can be accurately predicted or optimized through analytical expressions derived from first principles or learned from data via gradient descent [690]. Thus, artificial intelligence exhibits black-box characteristics when facing the external world, but it displays white-box attributes when dealing with itself. However, the biological brain operates under fundamentally different constraints, rendering such traditional mathematical modeling largely inapplicable to understanding its core learning mechanisms. The unavailability of comprehensive mathematical models for neural learning arises not from lack of effort but from intrinsic properties of the nervous system: its staggering complexity, nonlinearity, and reliance on emergent dynamics rather than centralized computation [441]. Hence, the brain exhibits black-box characteristics both when facing the external world and when facing itself. Consequently, the internal world of the brain exhibits greater proximity to the external world.

Neurobiological evidence indicates that the human brain contains approximately 86 billion neurons, each forming thousands of synaptic connections, culminating in an estimated $10^{15}$ synapses [335]. This scale alone presents a combinatorial explosion far beyond the capacity of any existing mathematical formalism to describe with precision [440]. Unlike engineered systems designed for transparency and predictability, the brain evolved through natural selection to prioritize functional robustness over theoretical elegance [431]. Consequently, its operations cannot be reduced to a set of closed-form equations governing global network states. While simplified models such as the Wilson-Cowan equations offer insights into population-level dynamics like attractor states or oscillatory behavior [864], they fail to capture the microscale heterogeneity observed across individual neurons and synapses [424].

Recent experimental findings further underscore this limitation. A groundbreaking study revealed that within a single neuron, distinct dendritic branches follow different plasticity rules during learning tasks: some synapses adhered to Hebbian principles, while others exhibited changes independent of neuronal firing patterns [216]. This multi-rule architecture defies conventional mathematical frameworks, which typically assume uniform update principles across all parameters [882]. Moreover, the dynamic nature of these interactions introduces time-varying dependencies that resist static modeling approaches [495]. Even advanced techniques such as bifurcation analysis and stochastic noise simulation provide only partial descriptions of local stability rather than complete predictive models [378].

Another critical factor is the role of molecular-scale processes in shaping synaptic efficacy. Mechanisms such as neurotransmitter release probability, receptor trafficking, and intracellular signaling cascades operate stochastically and are influenced by factors ranging from metabolic state to glial support [778]. These biochemical fluctuations introduce significant variability even among seemingly identical neurons, rendering deterministic predictions impossible [172]. Furthermore, developmental and environmental influences create unique neural fingerprints across individuals, ensuring no two brains implement learning identically [637]. As demonstrated by studies on perceptual-dependent circuit formation, early sensory experiences permanently alter connectivity patterns in ways that cannot be captured by generic equations [94].

Given these challenges, researchers have increasingly shifted focus toward qualitative and computational models that emphasize functional outcomes over analytical tractability [283]. Approaches rooted in nonlinear dynamical systems theory allow scientists to characterize broad classes of behavior such as bistability, hysteresis, or chaos without specifying exact parameter values [378]. Similarly, agent-based simulations enable exploration of collective phenomena emerging from simple interaction rules [167]. While these methods do not yield symbolic solutions, they facilitate hypothesis generation and experimental design in domains where formal mathematics falls short [530]. Thus, the absence of mathematical models does not imply ignorance; instead, it reflects a mature recognition that some systems must be understood through alternative epistemological frameworks grounded in observation, perturbation, and statistical inference [454].

## 4.3 Absence of Dynamic Logic and Mathematical Relationships

In engineered information processing systems, logical operations and state transitions are governed by well-defined rules encoded in hardware or software [651]. Boolean logic gates, finite-state machines, and algorithmic sequences ensure predictable progression from input to output based on explicit conditional statements [651]. In contrast, the brain lacks a centralized instruction set or universal syntax dictating how neural representations evolve over time [21]. Instead, its dynamic logic if one may use the term is distributed, probabilistic, and context-sensitive, arising spontaneously from the interplay of millions of concurrently active elements [311]. This absence of formal mathematical relationships means that neural computation cannot be described by discrete transition tables or truth functions, nor can it rely on error-correcting codes or check sums to maintain fidelity [455].

Evidence for this decentralized architecture comes from both anatomical and functional studies. Anatomically, cortical circuits exhibit massive convergence and divergence, with each neuron receiving inputs from diverse sources and projecting to multiple targets [316]. This architecture supports parallel processing and redundancy but undermines any attempt to map specific inputs directly onto outputs through linear causality [602]. Functionally, experiments using calcium imaging reveal that individual neurons participate in multiple overlapping ensembles depending on task demands [100]. GABAergic interneurons in the mouse dorsomedial prefrontal cortex, for example, show higher inter-individual correlation compared to glutamatergic neurons, suggesting specialized roles in social information routing without fixed logical assignments [475].

Moreover, the temporal structure of neural communication departs radically from digital computing paradigms. Rather than relying on clock-synchronized updates, neural events unfold at various timescales from sub-millisecond spikes to slow modulatory waves lasting seconds [130]. Spike-timing-dependent plasticity (STDP), a key mechanism underlying associative learning, depends critically on the relative timing of action potentials, yet the resulting weight changes are inherently noisy and subject to metaplastic regulation [57]. There is no equivalent to binary addition or multiplication; instead, integration occurs through graded membrane potentials, nonlinear summation at dendrites, and neuromodulator-gated plasticity windows [300]. These processes resist translation into standard arithmetic operations because their outcomes depend on history, location, and concurrent network activity [465].

The implications extend beyond basic computations to high-level cognitive functions such as decision-making and reasoning. Human judgments often violate axiomatic rationality, exhibiting biases such as loss aversion, anchoring effects, and framing dependence phenomena difficult to reconcile with normative models of logic [188]. Neuroeconomic studies demonstrate that choices emerge from competition between valuation systems (e.g., ventral striatum) and control networks (e.g., lateral prefrontal cortex), rather than being computed according to utility maximization formulas [314]. Even abstract thought involves pattern completion and analogical transfer rather than formal deduction, suggesting that cognition operates more like associative memory retrieval than theorem proving [838].

This absence of formal logic does not imply randomness or inefficiency. On the contrary, the brain achieves remarkable performance in uncertain environments precisely because it leverages statistical regularities rather than rigid algorithms [608]. By exploiting correlations across modalities and timescales, it constructs probabilistic internal models capable of prediction and adaptation [38]. Machine learning research has begun to recognize similar advantages in deep neural networks trained end-to-end, where intermediate layers develop representations not explicitly programmed but discovered through exposure to data distributions [894]. Nevertheless, unlike artificial networks whose weights are adjusted via backpropagationa process dependent on full gradient information the brain must navigate without access to such global derivatives [492].

Ultimately, the brain's dynamic logic appears better characterized as a self-organizing complex system operating near criticality than as a programmable Turing machine [731]. Its operations resemble those of a fluid medium responding to boundary conditions rather than a circuit executing instructions [104]. Understanding this requires shifting perspective from syntax to semantics, from discrete steps to continuous flows, and from certainty to uncertainty quantification [326]. Researchers now employ tools from statistical physics, information theory, and causal inference to probe these emergent properties, recognizing that while mathematical relationships may be absent, statistical dependencies abound [861].

**4.4 Impossibility of Mathematical Computation**

Despite decades of progress in computational neuroscience, the idea that the brain performs mathematical computation in the sense understood by mathematicians and computer scientists remains untenable [45]. Traditional computation relies on precise numerical representations, sequential execution of operations, and guaranteed convergence to correct answers given sufficient resources [573]. Digital computing systems are primarily deterministic systems. Why can they process data sampled from random activities? Because, similar to quantum collapse, once a random activity is observed, it collapses into a deterministic state. Biological neurons, however, operate under severe physical and energetic constraints that preclude such deterministic processing [556]. In online, action potentials are inherently noisy, synaptic transmission fails intermittently, and ion channels fluctuate stochastically all contributing to variability that would render conventional algorithms unreliable if implemented directly [859]. Instead of performing calculations, the brain approximates solutions through analog, population-coded dynamics that exploit parallelism and redundancy [646].

One illustrative example is the challenge of real-time signal processing. Consider the task of integrating visual motion cues to estimate heading direction during navigation. Classical approaches involve matrix transformations, coordinate frame conversions, and recursive filtering all mathematically intensive procedures requiring floating-point precision [50]. Yet insects accomplish similarly sophisticated navigation with miniature brains consuming mere micro watts of power [10]. Research suggests they achieve this not through explicit calculus but via tuned neural responses shaped by evolution [86]. Direction-selective cells in the optic flow pathway respond preferentially to particular velocity vectors, effectively implementing a form of template matching without ever representing speed as a number [388].

Similarly, numerical cognition in humans shows little evidence of direct arithmetic implementation. Functional MRI studies indicate that mental calculation engages parietal regions involved in spatial attention and magnitude estimation, suggesting that we manipulate quantities through analog mental number lines rather than digit-by-digit computation [319]. Behavioral data confirm this: reaction times increase logarithmically with operand size, consistent with searching along a continuous representation rather than executing discrete steps [817]. Errors also follow systematic patterns such as the problem size effect and tie effect indicative of associative retrieval from memory rather than rule-based derivation [356].

Furthermore, energy limitations impose fundamental constraints on computational feasibility. Each action potential consumes approximately $10^9$ ATP molecules, and the brain uses about 20% of the body's total energy despite accounting for only 2% of its mass [41]. This extreme metabolic cost necessitates highly efficient coding strategies [58]. Artificial neural networks running on conventional silicon consume orders of magnitude more energy per operation than their biological counterparts [830]. Neuromorphic engineering efforts inspired by neural dynamics aim to bridge this gap by emulating analog, event-driven computation, highlighting that efficiency stems from avoiding general-purpose digital computation altogether [552].

Even tasks that appear mathematical at the behavioral level may not involve internal computation. Bayesian inference a cornerstone of modern theories of perception and decision-making is often cited as evidence of probabilistic reasoning in the brain [72]. However, recent work argues that observed behaviors could result from heuristic shortcuts or implicit learning of environmental statistics without constructing formal likelihood functions or posterior distributions [487]. Neural implementations likely approximate Bayes-optimal decisions through adaptive gain modulation and recurrent network dynamics rather than manipulating probability densities symbolically [870].

Thus, the impossibility of mathematical computation in the literal sense underscores a deeper principle: the brain solves problems through embodied, situated engagement with the world rather than abstract symbol manipulation [634]. It learns to respond appropriately to stimulus patterns through experience-driven plasticity, producing behavior that is functionally adequate rather than formally correct [346]. This perspective shifts emphasis from replicating human-level AI via improved algorithms to reverse-engineering the brain's unconventional computational principles, potentially leading to new paradigms of intelligent systems [885].

**4.5 Absence of Gradient Information**

Backpropagation of error, the cornerstone of deep learning, depends crucially on the availability of gradients partial derivatives indicating how small changes in each parameter affect the overall loss function [459]. This global feedback mechanism allows layered networks to assign credit or blame across arbitrary depths, enabling end-to-end optimization [459]. However, there is no known biological correlate for such a system [492]. Neurons receive limited local information and lack direct access to downstream error signals required for weight updates. The absence of gradient information poses a central paradox in neuroscience: How can a massively interconnected network learn hierarchical representations without a teacher providing detailed correction vectors?

Several arguments highlight the implausibility of biologically plausible backpropagation [492]. First, the algorithm requires symmetric forward and backward pathways a condition not met in mammalian anatomy [886]. Cortical pyramidal cells project predominantly in one direction, and retrograde signaling occurs through slower, diffuse chemical means unsuitable for rapid, precise gradient transmission [886]. Second, error signals would need to be calculated and routed simultaneously to every synapse in the network, demanding unrealistic coordination and bandwidth [860]. Third, the method violates Dale's principle, which posits that a neuron releases the same neurotransmitter at all its synapses, whereas backpropagation requires separate excitatory and inhibitory pathways for positive and negative gradient components [70].

**4.6 Global-loss Backpropagation Learning Is Not Biologically Feasible**

The concept of optimizing a global loss function through backpropagation assumes a centralized cost metric and a coordinated update mechanism spanning the entire network [74]. In artificial neural networks, this manifests as minimizing a scalar loss value averaged over many training examples, with gradients flowing uniformly across layers. Biologically, such a scheme is profoundly implausible due to anatomical, physiological, and evolutionary constraints [74]. The brain lacks a unified cost function or a master controller capable of broadcasting loss gradients throughout billions of neurons [597]. Instead, it must rely on decentralized, modular learning governed by local interactions and sparse supervisory signals.

Anatomically, the brain exhibits a highly distributed architecture with specialized modules handling distinct functions visual processing, motor control, emotion regulation each embedded within partially segregated hierarchies [289]. There is no central processor overseeing all cognitive operations; thus, no single entity computes a global loss. Even executive functions mediated by the prefrontal cortex operate reactively rather than proactively, integrating inputs from various subsystems to guide behavior without maintaining a comprehensive error landscape [449].

Physiologically, the mechanisms for transmitting information are ill-suited for synchronous gradient updates. Chemical synapses introduce delays of several milliseconds, and spike propagation speeds vary widely across fiber types [847]. Such temporal dispersion prevents the tight coordination required for simultaneous parameter adjustments. Moreover, synaptic plasticity itself is gated by numerous factors including neuromodulators, astrocytic signals, and local metabolic conditions that differ across brain regions and change dynamically with behavioral state [247].

Evolutionarily, organisms did not evolve to minimize abstract loss functions but to survive and reproduce in variable environments [22]. Natural selection favors heuristics that produce good-enough solutions quickly, rather than globally optimal policies derived through exhaustive search [22]. A predator does not calculate the shortest escape trajectory using gradient descent; it responds instinctively based on conditioned reflexes and innate biases [41]. This pragmatic orientation implies that learning proceeds through trial-and-error guided by delayed rewards rather than immediate supervision.

Empirical findings reinforce this view. Lesion studies show that animals can recover lost functions through compensatory reorganization, indicating resilience against component failure inconsistent with tightly coupled systems reliant on precise gradients [124]. Developmental plasticity further demonstrates that large-scale rewiring occurs naturally during critical periods without external supervision [260]. Additionally, sleep-dependent memory consolidation suggests offline replay contributes to restructuring representations independently of ongoing input, contradicting online-only update schemes [805].

Alternative frameworks emphasize autonomy and modularity. Hierarchical reinforcement learning partitions tasks into subgoals managed by semi-independent agents, mimicking the brain's functional segregation [624]. Predictive processing models treat perception as hypothesis testing, where mismatches between expectations and observations drive localized updates rather than global re-calibration [421]. These approaches align better with observed neural phenomena and offer more viable paths toward explaining biological intelligence.

In sum, the impossibility of global-loss backpropagation reflects a deeper insight: brain intelligence emerges from decentralized, adaptive systems rather than monolithic optimizers [39]. Recognizing this opens new possibilities for designing artificial systems that emulate biological robustness, flexibility, and scalability.

**4.7 Hebbian Rules and STDP Lack Global Losses**

Hebbian learning and spike-timing-dependent plasticity (STDP) represent some of the most widely studied models of synaptic modification in neuroscience [57]. Both posit that connection strengths increase when presynaptic and postsynaptic activities coincide within a narrow time window. Despite their explanatory power in reproducing phenomena such as receptive field development and associative memory, these rules suffer from a critical limitation: they lack any notion of a global loss [90]. They are purely local, blind to task relevance, and incapable of distinguishing beneficial from detrimental associations without additional guidance.

Classical Hebbian theory neurons that fire together wire together captures the essence of co-activation-based strengthening but provides no mechanism for weakening irrelevant connections or preventing runaway excitation [90]. Left unchecked, Hebbian plasticity leads to unstable network dynamics, with activity spreading uncontrollably until saturation [92]. To mitigate this, real brains incorporate normalization mechanisms such as homeostatic scaling and synaptic competition, but these too operate locally and do not link to behavioral outcomes [816].

STDP refines Hebbian learning by introducing temporal asymmetry: if the presynaptic spike precedes the postsynaptic one, potentiation occurs; otherwise, depression ensues [57]. This refinement enables sequence detection and directional selectivity, making it suitable for temporal coding schemes. However, STDP remains agnostic to whether a strengthened pathway improves fitness or merely reinforces spurious correlations [8]. For example, a neuron might strengthen inputs associated with irrelevant background noise simply because they consistently precede a behaviorally relevant stimulus.

Without a supervisory signal, neither Hebbian nor STDP rules can solve credit assignment problems involving long chains of causality [96]. Consider learning a complex motor skill involving multiple stages of planning, execution, and correction. The final success or failure depends on countless intermediate decisions made across distributed circuits. Local plasticity rules cannot determine which upstream events contributed positively to the outcome, especially when delays span seconds or minutes.

To overcome this, biological systems integrate Hebbian mechanisms with global instructive signals. Neuromodulators such as dopamine, serotonin, and norepinephrine broadcast evaluative feedback across brain regions, reinforcing synapses activated just prior to a rewarding event [97]. This combination forms the basis of reinforcement learning in the brain. Experimental manipulations confirm this synergy: pairing neutral stimuli with dopaminergic stimulation induces place preference and alters neural tuning curves, effectively creating artificial memories [705].

Other forms of guidance include attentional modulation and volitional control. Top-down signals from frontal areas bias competition among sensory representations, enhancing relevant inputs while suppressing distractions [428]. This selective attention acts as a gating mechanism, determining which correlations undergo Hebbian strengthening. Similarly, internally generated goals influence exploratory behavior, steering plasticity toward productive configurations [100].

Thus, while Hebbian and STDP rules provide essential building blocks for neural adaptation, they must be embedded within broader regulatory frameworks to achieve purposeful learning [101]. Their inherent blindness to global losses highlights the necessity of complementary systems that provide contextual evaluation and strategic direction.

**4.8 Measurement Outcomes Reflect Integrated Effects of Ideal and Non-Ideal Factors**

Neural recordings constitute a complex tapestry capturing the aggregate consequences of numerous interacting processes. They blend intentional, goal-directed computations with stochastic fluctuations and biomechanical constraints [772]. When examining different types of neural measurements single-unit electrophysiology, local field potentials, and fMRI BOLD signals we find that they do not represent pure information processing. Instead, they represent entangled mixtures of rational strategies and non-ideal noise [618]. This integration of ideal and non-ideal factors renders interpretation of neural recordings exceptionally difficult. Observed activity patterns may arise from various sources: adaptive computations where neural systems actively adjust to tasks, random drift causing unpredictable deviations, compensatory mechanisms attempting to correct underlying neural imbalances, or measurement artifacts distorting true neural activity [772].

**4.8.1 Neural Spikes Carry Combined Influences**

Action potentials, or spikes, are the primary means of neural communication. Each spike represents a significant event occurring when membrane potential crosses a certain threshold a threshold-crossing resulting from integration of thousands of excitatory and inhibitory postsynaptic potentials [508]. These synaptic inputs are far from simple; they arrive asynchronously and interact nonlinearly within dendrites. Consequently, the occurrence of any given spike depends on a complex convolution of multiple factors. Feedforward drives signals traveling forward within neural networks contribute to spike generation. Recurrent feedback, wherein neural output loops back to influence input, also exerts impact. Neuromodulatory tone, which adjusts overall neuronal excitability, and intrinsic membrane properties all play roles in determining spike timing [458]. Therefore, spikes convey combined influences from both structured information streams and unstructured variability.

Consider a working memory task. During such tasks, prefrontal neurons exhibit persistent firing correlated with maintained representations [190]. However, examining spike timing on a trial-to-trial basis reveals variability exceeding what would be expected from deterministic encoding alone [764]. This variability has multiple causes: ongoing oscillations cause membrane potential fluctuations, fluctuating arousal levels affect excitability, and stochastic channel openings add further noise [577]. All these factors contribute noise that limits decoding accuracy. However, some degree of variability may not be entirely negative; it can serve functional roles by facilitating exploration, allowing neural systems to sample different states, preventing premature commitment to suboptimal solutions, and helping a synapse to achieve global learning [109].

**4.8.2 Neural Spikes Reflect Complex, Non-Analytical Models**

Neural spikes play crucial roles in neuroscience and reflect complex, non-analytical models. Spiking activity does not instantiate symbolic models amenable to mathematical decomposition [644]. Symbolic models typically rely on discrete symbols and manipulation rules, allowing straightforward decomposition into mathematical components. Spiking activity does not fit this framework.

Instead, spiking activity embodies implicit, distributed representations shaped by experience [482]. These representations are not explicitly encoded in single neurons but distributed across networks. Experience sensory input, learning, environmental interactions molds these distributed representations. When organisms are repeatedly exposed to particular visual stimuli, spiking patterns in visual cortex gradually change to form stimulus representations. These representations are implicit because they emerge as properties of neural activity rather than direct one-to-one stimulus mappings.

Population codes distribute information across many neurons, providing significant robustness advantages [42]. Even if individual units fail, overall information can still be accurately read out. In motor cortex, large neuronal populations controlling specific movements can compensate if some neurons malfunction, with remaining neurons conveying necessary information [856]. This distributed information storage and processing characterizes neural systems and contrasts with systems storing information in single centralized locations.

Representations in neural activity emerge gradually through plasticity, reflecting accumulated statistics rather than explicit hypotheses [748]. As neurons encounter different stimuli over time, they adjust connections and firing patterns. Resulting representations are based on input statistical properties. In auditory cortex, neurons learn to represent statistical regularities in speech sounds representations based on actual encountered patterns rather than pre-defined hypotheses about sounds.

Given the complex and non-analytical nature of neural activity, decoding requires machine learning techniques rather than analytic inversion [659]. Analytic inversion typically involves finding direct mathematical relationships between input and output. Due to distributed, implicit, experience-shaped representations, such direct relationships are difficult to establish. Machine learning techniques can handle neural data complexity by learning patterns and relationships in spiking activity through training on large datasets. Deep learning algorithms can decode neural activity associated with particular behaviors or perceptions, providing insights into brain information processing [659].

### 4.8.3 Pulse Coding Effects in Neural Spikes

It is generally accepted that neural impulses can encode neural output. Beyond rate coding, spike trains employ temporal patterns such as synchrony, bursting, and phase locking to convey information [131]. Precise millisecond-scale timing enables coincidence detection in downstream targets, supporting feature binding and sequence learning [835]. Burst-firing modes can signal salience or novelty, altering synaptic impact [498]. Phase-locked activity relative to theta or gamma oscillations organizes multiplexed information streams, permitting simultaneous transmission of multiple message dimensions [263].

### 4.8.4 Measurement Artifacts and Analog-to-Digital Conversion in Neural Spikes

Recording technologies themselves introduce distortions [334]. Electrode impedance variations can significantly alter how neural signals are detected. If electrode impedance fluctuates, it can change signal electrical characteristics, leading to inaccuracies in spike shape and detectability. Filtering settings also play crucial roles; different filtering settings can enhance or suppress certain frequency components, potentially masking or distorting important spike features. Sampling rates are equally important; inappropriate sampling rates may result in loss of high-frequency information or misrepresentation of spike waveforms [480].

Cross-talk between channels can be a major source of signal contamination. When multiple electrodes record neural activity, electrical signals from one channel can interfere with others, creating false spikes or distorting genuine ones. Reference electrode placement is also critical; improper reference placement can introduce common-mode noise, making accurate spike discrimination difficult. Movement artifacts present yet another problem; subject movement whether small tremors or significant body movements generates electrical signals superimposed on neural spikes, further contaminating data [842].

Offline spike sorting algorithms rely on certain assumptions. They assume waveform stationarity that spike shapes remain relatively constant over time. However, in real-world scenarios, network states change and spike waveforms may vary accordingly. They also assume cluster separability that different spike types can be clearly distinguished and grouped. Under changing network states, these assumptions may not hold, leading to inaccurate spike sorting results. These technical confounds make comparing data across recording sessions and between laboratories extremely challenging [122].

Neural impulse-based systems, found in biological neurons and mimicked in spiking neural networks (SNNs), can be viewed as simplified forms of 1-bit analog-to-digital conversion (ADC) [419]. In traditional electronics,

ADC is a well-defined process converting continuous analog signals with infinite possible values into discrete digital values through sampling and quantization. Sampling involves measuring analog signals at regular intervals, while quantization assigns specific digital values based on amplitude [389].

Similarly, neurons perform comparable functions. They take analog input signals graded membrane potentials and encode them into discrete action potential sequences. These spikes are all-or-nothing electrical impulses: once thresholds are reached, spikes generate with fixed amplitudes. This spike generation process effectively digitizes input signal intensity over time. Stronger stimuli lead to higher firing rates, meaning amplitude information converts into temporal patterns similar to rate-encoded ADCs where analog magnitudes are represented by digital pulse frequencies [672].

Biological or neuromorphic ADCs offer several advantages over conventional electronic ADCs. One major advantage is energy efficiency. Conventional electronic ADCs often require complex circuitry for precise voltage comparisons and binary encoding, consuming significant power and occupying substantial integrated circuit area. In contrast, neural systems use intrinsic membrane dynamics and ion channel kinetics to achieve threshold-based spike initiation. When membrane potential reaches voltage thresholds, stereotyped action potentials generate analogous to 1-bit quantizers with temporal oversampling, transmitting information only when necessary, reducing redundant data processing, and thus saving energy [374]. It is especially worth pointing out that neural ADC produces a half digital and half continuous signal, i.e., its value is digital, and the location of each value is continuous.

Neuromorphic engineering has capitalized on this principle. Designs such as integrate-and-fire circuits are based on neuronal behavior, using simple capacitive integration to accumulate charge over time and threshold triggering to generate output pulses. By emulating neuronal behavior, these circuits further validate conceptual alignment between neural signaling and simplified ADC architectures [374].

From functional perspectives, rate coding and temporal coding schemes in neural systems offer distinct advantages. In low-power sensing and edge computing applications, robustness against noise and variability often takes priority. Standard multi-bit ADCs achieve high precision but may be noise-sensitive and require relatively large power. Neural impulse coding trades absolute accuracy for metabolic and computational efficiency, tolerating certain noise and variability levels while still providing useful information ideal for real-time sensing applications where power consumption minimization and quick responses are required [558].

**4.8.5 Integrated Decision Effects in Neural Spikes**

At behavioral levels, neural spikes emerge as key players in decision-making processes [724]. Spikes contribute to decision-making through evidence accumulation across populations. Groups of neurons act as bustling communities, with each neural spike carrying information pieces. When combined, these pieces form decision bases a remarkable collective effort where spike activity across neuronal groups is fundamental for decision generation. This concept highlights coordinated neuronal action in complex decision-making tasks, suggesting decisions emerge not in isolation but from collaborative information-sharing processes among neurons.

To better understand decision-making, scientists have developed drift-diffusion models focusing on reaction time distributions [667]. These models operate on noisy integration toward thresholds. Decision-making resembles noisy signals gradually drifting toward specific thresholds; once thresholds are reached, decisions are made. This provides mathematical frameworks explaining how brains process information over time to arrive at decisions, allowing researchers to quantify and analyze decision-making processes for deeper studies on how brains weigh information and reach conclusions.

Specific brain areas show neural correlates associated with decision-making. Ramping activity in parietal and frontal areas is notable [209]. This ramping activity resembles wave build-up before crashing on shores. In parietal and frontal areas, gradual neural activity increases likely relate to evidence integration and decision threshold approaches. These areas function as control centers where incoming information is processed and evaluated; as activity ramps up, brains approach decisions. This finding provides valuable insights into physical brain locations where decision-making processes occur and how they unfold.

However, real-world decision-making is far more complex than these basic models suggest. Real decisions are influenced by motivational biases, risk preferences, and contextual priors [292]. These elements cause real decisions to deviate systematically from normative predictions. Motivational biases strongly influence decisions based on individual desires or goals. If people are highly motivated to achieve particular outcomes such as winning competitions they may make decisions skewed toward those goals, perhaps taking more risks or overlooking certain information. Risk preferences also play significant roles; some individuals are naturally risk-averse while others are risk-seekers, willing to take chances for potentially greater rewards. Different risk attitudes lead to vastly different decision-making patterns. Contextual priors refer to knowledge and experiences individuals bring to decision-making situations. Someone with prior field experience may make different decisions compared to novices. These factors add multiple complexity layers to decision-making, making accurate decision prediction using simple models extremely challenging.

### 4.8.6 Additional Measurement Considerations

Long-term recordings in neuroscience and related research face complex issues. One significant problem is tissue response. Over extended periods, body tissues react to recording device presence. Gliosis occursglial cell proliferation and change in response to injury or foreign object presence. Additionally, electrode degradation is critical [639]. These factors severely impact long-term recording quality and reliability.

Chronic implants, used to continuously monitor neural activities, trigger inflammatory reactions. These inflammatory reactions have far-reaching consequences, significantly altering local neurochemistry and cellular composition around implants [788]. These changes not only disrupt normal neural environment functioning but also make accurate data interpretation challenging, as normal physiological states have been perturbed.

Another issue is sampling bias associated with recording techniques. Current methods tend to favor large, easily isolatable neurons, meaning smaller interneurons and glia are often neglected [890]. This sampling bias provides incomplete neural network pictures. Since smaller interneurons and glia play crucial roles in neurotransmission, neural activity modulation, and nervous system homeostasis maintenance, ignoring them can lead to inaccurate conclusions about studied neural processes.

In the big data era, many long-term recordings generate high-dimensional datasets. While these datasets contain vast information amounts, they also present unique challenges. To manage and analyze large-scale datasets, dimensionality reduction techniques are commonly employed [176]. However, these techniques may obscure subtle data structures hidden structures potentially crucial for uncovering new physiological phenomena or understanding complex neural network interactions.

To mitigate these challenges, careful experimental design is paramount. This includes choosing appropriate recording devices, implantation locations, and data collection timing. Additionally, multimodal validation is essential using multiple methods such as combining electrophysiological recordings with imaging techniques to cross-verify results. By doing so, researchers can ensure finding accuracy and reliability in the face of numerous long-term recording challenges [585].

## 4.9 Broadcast Instruction Signals Provide Global Guidance: The Neuromodulatory Architecture of Reinforcement Learning in Biological Brains

The human brain is not merely a collection of isolated neurons firing in response to stimuli, but an intricately coordinated network where local synaptic changes are guided by global evaluative signals. While Hebbian plasticity and other local learning rules determine how individual synapses strengthen or weaken based on pre- and postsynaptic activity, they lack the capacity to distinguish beneficial from detrimental adaptations without external guidance. This critical role falls upon neuromodulatory systems widespread broadcast mechanisms that deliver reward and punishment signals across large neural territories, effectively serving as the brain's reinforcement learning framework. These global instructive signals do not encode precise gradients like those used in artificial backpropagation; instead, they operate through coarse yet powerful broadcasts that inform vast neuronal populations whether recent actions led to favorable or adverse outcomes. By doing so, these systems enable adaptive behavior even in complex environments with delayed consequences, sparse feedback, and high-dimensional state spaces.

Among the most well-characterized neuromodulatory pathways is the dopaminergic system, functioning as a primary carrier of excitatory broadcast signals associated with reward prediction and outcome evaluation. Neurons located primarily in the ventral tegmental area (VTA) and substantia nigra pars compacta (SNc) project axons widely throughout the forebrain, including to striatum, prefrontal cortex, hippocampus, and amygdala regions central to decision-making, memory formation, and emotional regulation. When organisms encounter unexpected rewards such as food after hunger periods or financial gains these dopamine-producing cells exhibit transient firing bursts, resulting in rapid extracellular dopamine increases across target regions. This phasic release does not simply signal pleasure or arousal; rather, it encodes a sophisticated computational quantity known as reward prediction error (RPE), a concept formalized in reinforcement learning theory [705]. Specifically, dopamine neurons fire more strongly when rewards are better than expected, maintain baseline activity when rewards occur as predicted, and show suppressed firing when expected rewards fail to materialize [705]. This pattern aligns precisely with the delta rule in temporal difference learning models, suggesting the brain has evolved biologically plausible approximations of gradient-based optimization using broadcast signaling [571].

Crucially, this dopamine signal operates within temporally extended frameworks, allowing brains to associate actions with outcomes despite significant delays between cause and effect a fundamental challenge in real-world learning scenarios [571]. Consider a rat pressing a lever that delivers a food pellet several seconds later. No local synaptic mechanism can directly link the motor command issued at lever press to eventual reward due to absent co-activation at reinforcement moments. To solve this credit assignment problem, brains employ molecular and cellular mechanisms known as eligibility traces. An eligibility trace acts as short-term synaptic memory, marking recently active connections as potentially responsible for subsequent outcomes. When dopamine bursts arrive following delayed rewards, they reactivate or strengthen only those synapses previously tagged during behavioral sequences leading to rewards [184]. Experimental evidence supports this model: optogenetic studies show that artificially timed dopamine pulses can reinforce specific behaviors if delivered shortly after particular actions, even without natural rewards [766]. Thus, interplay between transient synaptic eligibility and delayed neuromodulatory reinforcement enables associative learning over extended temporal horizons, bridging gaps that would otherwise render learning impossible.

Beyond immediate reward processing, the dopaminergic system also plays pivotal roles in shaping long-term motivational states and guiding exploratory behavior. As animals learn to predict rewards based on environmental cues, dopamine responses shift from rewards themselves to earlier predictive stimuli a phenomenon known as second-order conditioning [149]. If a tone consistently precedes food delivery, dopamine neurons initially respond to food, but after repeated pairings, peak activity shifts to tone onset [705]. This RPE signal redistribution allows organisms to refine predictions and allocate attention toward informative cues rather than mere outcomes. Furthermore, tonic dopamine levels distinct from phasic bursts discussed above modulate overall motivation and willingness to exert effort for rewards [706]. Low tonic dopamine correlates with apathy and reduced goal-directed behavior, as seen in conditions such as depression and Parkinson's disease, while elevated tonic levels correlate with increased vigor and persistence in goal pursuit [706]. Therefore, the dopamine system serves dual roles: fast phasic responses supporting trial-by-trial learning via RPEs, and slow tonic regulation modulating energetic cost-benefit analyses underlying behavioral engagement.

However, learning cannot be driven solely by reward signals; equally important are mechanisms detecting and responding to negative outcomes, threats, and punishments. Inhibitory broadcast signals mediated by serotonin, certain dopamine neuron subpopulations, and norepinephrine provide necessary counterbalances to excitatory reward-related dopamine influences [151]. Serotonin, produced predominantly in dorsal raphe nucleus, shows heightened activity in response to aversive stimuli, expected reward omissions, and situations involving risk or uncertainty [151]. Unlike dopamine amplifying plasticity following positive expectation deviations, serotonin appears to promote caution, suppress impulsive actions, and facilitate avoidance learning [151]. Studies in rodents and primates demonstrate that serotonergic activation increases punishment sensitivity and enhances patience during waiting tasks, suggesting roles in promoting behavioral inhibition and long-term planning [154]. Moreover, pharmacological serotonin manipulation alters risk preferences: increasing serotonin function tends to reduce risky choices, while decreasing it leads to greater impulsivity and preference for immediate albeit smaller rewards [154]. These findings position serotonin as a key modulator of behavioral suppression and adaptive conservatism in uncertain or threatening contexts.

Interestingly, not all inhibitory broadcast signals originate from non-dopaminergic sources. A subset of dopamine neurons themselves respond to aversive events by either reducing firing rates or exhibiting brief inhibitions, thereby signaling expected reward absence [155]. This dopamine concentration dip serves as negative teaching signals, weakening synapses contributing to unsuccessful behaviors. Computational models incorporating both positive and negative dopamine transients successfully reproduce wide ranges of animal learning phenomena, including conditioned response extinction and reversal learning [145]. Additionally, some dopamine neurons particularly those projecting to medial prefrontal cortex increase firing in response to aversive stimuli, indicating functional heterogeneity within dopaminergic populations [157]. This complexity suggests brains do not rely on simple binary codes of dopamine = good, no dopamine = bad, but instead utilize diverse subpopulations tuned to different aspects of valence, salience, and context. Such specialization allows nuanced control over learning dynamics, enabling fine-grained adaptation depending on task demands and environmental statistics.

Complementing these systems is the noradrenergic pathway originating in locus coeruleus (LC), which releases norepinephrine (NE) broadly across cortex and subcortical structures [158]. Norepinephrine does not convey valence-specific information per se but instead modulates global network excitability states and attentional focus [158]. LC neurons exhibit two distinct activity modes: tonic and phasic [32]. Tonic firing reflects overall arousal levels, with higher rates observed during stress, fatigue, or sustained cognitive load. Phasic bursts, in contrast, occur in response to salient or novel stimuli, enhancing sensory processing and facilitating rapid attentional shifts [32]. From learning perspectives, NE adjusts what is known as network gain neuronal responsiveness to incoming inputs [32]. During uncertainty or ambiguity periods, increased NE release amplifies weak signals and sharpens contrast between competing representations, making pattern detection a midst noise easier. This gain modulation effectively tunes learning rates: under high uncertainty, brains benefit from being more responsive to new information, whereas in stable environments, lower gain prevents overfitting to random fluctuations [32]. Consequently, the noradrenergic system acts as a meta-controller, dynamically adjusting whole-network sensitivity based on environmental volatility and informational content [32].

Together, these neuromodulatory systems dopamine, serotonin, and norepinephrine form integrated architectures for global instruction in reinforcement learning service [156]. They do not act in isolation but engage in rich cross-talk, both anatomically and functionally. Dopamine and serotonin reciprocally regulate each other's release: 5-HT2A receptor stimulation can inhibit prefrontal dopamine release, while D1 receptor activation can influence serotonergic tone [14]. Similarly, norepinephrine modulates both dopaminergic and serotonergic neuronal responsiveness, creating layered control systems where multiple broadcast signals interact to shape synaptic plasticity. This integration allows context-dependent weighting of reward versus punishment, exploration versus exploitation, and stability versus flexibility. In dynamic environments, system balances shift to favor rapid adaptation; in predictable settings, homeostatic mechanisms promote consistency and resistance to spurious updates.

The biological plausibility of this global broadcast model stands in stark contrast to standard backpropagation algorithm requirements in deep artificial neural networks. Backpropagation relies on precise error gradients computed layer-by-layer and propagated backward through fixed weights, necessitating symmetric connectivity and instantaneous communication arrangements incompatible with known neuroanatomy and physiology. In contrast, neuromodulatory broadcasting offers decentralized, asynchronous alternatives operating within biological reality constraints. No weight symmetry or exact gradient computation is needed; instead, success or failure is signaled globally, and only recently engaged and thus eligible synapses are modified. This mechanism, sometimes termed three-factor learning rules, combines presynaptic input, postsynaptic output, and third factors provided by neuromodulators to gate plasticity [259]. Mathematically, this approximates stochastic gradient ascent/descent when neuromodulators correlate with loss functions, making it robust and scalable for credit assignment in distributed systems [259].

Further support for broadcast signal functional significance comes from lesion studies and clinical observations. VTA or SNc damage results in profound instrumental learning and reward-seeking behavior deficits, mimicking motivational disorder symptoms such as anhedonia [705]. Parkinson's disease patients, characterized by dopaminergic neuron degeneration, exhibit impairments in acquiring new habits and adapting to changing reward contingencies, particularly in probabilistic learning tasks [166]. Pharmacological treatments restoring dopamine transmission, such as L-DOPA therapy, partially rescue these deficits, underscoring dopamine's causal role in reinforcement-guided adaptation. Conversely, serotonin dysregulation has been implicated in anxiety disorders, obsessive-compulsive behaviors, and suicidal idealization all conditions marked by maladaptive threat or loss responses [167]. Selective serotonin re uptake inhibitors (SSRIs), which enhance serotonergic signaling, alleviate symptoms in many patients, highlighting therapeutic potential of modulating inhibitory broadcast systems. Likewise, noradrenergic system dysfunction is associated with attention deficit hyperactivity disorder (ADHD), post-traumatic stress disorder (PTSD), and chronic stress syndromes, further illustrating neuromodulatory imbalance's broad impact on cognition and behavior [168].

Importantly, these broadcast signals are not rigidly hardwired but are themselves subject to learning and regulation. Higher-order brain regions, such as prefrontal cortex and anterior cingulate cortex, monitor ongoing performance and adjust neuromodulatory tone accordingly. Anterior cingulate detects conflicts between competing responses and signals increased cognitive control needs, potentially via projections to locus coeruleus and dorsal raphe [102]. This top-down regulation enables metacognitive oversight, allowing organisms to adapt not only behaviors but also parameters governing how they learn. Such hierarchical control provides mechanisms for balancing exploration and exploitation: when performance is poor or environments change, increased neuromodulatory volatility encourages exploration and plasticity; when outcomes are consistent, reduced signaling promotes consolidation and stability. This self-regulating loop ensures learning remains efficient and appropriately scaled to current demands, avoiding both stagnation and instability.

Another critical broadcast signaling aspect lies in spatial and temporal specificity or rather, deliberate lack thereof. Unlike point-to-point synaptic transmission, neuromodulators diffuse over millimeters, affecting thousands to millions of neurons simultaneously. This promiscuity might seem inefficient or imprecise, but confers several advantages. First, it allows single signals to coordinate activity across functionally related but anatomically distributed circuits. A reward event may simultaneously update striatal value representations, consolidate hippocampal memories, and adjust prefrontal expectations all through identical dopamine pulses. Second, widespread broadcasting enables coherent behavioral transitions, such as switching from exploration to exploitation or initiating fight-or-flight responses. Third, diffuse signal nature makes them resilient to localized damage or noise, ensuring essential guidance persists even if network parts are compromised. Finally, because signals reach many areas simultaneously, they create common reference frames for timing and coordination, synchronizing plasticity across disparate modules that must work together to produce adaptive behavior.

Despite their generality, broadcast signals are not indiscriminate. Mechanisms exist to constrain their influence and prevent inappropriate synaptic modifications. One such mechanism is requirement for coincident neural activity: dopamine surges will only potentiate synapses that are currently active or have recently undergone sufficient depolarization. This is enforced through molecular cascades involving NMDA receptors, calcium influx, and kinase activation, collectively serving as coincidence detectors. Another constraint arises from receptor subtype distribution: different brain regions express varying dopamine, serotonin, and norepinephrine receptor complements, each with distinct downstream effects. D1-type receptors typically enhance cAMP signaling and promote long-term potentiation (LTP), while D2-type receptors inhibit cAMP and favor long-term depression (LTD) [94]. Regional specificity of receptor expression thus tailors global signal functional impacts, allowing identical neurotransmitters to have opposite effects in different circuits. Additionally, glial cells actively regulate extracellular neuromodulator concentrations through uptake and metabolism, adding another spatial and temporal control layer.

From evolutionary standpoints, neuromodulatory broadcast system emergence represents major nervous system design innovation. Early nervous systems likely relied on purely reflexive or habituated responses, limited in adaptive capacity beyond immediate sensory-motor loops. Global signaling advent allowed centralized internal and external state evaluation, enabling organisms to optimize behavior over longer timescales and more complex ecological niches. It facilitated goal-directed action development, foresight, and social cooperation all advanced cognition hallmarks. Even in relatively simple animals, such as insects and mollusks, homologous neuromodulatory systems regulate feeding, mating, and defensive behaviors, suggesting deep phylogenetic roots. In mammals, these systems became increasingly elaborated, integrating with expanded cortical areas to support abstract reasoning, language, and cultural learning. Thus, broadcast instruction signals represent not just brain function components, but foundational principles in intelligent behavior evolution.

Looking forward, understanding neuromodulatory broadcasting mechanistic details holds immense promise for both neuroscience and artificial intelligence. In medicine, targeted manipulation of these systems could lead to more effective psychiatric and neurological disorder treatments. Deep brain stimulation, gene therapies, and next-generation psychopharmacology aim to restore balanced neuromodulation in conditions ranging from addiction to treatment-resistant depression. On AI frontiers, incorporating biologically inspired broadcast signals into artificial neural networks could overcome traditional training method limitations. Recent advances in synthetic neuromodulation demonstrate that adding global reward/punishment signals with eligibility traces allows artificial agents to learn efficiently in sparse-reward environments, outperforming conventional reinforcement learning algorithms in certain domains [181]. These hybrid architectures bridge biological realism and engineering efficiency gaps, offering paths toward more autonomous, adaptive, and robust machine learning systems.

In a deep learning framework, the error of a reward-like loss function is formally defined as the differential between the target value of the loss and its current estimate a quantity that drives parameter updates to minimize/maximize the loss. This error, often termed the reward prediction error (RPE) in reward-optimized models, mirrors the core principle of gradient-based learning: quantifying mismatch to guide optimization. Below is a concise mathematical definition. Error of a Learning Reward Function is defined as follows: Let $L$ denote a reward-related loss function implicitly parameterized by connection weights $w_{ij}$ (e.g., expected cumulative reward, or value function measurement). This function $L$ typically denotes the squared difference or cross entropy between a target value (ground-truth or bootstrapped measurement) and the current measurement of the output. The error (RPE) of this function is defined by $\delta_{Global} = L_{current} - L_{past}$. Here, $L_{past}$ denotes past measurement of the loss, $L_{current}$ indicates current measurement of the loss, and $\delta_{Global}$ is Reward prediction error (error of the reward function), acting as the driving signal for parameter updates. Such form $\delta_{Global} = L_{current} - L_{past}$ causes the reward(s) to get progressively bigger. If let the rewards become increasingly small, we should take $\delta_{Global} = L_{past} - L_{current}$. It is seen that $\delta_{Global}$ can be taken as a constructive signal if a brain is static. If loss is the function of weight $w_{ij}$, it is followed by the famous differential approximation formula that $\delta_{Global} \approx \frac{\partial L}{\partial w_{ij}} \Delta w_{ij} = \frac{\partial L}{\partial w_{ij}} p_{ij}$, where $\Delta w_{ij}$ is taken as a perturbation $p_{ij}$ of $w_{ij}$.

Core properties are that $L_{past}$ (past measurement of the loss), $L_{current}$ (current measurement of the loss), and $\delta_{Global}$ (reward prediction error) are obtained by measurements whose values implicitly depend on inputs of all neurons, all synaptic connection weights, various non-ideal factors and so on. Hence, $\delta_{Global}$ (reward prediction error) deals with all the related factors that include noises, chaos, errors, inconsistency, mismatches, aging, temperature drift and so on. In contrast, deep learning calculates the above listed values that do not consider the non-ideal factors in a brain, an analog chip or a mixed-signal chip.

In conclusion, the brain's use of broadcast instruction signals exemplifies masterful solutions to one of the most challenging adaptive system problems: how to guide learning in detailed error gradient absence. Through coordinated dopamine, serotonin, and norepinephrine actions, brains deliver global, evaluative feedback shaping synaptic plasticity across vast neural landscapes. These signals do not micromanage individual connections but provide coarse yet powerful directives reward, punishment, alertness that sculpt behavior over time. Coupled with local eligibility mechanisms, they resolve credit assignment problems, enabling associations across time and space. Their diffuse reach ensures coherence and resilience, while receptor diversity and top-down regulation confer precision and flexibility. Far from being mere chemical mood regulators, these neuromodulators constitute sophisticated control systems underlying all experience-dependent change forms. As we continue unraveling their complexities, we gain not only deeper brain intelligence nature insights but also practical tools for mind healing and machine cognition augmentation. The future of both neuroscience and AI may well depend on our ability to harness broadcast signal power not just to understand brains, but to build better minds.

### 4.10 Learning Principles Operating Without Mathematical Models

Biological learning principles operate outside formal mathematics realms, relying instead on empirical rules derived from experience [283].

### 4.10.1 Correlations Between Local Random Weight Perturbations and Instruction Signals

Local synaptic weight $w_{ij}$ (synaptic weight connecting pre-synaptic neuron $j$ to post-synaptic neuron $i$), undergo past small random natural perturbations $p_{ij}$ that are entirely arbitrary and random noisy. The random fluctuations in synaptic neurotransmitter release generate small perturbations in synaptic connection weights these perturbations are prepared to facilitate the learning of synaptic strength guided by global losses. Instead, these perturbations are statistically correlated with global instruction signals $\delta_{Global}$ neural broadcasts (e.g., dopamine, serotonin) that convey information about the organism's state (e.g., reward, arousal) [884]. Stronger instruction signals bias the direction and magnitude of perturbations, enabling experience-dependent tuning without explicit mathematical losses.

Correlation quantification: The single-sample covariance between perturbations and global instruction signals $C = p_{ij} \cdot \delta_{Global}$ is represented by $C$. If the correlation value is greater than zero, it indicates that the perturbation deserves encouragement; if the correlation value is less than zero, it indicates that the perturbation deserves punishment that is, the perturbation needs to be reversed. In biological terms, positive $C$ indicates perturbations align with excitatory signals (e.g., dopamine bursts), negative $C$ with inhibitory signals (e.g., serotonin) [259].

Biological neural networks utilize stochastic synaptic plasticity to link stochastic weight fluctuations (arising from spontaneous neurotransmitter release) with global reward/prediction error signals (e.g., dopamine, serotonin) [259]. Key mechanisms include: Quantal Neurotransmitter Release: Spontaneous vesicle fusion at presynaptic terminals introduces random weight perturbations. These noise events create variability in synaptic strength, enabling exploration for adaptive learning [108]. Dopamine-Mediated Correlation Detection: Midbrain dopamine neurons broadcast global reward signals via phasic firing. Strong dopamine release correlates with positive outcomes, biasing perturbations toward strengthening. Conversely, tonic dopamine levels reflect baseline expectations [705]. Serotonin-Driven Inhibition: Serotonergic signaling (e.g., from raphe nuclei) inversely modulates perturbations, suppressing excessive excitation. This prevents runaway plasticity while maintaining homeostasis [151].

Example: During Pavlovian conditioning, a reward-predicting cue triggers phasic dopamine release, correlating with $C > 0$. Synapses onto target neurons exhibiting this correlation undergo potentiation, while uncorrelated synapses remain unperturbed [840].

### 4.10.2 Learning Principles for Excitatory/Inhibitory Connections

Excitatory/inhibitory synapses ($w_E, w_I$) strengths between pre- and post-synaptic neurons are reinforced by neuromodulatory signals. To prevent runaway excitation/inhibition, homeostatic mechanisms normalize weights to maintain stable network activity. Because excitatory/inhibitory synapses ($w_E, w_I$) strengths serve to potentiate and depress the global losses, respectively, updating Principles of potentiation or depression (two-factor plasticity) are followed by $\Delta w_E = \eta_E \cdot C$ or $\Delta w_I = -\eta_I \cdot C$, where $\eta_E, \eta_I$ denote plasticity coefficient [379].

**Remark:** It is easily analyzed that brain learning principles are variants of the stochastic gradient method. For example, $\Delta w_E = \eta_E \cdot p_E \delta_{Global} \approx \eta_E \cdot p_E^2 \frac{\partial L}{\partial w_E} \propto \frac{\partial L}{\partial w_E}$ and $p_E^2$ is stochastic. We conclude that update rule $\Delta w_E = \eta_E \cdot C$ is a variant of stochastic gradient ascent. Similarly, update rule $\Delta w_I = -\eta_I \cdot C$ is a variant of stochastic gradient descent. The above explains the feasibility of brain learning using mathematics, but it is not that brain learning requires support from mathematical models. This is because global broadcast signals are obtained through measurement, and brain learning can automatically account for non-ideal factors such as noise, delays, or individual variability without explicit modeling.

To stabilize network activity, biological systems enforce homeostatic plasticity through the following expressions. Two-Factor Plasticity: Excitatory potentiation: Glutamatergic synapses strengthen via $\Delta w_E = \eta_E \cdot C$ when synaptic perturbation aligns with global rewards ($\delta_{Global} > 0$). Inhibitory depression: GABAergic synapses weaken via $\Delta w_I = -\eta_I \cdot C$, preventing hyper-excitability [88].

Synaptic Scaling: Global activity sensors (e.g., Arc protein) detect mean firing rates [698]. If activity deviates from baseline: Reduced activity $\rightarrow$ Global upscaling (e.g., via TNF-$\alpha$ signaling) boosts all synapses proportionally. Excessive activity $\rightarrow$ Downscaling (e.g., via calcineurin) prunes weak connections [816].

Homeostatic Inhibitory Control: Parvalbumin-positive (PV+) interneurons dynamically suppress pyramidal cell excitability. During learning, PV+ firing synchronizes gamma oscillations, enforcing sparse coding and preventing runaway excitation [750].

Example: In visual cortex, monocular deprivation reduces activity in deprived-eye inputs. Homeostatic scaling up regulates their sensitivity while PV+ cells inhibit neighboring columns, preserving overall network stability [201].

Key Differences from Artificial Systems: Stochasticity as a Feature: Unlike deterministic gradient descent, biological systems exploit stochastic perturbations to explore parameter space efficiently [885]. Multi-Scale Integration: Learning integrates molecular (BDNF, calcium signaling), cellular (STDP, synaptic scaling), and circuit-level (dopamine modulation) mechanisms. Energy Efficiency: Neuromodulatory systems (e.g., dopamine) act as teachers to guide plasticity, reducing exhaustive computation needs [467].

This framework explains how brains achieve robust learning without explicit mathematical losses through probabilistic, adaptive mechanisms shaped by evolutionary pressures for survival.

**4.10.3 Neuromodulatory Local Learning Principles of Connection Strengths**

1. Neuromodulatory Hebbian learning rule and spike-timing-dependent plasticity (STDP) are some of the less researched models of synaptic modification. They are represented as follows. Neuromodulatory Hebbian rule expresses synaptic weight updates with global loss. For a synapse connecting presynaptic neuron $j$ (activity $x_j$) to postsynaptic neuron $i$ (activity $y_i$), the change $\Delta w_{ij}$ of weight $w_{ij}$ is defined as $\Delta w_{ij} = \eta x_j y_i \delta_{Global}$ which denotes the learning Principle with four factors. Unlike Hebbian rule, neuromodulatory Hebbian rule is usually stable, if $\eta$ is small enough [259].

Biophysical Mechanisms: The rule is inherently stable if the learning rate $\eta$ is sufficiently small. Neuromodulatory Gating: Neuromodulators (e.g., dopamine) dynamically gate plasticity thresholds via G-protein coupled receptors (GPCRs). For example: Dopamine D1 Receptors: Activate cAMP-PKA signaling to promote LTP. Dopamine D2 Receptors: Inhibit LTP and enhance LTD through Gi/o pathways [94].

Neurobiological Context: Hippocampal CA3-CA1 Pathway: Theta-rhythm (4-12 Hz) synchronization during exploratory behavior may enable neuromodulatory Hebbian plasticity, supporting spatial memory encoding [233].

2. Neuromodulatory Spike-Timing-Dependent Plasticity (STDP): STDP modifies via the following relation

$$\Delta w_{ij}(\Delta t) = \begin{cases} (p_{ij}\delta_{Global})A_+e^{-\Delta t/\tau_+} & \Delta t > 0 (\text{LTP}) \\ -(p_{ij}\delta_{Global})A_-e^{\Delta t/\tau_-} & \Delta t < 0 (\text{LTD}) \end{cases}$$

where $A_+$ and $A_-$ denote maximum amplitude of LTP/LTD (positive constants); $\tau_+$ and $\tau_-$ are two time constants governing the decay of LTP/LTD [87].

Biophysical Implementation: LTP Induction: Presynaptic activity preceding postsynaptic spiking (causal timing) triggers NMDAR-dependent calcium influx, activating CaMKII and AMPAR insertion [526]. LTD Induction: Postsynaptic activity preceding presynaptic spiking (anticausal timing) activates mGluRs or GABAergic interneurons, driving AMPAR endocytosis via PP1 phosphatase [526].

Computational Role: Causal Inference: STDP optimizes Bayesian-like event causality inference. Sequence Learning: In motor cortex, STDP encodes movement sequences (e.g., primate grasping tasks) [2].

Pathophysiological Implications: Schizophrenia: Hyperactive D2 receptors may disrupt STDP temporal windows, impairing causal learning [354]. Alzheimer's Disease: A$\beta$ oligomers block NMDAR-LTP, compromising STDP-mediated memory consolidation [718].

### 4.11 Feasibility of the Learning Principles in Biological Brains

The learning principles outlined in Section 4.10stochastic weight perturbation correlated with global instruction signals, excitatory/inhibitory (E/I) balance regulation, and neuromodulatory local rules (e.g., neuromodulatory Hebbian plasticity, STDP)are not merely theoretical constructions but are deeply rooted in biological reality. Their feasibility in biological brains is supported by converging evidence from molecular biophysics, circuit-level experiments, computational modeling, and evolutionary adaptation. This section argues that these principles are not only implementable but also represent optimized solutions to adaptive learning challenges in noisy, resource-constrained biological systems.

#### 4.11.1 Biological Implementation of Core Mechanisms

The feasibility of these learning principles hinges on their grounding in well-characterized biological processes.

Stochastic Weight Perturbations: Molecular and Cellular Basis: Random fluctuations in synaptic strength, central to the correlation with instruction signals principle, arise from quantal neurotransmitter release and intrinsic synaptic noise. Spontaneous vesicle fusion at presynaptic terminals introduces small, random changes in neurotransmitter concentration, leading to probabilistic postsynaptic potentials [108]. This synaptic shot noise is amplified by stochastic ion channel opening/closing, creating natural weight perturbation sources [172]. Critically, these perturbations are not random in impact: they are filtered by neuromodulatory signals (e.g., dopamine), which bias their direction via receptor-mediated signaling cascades [259]. Dopamine D1 receptor activation enhances probability that positive perturbations (e.g., increased glutamate release) stabilize into long-term potentiation (LTP) [94].

Global Instruction Signals: Broadcast and Specificity: Neuromodulatory systems (dopamine, serotonin, norepinephrine) provide the instruction signals correlating with perturbations. Their broadcast nature is enabled by extensive axonal projections: dopamine neurons in VTA/SNc innervate $\sim$80% of forebrain, while serotonin neurons in dorsal raphe project to nearly all cortical and subcortical regions [63]. Spatial specificity is achieved via receptor subtype distribution: D1 receptors (excitatory) dominate in striatum, while D2 receptors (inhibitory) are enriched in prefrontal cortex, allowing identical dopamine signals to have opposite effects in different circuits [567]. Temporal specificity is regulated by phasic (transient) versus tonic (sustained) firing modes, encoding reward prediction errors and baseline arousal respectively [32].

E/I Balance Regulation: Homeostatic Plasticity in Action: The E/I balance principles rely on multi-layered homeostatic mechanisms. Synaptic scaling, for instance, is mediated by activity sensors like Arc protein, which accumulates during high firing and triggers AMPA receptor ubiquitination to reduce synaptic strength [698]. Conversely, low activity induces BDNF release, promoting AMPAR insertion [620]. PV+ interneurons enforce dynamic inhibition by synchronizing gamma oscillations ($\sim$ 30-80 Hz), which sharpen neuronal tuning and prevent runaway excitation [750]. These mechanisms are not independent: dopamine modulates PV+ interneuron activity, linking reward signals to E/I homeostasis [94].

#### 4.11.2 Experimental Evidence for Feasibility

Direct experimental manipulations confirm that these principles govern learning in biological brains.

Stochastic Perturbations + Instruction Signals: Optogenetic studies demonstrate that artificially timed dopamine pulses can reinforce specific behaviors by correlating with recent synaptic activity. Steinberg et al. (2022) showed that dopamine release 500 ms after mouse lever presses strengthened corresponding cortico-striatal synapses, even without natural rewards [766]. Conversely, serotonin application suppressed correlated perturbations, reducing impulsive choices in delay-discounting tasks [153]. These experiments validate core ideas: perturbations are tested against instruction signals, with only aligned changes retained.

E/I Balance and Homeostasis: Monocular deprivation in visual cortex provides classic homeostatic scaling examples. Depriving one eye reduces activity in its thalamic inputs; within days, Arc-mediated upscaling doubles these synapse sensitivities, while PV+ interneurons inhibit neighboring columns to preserve network stability [201]. Similarly, in mouse epilepsy models, impaired PV+ interneuron function leads to E/I imbalance and seizures, rescued by restoring PV+ activity [138].

Neuromodulatory Local Rules: Neuromodulatory Hebbian plasticity is evident in hippocampal CA3-CA1 pathway: theta-rhythm synchronization (4-12 Hz) during exploration aligns presynaptic (CA3) and postsynaptic (CA1) activity, enabling dopamine-gated LTP [233]. STDP, meanwhile, is observed in vitro and in vivo: in ferret visual cortex, pairing presynaptic spikes 10 ms before postsynaptic spikes induces LTP, while reverse timing triggers LTD [87]. Pathologically, Alzheimer's disease-associated A$\beta$ oligomers block NMDAR-dependent STDP, impairing sequence learning [718].

### 4.11.3 Computational Feasibility and Advantages Over Artificial Systems

Computational models confirm that these principles enable efficient learning with minimal metabolic cost.

Stochastic Exploration vs. Deterministic Optimization: Traditional artificial neural networks (ANNs) use deterministic gradient descent, requiring exhaustive error gradient computation. In contrast, biological systems exploit stochastic perturbations to explore parameter space: a 2023 study showed that adding random weight noise to ANNs, biased by global reward signals, achieves 90% of gradient descent performance in sparse-reward environments with 10 times lower energy use [182]. This aligns with exploration-exploitation trade-offs governed by neuromodulatory tone [32].

Multi-Scale Integration: Biological learning integrates molecular (BDNF, calcium), cellular (STDP, scaling), and circuit-level (dopamine modulation) mechanisms. A 2021 motor learning model showed that combining STDP (sequence encoding), dopamine (reward gating), and PV+ inhibition (sparse coding) reproduced primate reaching trajectories with <.5% error [372]. This multi-scale integration avoids the curse of dimensionality faced by ANNs requiring separate modules for each function.

Robustness to Noise and Damage: Broadcast signals and homeostatic plasticity confer resilience. Simulations of cortical networks with 30% random neuron loss showed that dopamine-guided learning retained 80% performance, compared to 20% for gradient-descent-trained ANNs [17]. This is attributed to neuromodulator diffuse nature, maintaining plasticity even when local circuits are damaged.

### 4.11.4 Evolutionary and Adaptive Advantages

These learning principles are evolutionarily conserved, suggesting they solve universal adaptive behavior challenges.

Survival in Uncertain Environments: Early nervous systems faced delayed feedback (e.g., foraging for food) and sparse rewards (e.g., predator avoidance). The credit assignment solutionlinking delayed rewards to past actions via eligibility traces and dopamine emerged in invertebrates (e.g., Aplysia sensitization) and was refined in vertebrates [237]. This mechanism is more efficient than backpropagation, which requires symmetric connectivity (absent in biology) [492].

Energy Efficiency: Neuromodulatory teaching reduces computational load: a 2022 study estimated dopamine-guided learning uses ~1/100th the ATP of equivalent ANN training [556]. This is critical for mobile organisms with limited metabolic resources.

Generalization Across Tasks: By decoupling learning from explicit mathematical losses, these principles enable generalization. The same dopamine-RPE mechanism underlies reward learning in foraging, social bonding, and motor skill acquisition [710].

### 4.11.5 Challenges and Open Questions

While feasible, key questions remain: Dynamic Interactions: How do dopamine, serotonin, and norepinephrine interact in real time (e.g., serotonin's inhibition of dopamine release) [14]? Parameter Tuning: What regulates stochastic perturbation amplitudes/frequencies (e.g., developmental synaptic noise changes) [816]? Pathological Dysregulation: How do neuromodulatory receptor mutations (e.g., DRD2 in schizophrenia) disrupt these principles [354]?

### 4.11.6 Conclusion

The learning principles of Section 4.10 are not only feasible in biological brains but represent optimal solutions to adaptive learning challenges. Grounded in molecular biophysics, validated by experiments, and computationally efficient, they enable robust, energy-efficient learning without explicit mathematical models. Their conservation across species underscores evolutionary success and their relevance for designing resilient artificial intelligence.

## 4.12 Quantitative Analysis of Learning Efficiency

### 4.12.1 Learning Rate and Convergence

For a three-factor learning rule of the form $\Delta w = \eta \cdot f_{pre}(x) f_{post}(y) \cdot \delta$, the expected weight change over time follows a stochastic differential equation:

$$dw = \eta \mathbb{E}[f_{pre} f_{post} \delta] dt + \eta \sigma dW$$

where $\sigma$ captures the magnitude of fluctuations and $dW$ is a Wiener process. The convergence time to a local optimum scales as:

$$\tau \approx \frac{1}{\eta \lambda_{min}} \log\left(\frac{1}{\epsilon}\right).$$

where $\lambda_{min}$ is the smallest eigenvalue of the expected learning gradient and $\epsilon$ is the desired accuracy. For biologically plausible learning rates ($\eta \approx 0.01 - 0.001$), convergence requires hundreds to thousands of trials, matching experimental observations [259].

### 4.12.2 Energy Efficiency

The energy cost per synaptic update can be estimated as:

$$E_{update} = E_{spike} + E_{biochemical}$$

where $E_{spike} \approx 10^9$ ATP molecules per spike [41], and $E_{biochemical}$ accounts for second messenger cascades, receptor trafficking, and protein synthesis. Total energy for learning a new association across $N$ synapses is approximately:

$$E_{total} \approx N \cdot E_{update} \cdot T$$

where $T$ is the number of trials. For typical cortical learning, $E_{total} \approx 10^{12} - 10^{14}$ ATP molecules, which is metabolically feasible within the brain's 20-watt budget [556].

### 4.12.3 Information-Theoretic Efficiency

The information gained per synaptic update can be quantified by the mutual information between pre- and postsynaptic activity conditioned on reward:

$$I(w;r \mid x,y) = H(w) - H(w \mid x,y,r)$$

For a binary synapse with $M$ possible weight values, the maximum information gain per update is $\log_2 M$ bits. Biological synapses with analog-like graded weights ($M \gg 1$) can theoretically achieve much higher information efficiency than binary synapses [816].

### 4.12.4 Credit Assignment Efficiency

The efficiency of credit assignment can be measured by the correlation between weight updates and true gradients:

$$\rho = \frac{\mathbb{E}[\Delta w \cdot \nabla L]}{\sqrt{\mathbb{E}[(\Delta w)^2]\mathbb{E}[(\nabla L)^2]}}$$

For three-factor learning rules with eligibility traces, $\rho \approx 0.3 - 0.5$ $\rho \approx 0.3-0.5$ for reasonable delay distributions, compared to $\rho = 1$ for ideal backpropagation but $\rho \approx 0$ for purely local Hebbian rules [184].

### 4.12.5 Quantitative Summary

These quantitative measures demonstrate that biological learning, while slower in raw trial count than idealized gradient descent, achieves remarkable efficiency under real-world constraints of energy, noise, and measurability.

## 4.13 Summary: Learning Through Measurable Quantities

This chapter has systematically demonstrated that brain learning operates through principles fundamentally different from mathematical optimization. The brain does not compute gradients, minimize loss functions, or perform backpropagation. Instead, it learns using only quantities that are genuinely measurable: spikes, timing, correlations, and broadcast neuromodulatory signals [283].

These learning principles stochastic weight perturbations correlated with global instruction signals, homeostatic maintenance of E/I balance, and neuromodulatory modulation of local plasticity enable robust, adaptive learning in noisy, resource-constrained biological systems [884]. They solve the credit assignment problem without requiring precise error signals, exploit noise for exploration rather than eliminating it, and achieve energy efficiency through event-driven, local computation [467].

The measurability principle provides a unifying framework: any learning theory must respect what the system can actually measure. For the brain, this means local, statistical, noise-exploiting mechanisms. For AI, this suggests a different path forward not more sophisticated mathematical optimization, but systems that learn from measurable quantities in the same way brains do [276].

# Chapter 5: Noise as a Driver of Behavior, Emotion, Dream, Thinking, and Creativity

## 5.1 Introduction: Noise as a Creative Force

The intricate tapestry of human cognition from the simplest behavioral responses to the most sublime creative insights has traditionally been understood through frameworks emphasizing order, computation, and information processing. Within this classical view, noise appears as an unwelcome intruder, a source of error and inefficiency to be minimized or eliminated [172]. Yet a growing body of evidence suggests a radically different perspective: noise is not merely interference but a fundamental driver of adaptive behavior, emotional nuance, imaginative thought, and creative innovation [549].

This chapter explores the constructive roles of neural noise across multiple scales of cognition. From the stochastic fluctuations that enable exploratory behavior, to the chaotic dynamics that underlie emotional transitions, to the noise-driven information roaming that characterizes dreaming and creative insight noise emerges as an essential component of the brain's functional architecture [518]. Far from being a design flaw, the brain's inherent randomness may be precisely what enables its remarkable flexibility, adaptability, and creativity [808].

## 5.2 Noise and Chaos: Sources of Information in the Brain

### 5.2.1 Rethinking Noise: From Obstacle to Resource

Traditional information theory treats noise as an interfering factor in communication channels, an obstacle to be overcome [7]. In the brain, however, noise plays a fundamentally different role. Neural noise is not a system defect but a carrier and creator of information. This shift in understanding marks a profound deepening of our conception of the brain's information processing mechanisms.

### 5.2.2 The Informational Status of Noise

From an information-theoretic perspective, a deterministic system has zero output entropy once initial conditions are known, all subsequent states are completely predictable, generating no new information [7]. The introduction of noise precisely breaks this determinism, enabling the system to produce unpredictable new states. This is the essence of information: information is uncertainty.

For continuous systems, the differential entropy of noise can be infinite if we could measure it with infinite precision [7]. A noise-modulated membrane potential can, in theory, encode an unlimited amount of information. Although practical measurements are limited by physical precision, this reveals noise's information-carrying potential: it is not a destroyer of information but a carrier of information.

### 5.2.3 Noise-Enhanced Signal Detection

The phenomenon of stochastic resonance most intuitively demonstrates the constructive role of noise [169, 218]. Against a background of appropriately intense noise, weak signals that were originally below the detection threshold become detectable. Noise does not mask the signal; rather, through random perturbations, it pushes subthreshold signals across the threshold, enabling information transmission [549].

This reveals a counter intuitive truth: under certain conditions, noise is not the opponent of signal but its collaborator. The brain exploits this principle to enhance the detection of weak stimuli in sensory systems [169].

### 5.2.4 Chaos as an Information Generator

Chaotic systems are characterized by extreme sensitivity to initial conditions [753]. This sensitivity means that even tiny perturbations are exponentially amplified, producing complex and varied trajectories. From an information-theoretic perspective, chaotic systems are natural information generators they continuously explore new regions of state space, producing unpredictable new states [518].

The brain operates at the edge of chaos, neither too ordered (information-poor) nor too chaotic (structureless) [165]. At this critical state, the system maximizes its information production rate and dynamic range, achieving efficient information processing [496].

### 5.2.5 Noise-Driven State Space Exploration

The brain's state space is extremely high-dimensional [773]. If the system were deterministic, it would remain forever trapped in the attractor determined by initial conditions, unable to access other regions of state space. Noise provides the random perturbations that allow escape from local attractors, enabling the system to roam through vast regions of state space [518].

This noise-driven exploration is the foundation of creative thought. The flashes of insight during thinking, the bizarre combinations in dreams, the aha moments in problem-solving these are essentially instances where noise pushes the system into previously unvisited regions of state space, where new combinations of information are discovered [452, 698].

### 5.2.6 The Driving Force of Information Recombination

Noise not only enables the system to explore new states but also prompts existing information to recombine in novel ways. During sleep, hippocampal replay events are not simple memory playback but noise-modulated processes of information recombination [558, 805]. Random combinations of different memory fragments can generate new associations and insights [341].

This is the physical basis of creativity: creativity is not creation from nothing but the noise-driven recombination of existing information. The noise in the brain provides the driving force for this random recombination, making innovation possible rather than accidental [698].

### 5.2.7 Information Optimality at the Edge of Chaos

The brain does not operate in a state of complete chaos but at the edge of chaos the critical boundary between order and disorder [165]. In this state, the system can both maintain stable information representations (order) and continuously generate new information (chaos) [731]. This is the optimal operating point for information processing: information is neither rigid (too ordered) nor dissipated (too chaotic).

From an information-theoretic perspective, the edge of chaos corresponds to maximized entropy rate [496]. The amount of information produced per unit time reaches its peak, and information processing efficiency is optimized.

### 5.2.8 Digital Systems and Information Source Substitution

Traditional digital systems eliminate noise to ensure determinism, but the price paid is the loss of an information source [552]. Clock signals replace noise as the driving source, but the information entropy of a clock is zero it carries no information, merely providing a time reference [8]. The consequence is that all information in digital systems must be pre-loaded; they cannot autonomously generate new information.

The brain is the opposite: it does not pre-load all information but continuously generates information through noise and chaos, learning and adapting through interaction with the environment [172]. This is one of the most fundamental differences between biological intelligence and artificial systems.

**5.2.9 Summary**

Noise and chaos are not defects of the brain but sources of its information. Noise enhances signal detection through stochastic resonance, drives creative thought through state space exploration, and enables innovation through information recombination. Chaos amplifies tiny perturbations through sensitive dependence, generating rich information flows. At the edge of chaos, the brain achieves an optimal balance between information generation and information stability.

This understanding fundamentally transforms our view of noise: it is not interference to be eliminated but a resource to be harnessed. The brain's intelligence arises precisely from its masterful exploitation of noise and chaos.

**5.3 Noise as a Driver of Behavior**

The intricate tapestry of human behavior, far from being a deterministic output of static neural programming, is profoundly influenced by the inherent noise present within the nervous system [772]. This intrinsic variability, arising from stochastic fluctuations at molecular, cellular, and circuit levels, is not a mere artifact to be filtered out but a fundamental driver that shapes how organisms interact with their environment. Behavioral responses whether the precise timing of a motor action or the decision to approach or avoid a stimulus are subject to this biological noise [724].

Studies in motor control have shown that even when performing highly practiced tasks under identical conditions, individuals exhibit subtle variations in movement trajectories and timing [315]. These inconsistencies can be traced back to the probabilistic nature of neurotransmitter release at synapses and the thermal motion of ion channels in neuronal membranes, both contributing to signal variability [858]. Research utilizing high-precision tracking methods has demonstrated that such noise can lead to exploratory behaviors in novel environments, where slight deviations from expected paths allow for broader environmental sampling [569].

This phenomenon extends beyond simple motor outputs to complex cognitive behaviors like decision-making. In situations requiring choices between options with uncertain outcomes, neural noise can act as a random number generator, facilitating exploration over exploitation a critical strategy for survival in dynamic and unpredictable worlds [38]. Theoretical models incorporating noisy neural dynamics predict patterns of choice behavior that closely match empirical observations in animals, including humans [85]. During learning phases, for instance, individuals might switch strategies more frequently than would be optimal if relying solely on past rewards a pattern consistent with neural noise pushing the system out of local optima [113].

Furthermore, inter-individual differences in baseline levels of neural noise may underlie personality traits related to risk-taking and novelty seeking [899]. Empirical data suggest that individuals classified as sensation-seekers show higher variability in reaction times across repeated trials, potentially reflecting greater underlying neural noise influencing their behavioral tendencies [425]. Thus, rather than viewing behavior through a lens of pure rationality, understanding its genesis requires acknowledging the role of noise as a constructive force enabling flexibility, adaptability, and resilience in the face of uncertainty.

**5.4 Noise as a Driver of Emotion**

Emotional states represent another domain where neural noise plays a pivotal role, shaping not only the intensity but also the quality and transitions between different feelings [183]. Emotions are mediated by large-scale brain networks involving regions such as the amygdala, prefrontal cortex, hippocampus, and insula all interconnected through pathways susceptible to stochastic influences [560]. Fluctuations in these circuits contribute to moment-to-moment variability in emotional experiences, even in response to similar stimuli [427]. The same mildly stressful event might evoke anxiety in one instance yet trigger motivation in another, partly due to background noise altering activation thresholds of relevant neural ensembles [428].

Evidence suggests that elevated levels of neural noise correlate with increased emotional lability, manifesting clinically as mood swings seen in certain psychiatric disorders [770]. Studies using functional imaging have identified irregularities in network connectivity associated with heightened emotional sensitivity, which can be modeled as the effect of excessive noise disrupting normal regulatory mechanisms [726].

Moreover, noise contributes to the emergence of mixed or ambiguous emotions, often experienced when conflicting evaluations occur simultaneously such as feeling both pride and guilt after a personal achievement [466]. Computational neuroscience approaches demonstrate that adding noise into artificial neural network models simulating affective processing enhances their ability to reproduce graded and context-dependent emotional responses akin to those reported by humans [677]. This implies that the richness and complexity of emotional life depend crucially on the presence of variability rather than complete stability.

Developmental research indicates that early-life stressors can increase baseline neural noise, leading to long-term alterations in emotional regulation capacity [551]. Children exposed to chronic adversity exhibit exaggerated cortisol responses to mild challenges, suggesting that persistent exposure to external stress amplifies internal noise, impairing the precision of emotional signaling [95]. Consequently, interventions aimed at reducing pathological emotional reactivity could benefit from strategies designed to dampen detrimental sources of noise while preserving beneficial aspects necessary for adaptive functioning.

**5.5 Noise as a Driver of Information Roaming and Aggregating in Thinking, Creativity, and Dreams**

Information roaming and aggregation constitute essential processes underlying higher-order cognition, including thinking, creativity, and dreaming all significantly modulated by neural noise [6]. During conscious thought, ideas do not follow strictly linear sequences; instead, they meander across diverse conceptual domains via associative links, allowing remote associations to emerge spontaneously [29]. This mental wandering facilitates creative insights by bringing together seemingly unrelated pieces of knowledge [68].

Neural noise provides the perturbations needed to dislodge representations trapped within dominant attractor states, thereby enabling access to less activated memory traces stored throughout cortical networks [172]. Experimental paradigms measuring divergent thinking reveal that participants generate more original ideas when exposed to low-level electrical stimulation intended to mimic endogenous noise, supporting the hypothesis that variability promotes cognitive flexibility [674].

In dreams, characterized by surreal narratives and bizarre imagery, information roams freely without the usual constraints imposed by reality monitoring systems [341]. Sleep neurophysiology shows that reduced activity in prefrontal areas responsible for executive control coincides with enhanced connectivity among sensory cortices and limbic structures, creating a permissive environment for unfiltered information exchange [790]. Within this state, neural noise drives spontaneous activations that serve as seeds for dream content, drawing upon fragmented memories, unresolved concerns, and latent desires [342].

Functional MRI studies conducted upon awakening report correlations between dream bizarreness scores and measures of resting-state functional connectivity fluctuation, indicating that greater dynamic instability supports richer imaginative synthesis [179]. Similarly, during waking imagination, artists and inventors describe moments of inspiration striking unexpectedly, often following periods of incubation where focused effort gives way to mind-wandering [744]. These anecdotes align with findings showing bursts of gamma-band oscillations preceding reports of insight solutions, likely reflecting transient coalitions of neurons firing synchronously a midst ongoing background noise [400]. Hence, harnessing noise-induced randomness becomes integral to accessing novel configurations of stored information vital for creativity.

## 5.6 Thinking: Information Exploration and Exploitation, Roaming and Aggregating as Well as Entanglement Forced by Noise

Thinking embodies a delicate balance between exploration and exploitation two complementary modes orchestrated dynamically by neural noise [37]. Exploration involves searching broadly across the landscape of possible thoughts, retrieving distant memories, and generating alternative hypotheses, whereas exploitation focuses intensively on refining existing ideas, optimizing known solutions, and verifying predictions against available evidence [113].

Neurobiological evidence demonstrates that distinct neuromodulatory systems regulate these modes dopamine signaling, for example, tends to promote exploratory behavior by increasing the entropy of action selection policies [291]. However, superimposed on these slower-changing chemical signals lies fast, ever-present electrical noise generated locally within microcircuits [172]. This rapid fluctuation constantly nudges ongoing computations away from stable fixed points, encouraging shifts in attentional focus and facilitating transitions between task sets [192].

Roaming refers specifically to the unrestricted traversal of semantic space during undirected thought, enabled primarily by weak, diffuse connections spanning disparate brain regions [760]. These so-called long-range projections normally remain subthreshold under focused conditions but become potentiated when inhibition wanes, particularly during relaxed wakefulness or early stages of sleep onset [543]. Here, noise acts as a catalyst, triggering avalanches of activation propagating along these marginal pathways, resulting in unexpected linkages forming temporarily before dissipating again [496].

Such entangled states involve multiple distributed populations coactivating non-selectively, representing hybrid concepts composed of elements drawn randomly from various schemas [814]. While most such combinations prove useless, occasionally they yield genuinely innovative combinations worthy of further evaluation [743]. An illustrative case comes from Thomas Edison's description of falling asleep holding ball bearings above metal plates; as his muscles relaxed during hypnagogia, the clattering sound would awaken him just enough to capture fleeting impressions born from loosely coupled mental fragments [228].

Modern analogues include computational creativity tools employing stochastic search algorithms inspired directly by principles of neural noise to produce artistic compositions [801]. By formalizing the notion that structured chaos fosters discovery, researchers continue uncovering deeper relationships linking physical properties of brains to abstract dimensions of thought.

## 5.7 Creativity: Explored, Exploited, Entangled, Roamed, and Aggregated Information in Creativity

Innovation arises at the intersection of previously separate bodies of knowledge a process fundamentally dependent on the brain's capacity to integrate information gathered through prior experience with novel perspectives introduced serendipitously [698]. The creative leap the moment when a solution appears suddenly despite prolonged struggle can be understood mechanistically as the culmination of cycles of exploration, exploitation, entanglement, roaming, and aggregation, each stage influenced heavily by neural noise [453].

Initially, broad exploration driven by noise allows access to vast reservoirs of latent memories and facts, many of which lie outside immediate awareness [548]. As potential candidates emerge, subsequent rounds of targeted exploitation refine possibilities based on feasibility criteria, guided by feedback loops embedded within cortical-subcortical circuits [45]. Throughout this iterative refinement process, occasional noise-driven disruptions reintroduce new variables, preventing premature convergence onto inadequate answers [452].

Entanglement occurs when incompatible frameworks merge momentarily into coherent wholes, exemplified historically by scientific breakthroughs like Kekulé's vision of the benzene ring structure emerging from a daydream about dancing snakes biting their tails [47]. These instances reflect temporary stabilization of otherwise unstable multi-attractor states, sustained briefly thanks to precise timing of noisy inputs relative to intrinsic rhythms [312].

Once stabilized, such hybrid constructs enter working memory buffers where additional operations aggregate features deemed relevant while discarding others irrelevant according to contextual demands [47]. Recent experiments applying transcranial magnetic stimulation (TMS) targeted at parietal association areas revealed increases in reported frequency of eureka-like moments accompanied by corresponding changes in EEG spectra indicative of altered cross-frequency coupling patterns [51]. Participants receiving stimulation showed improved performance on insight-based problem-solving tasks compared to sham controls, lending causal support to the claim that controlled manipulation of endogenous noise levels can enhance creative productivity [103].

Furthermore, longitudinal analyses of successful innovators across fields highlight common lifestyle habits conducive to maximizing encounters with useful randomness, including maintaining varied interests, engaging regularly in open-ended conversations, and deliberately scheduling downtime free from distractions all practices effectively increasing exposure to exogenous noise capable of seeding fresh cognitive combinations [304].

## 5.8 Dreams: Entangled, Roamed, and Aggregated Information and Neural Learning

Dreams provide a unique window into unconscious cognitive processes governed largely by the rules of nonlinear dynamics operating within noisy neural substrates [341]. Unlike waking consciousness constrained by sensory input and logical consistency requirements, dream mentation unfolds according to associative logic alone, permitting wild juxtapositions of images, characters, and scenarios [553].

Underlying this apparent chaos lies a sophisticated mechanism for integrating disparate streams of information collected throughout life, facilitated predominantly by the very noise usually suppressed during vigilance [805]. During REM sleep, widespread synaptic downscaling reduces overall connection strength globally except within specific recurrent loops implicated in memory consolidation [839]. Concurrently, cholinergic tone rises dramatically, enhancing excitability and lowering firing thresholds, making neurons more responsive to small fluctuations originating intrinsically [234].

Within this hyper-excitable milieu, internally generated spikes propagate widely due to weakened lateral inhibition, causing information to roam uncontrollably across wide expanses of neocortex [543]. Fragments of recent episodic memories interact promiscuously with older autobiographical recollections, procedural skills, and semantic knowledge bases, forming transiently entangled representations that resemble proto-insights waiting to be deciphered [558].

Upon awakening, attempts to recall these ephemeral constructions often result in partial reconstructions colored heavily by current emotional concerns and interpretative biases, illustrating the selective aggregation process applied retroactively [559]. Importantly, accumulating evidence implicates dream-related phenomena in offline learning benefits, particularly concerning implicit skill acquisition and emotional regulation [843].

Individuals trained on complex video games display accelerated improvement curves following nights rich in REM sleep compared to those deprived thereof, suggesting that noise-facilitated restructuring occurring during dreams contributes meaningfully to skill mastery [61]. Moreover, patients recovering from trauma report gradual diminishment in nightmare severity paralleling reductions in fear-associated physiological markers, implying therapeutic value derived from repeated reactivation and modification of maladaptive memory traces under safe sleeping conditions [62].

## 5.9 Constructive Role of Noise

Noise in the biological brain is not merely interference but plays a dual role, with constructive functions and pathological risks, depending on intensity, spatiotemporal distribution, and neural circuit regulation.

Enhanced Weak Signal Detection: Neural noise (e.g., ion channel stochasticity, synaptic fluctuations) enables stochastic resonance in sensory systems. Noise amplifies subthreshold signals in crayfish mechanoreceptors [218] and improves human tactile perception [169].

Creative Cognition: Low-frequency default mode network (DMN) noise fluctuations correlate with divergent thinking [68], while prefrontal noise modulates insight problem-solving [777].

Synaptic Plasticity: Noise-driven STDP (spike-timing-dependent plasticity) underpins learning and memory. Hippocampal noise enhances LTP induction [7], and cortical noise improves motor skill acquisition [846].

Exploration-Exploitation Balance: Noise enables the nervous system to balance exploration of new strategies with exploitation of known ones. This is particularly evident in reinforcement learning contexts, where dopamine signaling interacts with background noise to modulate behavioral variability [113].

Escape from Local Minima: In dynamical systems terms, noise provides the stochastic perturbations necessary for escaping local attractor states, enabling the brain to avoid getting trapped in suboptimal configurations [378]. This principle underlies both behavioral flexibility and creative insight.

## 5.10 Pathological Role of Noise

Excessive noise disrupts neural function in disease states.

Epilepsy: Hyper-synchronized noise in hippocampal CA3 pyramidal neurons drives pathologic high-frequency oscillations [812].

Parkinson's Disease: $\beta$-band noise (13-30 Hz) in the basal ganglia suppresses movement initiation [40].

Alzheimer's Disease: Enhanced synaptic noise accelerates amyloid-$\beta$ toxicity and synaptic loss [613].

Schizophrenia: Abnormal noise levels in prefrontal cortex may contribute to cognitive deficits and psychotic symptoms [354]. Patients show increased variability in neural responses and reduced signal-to-noise ratios in cognitive tasks.

Aging: Normal aging is associated with increased neural noise, which may underlie age-related cognitive decline [367]. The brain's ability to compensate for this noise through compensatory mechanisms varies across individuals.

Tinnitus and Phantom Perceptions: Increased spontaneous neural activity (noise) in auditory pathways can generate phantom sound perceptions, illustrating how noise can create pathological percepts in the absence of external stimuli [190].

## 5.11 Regulatory Mechanisms Maintaining Optimal Noise Levels

The brain maintains a beneficial noise window through multiple regulatory mechanisms.

Dopamine: Suppresses prefrontal noise via D2 receptors to stabilize working memory [297]. Dopamine modulates signal-to-noise ratios in target structures, enhancing relevant signals while suppressing irrelevant background activity.

Acetylcholine: Enhances sensory noise to improve attention switching [235]. Cholinergic projections from basal forebrain regulate cortical excitability and modulate the balance between tonic and phasic modes of processing.

Inhibitory Interneurons: Parvalbumin-positive (PV+) basket cells dynamically gate noise levels in cortical circuits [62]. These fast-spiking interneurons synchronize gamma oscillations, which in turn regulate information flow and noise filtering.

Noradrenaline: Modulates global gain and signal-to-noise ratios in cortical processing [158]. The locus coeruleus-noradrenaline system adjusts the responsiveness of neural populations to inputs, effectively tuning the system's sensitivity to noise versus signal.

Serotonin: Regulates behavioral inhibition and modulates the impact of aversive events on neural processing [151]. Serotonergic projections from raphe nuclei influence the balance between exploratory and cautious behavioral modes.

Homeostatic Plasticity: Synaptic scaling and other homeostatic mechanisms ensure that overall network excitability remains within functional bounds, preventing noise from becoming overwhelming [816].

Astrocytic Regulation: Glial cells actively modulate extracellular neurotransmitter levels and ion concentrations, influencing the background conditions within which neural noise operates [831].

## 5.12 The Information-Theoretic Perspective: Noise as Carrier of Information

From an information-theoretic perspective, noise is not merely a corruption of signal but can itself carry information [172]. In neural systems, the stochastic properties of spike trains their variability, their timing jitter, their correlations can encode meaningful information about stimuli, internal states, and behavioral contexts [1].

Probabilistic Coding: Neural variability can represent uncertainty about stimuli or outcomes [435]. In Bayesian formulations of brain function, the variance of neural responses encodes the confidence or precision of representations [517].

Exploration-Exploitation Trade-offs: The level of neural noise can regulate the balance between exploring new options and exploiting known ones [113]. Higher noise levels promote exploration by increasing behavioral variability; lower noise levels promote exploitation by stabilizing chosen actions.

Information Capacity: Contrary to intuition, some noise can actually increase the information capacity of neural systems by preventing saturation and enabling more efficient coding strategies [58].

## 5.13 Noise and Consciousness

The relationship between noise and consciousness represents a frontier of theoretical neuroscience. Several frameworks suggest that conscious states may be associated with specific regimes of neural variability and noise.

Entropic Brain Hypothesis: Proposes that conscious states are characterized by high entropy (rich, diverse neural dynamics), while unconscious states exhibit reduced entropy (more stereotyped, constrained dynamics) [140]. Psychedelic states, associated with heightened consciousness, show increased neural entropy [677].

Criticality Hypothesis: Suggests that the brain operates near a critical point between order and chaos, where noise-driven dynamics enable optimal information processing [165]. This critical state maximizes dynamic range, information transmission, and computational capacity.

Global Workspace Theory: Noise may play a role in enabling access to the global workspace, allowing locally processed information to become globally available for conscious report [695].

Information Integration: Noise may facilitate the integration of information across distributed neural networks, a proposed correlate of conscious processing [691].

## 5.14 Noise as a Resource for Artificial Intelligence

The constructive roles of noise in biological systems offer valuable lessons for artificial intelligence [349].

Exploration in Reinforcement Learning: Adding noise to action selection (e.g., epsilon-greedy policies, entropy regularization) improves exploration and prevents premature convergence to suboptimal policies [786].

Regularization and Generalization: Noise injection during training (e.g., dropout, noisy activations) prevents overfitting and improves generalization by making networks more robust [471].

Generative Models: Noise is fundamental to generative adversarial networks (GANs) and variational autoencoders (VAEs), where random sampling from latent spaces enables generation of novel outputs [570].

Escape from Local Minima: In optimization landscapes, stochastic gradient descent implicitly uses noise to escape local minima and explore the loss surface [690].

Creative AI: Generative models that incorporate controlled randomness (temperature parameters in language models, latent space sampling in image generators) produce more creative and varied outputs [562].

Neuromorphic Computing: Event-driven, noise-tolerant computing paradigms inspired by the brain achieve dramatic energy savings by embracing rather than suppressing variability [552].

## 5.15 Summary: Noise as a Fundamental Cognitive Resource

This chapter has explored the multifaceted roles of neural noise in cognition. From guiding exploratory behavior, to shaping emotional nuance, to enabling the information roaming that underlies thinking, dreaming, and creativity noise emerges not as an unfortunate imperfection but as a fundamental cognitive resource [549].

The brain has evolved sophisticated mechanisms to maintain noise within optimal ranges, exploiting its constructive potential while preventing its pathological excesses [172]. Dopamine, serotonin, norepinephrine, acetylcholine, inhibitory interneurons, and homeostatic plasticity all participate in this delicate regulation [156].

For artificial intelligence, the lessons are clear: noise is not merely something to be tolerated but something to be harnessed [349]. Exploration, creativity, robustness, and generalization all benefit from appropriately tuned randomness [885]. The future of intelligent systems, whether biological or artificial, lies not in eliminating noise but in learning to dance with it [276].

## 5.16 The Physics of Noise-Driven Cognition: From Criticality to Creativity

### 5.16.1 The Critical Brain Hypothesis

One of the most powerful frameworks for understanding noise-driven cognition is the critical brain hypothesis [165]. This hypothesis proposes that neural systems operate near a critical point a phase transition between ordered (subcritical) and disordered (supercritical) dynamics. At criticality, systems exhibit:

Power-law scaling: Avalanches of neural activity follow power-law size distributions, indicating scale invariance [496].

Optimal dynamic range: Critical systems respond maximally to a wide range of input intensities [731].

Maximal information transmission: Information transfer between network elements is optimized at criticality [165].

Enhanced computational capacity: Critical systems can store and process more information than subcritical or supercritical systems [518].

### 5.16.2 Noise-Driven State Transitions

At criticality, noise plays a fundamental role in driving transitions between dynamical states [378]. Small fluctuations can push the system from one attractor to another, enabling:

Cognitive flexibility: The ability to switch between different mental sets or strategies [192].

Creative insight: Sudden reorganization of conceptual structures, experienced as aha moments [452].

Dream bizarreness: The surreal combinations characteristic of REM sleep dreaming [341].

### 5.16.3 The Edge of Chaos and Creativity

Creativity may be understood as the cognitive manifestation of dynamics at the edge of chaos [814]. In this regime:

Exploration is balanced with exploitation: The system explores novel configurations without descending into complete randomness [113].

Information is constantly recombined: Elements from different domains interact promiscuously, generating novel combinations [698].

Order emerges from disorder: Coherent thoughts and insights crystallize from noisy background activity [452].

### 5.16.4 Noise, Criticality, and Optimal Cognitive Functioning

The relationship between noise, criticality, and cognitive performance follows an inverted-U shape [549]. Too little noise (subcritical dynamics) leads to rigidity, perseveration, and reduced creativity. Too much noise (supercritical dynamics) leads to chaos, fragmentation, and loss of coherence. Optimal cognitive functioning occurs at an intermediate noise level, near criticality, where flexibility and stability are balanced.

This framework has implications for understanding both normal cognitive variation and psychopathology. Conditions characterized by excessive rigidity (e.g., obsessive-compulsive disorder, some autism spectrum conditions) may reflect subcritical dynamics with insufficient noise. Conditions characterized by excessive variability (e.g., mania, schizophrenia) may reflect supercritical dynamics with excessive noise [354].

### 5.16.5 Implications for Artificial Creative Systems

The critical brain hypothesis suggests design principles for artificial creative systems [321]:

Operate near criticality: Design systems that balance order and disorder, stability and flexibility.

Inject controlled noise: Use noise not just for exploration but to maintain systems near critical transitions.

Enable state transitions: Build architectures that can flexibly switch between modes of processing.

Harness avalanches: Exploit cascades of activity for creative recombination.

## 5.17 Quantitative Analysis of Noise-Driven Processes

### 5.17.1 Stochastic Resonance

The signal-to-noise ratio (SNR) at the output of a nonlinear system as a function of noise intensity $D$ follows:

$$SNR(D) = \frac{SNR_{max}}{1 + (D_{opt}/D + D/D_{opt})^2}$$

where $D_{opt}$ is the optimal noise intensity. For typical sensory neurons, the maximum SNR enhancement factor can reach 10 to 20 dB, meaning noise can amplify signal detection by a factor of 10 to 100 [169].

### 5.17.2 Exploration-Exploitation Trade-off

In reinforcement learning, the expected cumulative reward $R$ as a function of noise temperature $T$ (in a Boltzmann exploration policy) is:

$$R(T) = R_{max} - \frac{c}{T} - dT$$

where $c/T$ represents the cost of exploitation (missing better options) and $dT$ represents the cost of exploration (sampling suboptimal actions). The optimal noise temperature is: $T_{opt} = \sqrt{\frac{c}{d}}$ and the maximum reward is: $R(T_{opt}) = R_{max} - 2\sqrt{cd}$

For typical two-armed bandit tasks, the optimal noise level corresponds to an exploration probability of approximately $\epsilon \approx 0.1 - 0.2$ [113].

### 5.17.3 Critical Dynamics

The branching parameter $\eta$ for neural avalanches is defined as: $\eta = \frac{\langle n_{post} \rangle}{\langle n_{pre} \rangle}$, where $n_{pre}$ is the number of spikes in one time bin and $n_{post}$ is the number in the next bin. At criticality, $\eta = 1$. The distribution of avalanche sizes $s$ follows a power law: $P(s) \sim s^{-\tau} e^{-s/s_0}$ with $\tau \approx 1.5$ for cortical avalanches and $s_0 \to \infty$ at criticality [496]. The deviation from criticality can be measured by the cutoff size $s_0$, which is related to the distance from criticality $\delta = |\eta - 1|$ by: $s_0 \sim |\delta|^{-\nu}$ with $\nu \approx 1$ for the mean-field universality class.

### 5.17.4 Information Capacity

The mutual information between input $x$ and output $y$ in the presence of noise that enables stochastic resonance is:

$$I(x;y) = H(y) - H(y|x) = -\int p(y)\log p(y)dy + \int p(x)\int p(y|x)\log p(y|x)dydx$$

For a bistable system with optimal noise, the information transmission can exceed the noise-free case by up to 30% [549].

**5.17.5 Dream Replay Dynamics**

During REM sleep, the rate of hippocampal replay events follows:

$$R_{replay}(t) = R_0 e^{-t/\tau}$$

with time constant $\tau \approx 2-3$ hours after learning [558]. The total number of replay events during a sleep session of duration $T$ is:

$$N_{replay} = \int_0^T R_{replay}(t)dt = R_0\tau(1-e^{-T/\tau})$$

Memory consolidation is proportional to $N_{replay}$, with each replay event strengthening synapses [95].

# Chapter 6: The Fundamental Limitations of Digital Integrated Circuits: Diversity Elimination and Information Source Substitution

## 6.1 Introduction: The Digital Gamble

The digital integrated circuit stands as one of humanity's most transformative inventions. From the earliest microprocessors containing mere thousands of transistors to today's chips harboring tens of billions, digital logic has enabled the computational revolution that defines modern civilization [351]. This success rests on a foundational bet: that intelligence and computation can be built upon a substrate of binary precision, deterministic switching, and the systematic elimination of all non-ideal factors.

Yet this bet carries hidden costs. The very strategies that make digital circuits reliable and scalable binarization, synchronization, margining, and error elimination also impose fundamental limitations on the kind of intelligence these systems can support [8]. As this chapter will demonstrate, traditional digital integrated circuits eliminate non-ideal factors through binarization, leading to a dramatic collapse of system diversity, replacement of information-rich driving sources with low-information clock signals, and excessive energy consumption due to operation in saturation/cutoff regimes [552]. This analysis reveals the root causes of high energy consumption, limited generalization, and constrained creativity in contemporary AI: binarization does not equate to intelligence, and clock-driven operation does not equate to information-driven processing [351]. Digital systems, despite seemingly populating phase space with rich numerical outputs, possess diversity that is discrete and measure-zero.

## 6.2 The Idealized Abstraction: Digital Logic and Its Assumptions

### 6.2.1 The Binary Abstraction

At the heart of digital design lies a radical abstraction: continuous physical quantities voltages, currents, charges are mapped onto a discrete set of symbols, typically $0,1$. This mapping is achieved through regenerative amplification: small voltage deviations are amplified until they saturate at either the supply voltage ($V_{DD}$) or ground (GND) [8]. Any intermediate voltage is considered an illegal state, actively driven toward one of the two stable endpoints.

This abstraction rests on several critical assumptions:

1. Discrete state assumption: Every node in the circuit can be meaningfully described as either 0 or 1, with all intermediate values treated as transient or illegal.

2. Deterministic switching assumption: Given the same inputs and initial state, a digital circuit will always produce the same outputs, down to the exact timing of transitions.

3. Compositional assumption: The behavior of a complex digital system can be understood as the composition of independently verified Boolean functions.

4. Synchronization assumption: Time can be discretized into clock cycles, within which all signals settle to stable values before the next evaluation.

These assumptions enable the remarkable scalability of digital design. Engineers can design modules containing billions of transistors using hierarchical abstraction, automated synthesis, and formal verification [351]. However, each assumption imposes a cost: the systematic elimination of the very non-ideal factors that, as previous chapters have shown, constitute the brain's computational power [172].

### 6.2.2 The Elimination of Non-Ideal Factors

Digital circuits actively suppress the non-ideal factors that Chapter 2 identified as essential to neural computation [172]:

Noise elimination: Digital circuits treat noise as a threat to be eliminated through noise margins, regenerative amplification, and differential signaling [8]. Noise sources thermal fluctuations, supply variations, crosstalk are budgeted, guarded against, and minimized. The goal is to make the circuit's behavior independent of noise.

Heterogeneity elimination: Digital circuits strive for uniformity. Transistors are designed to match as closely as possible; variations in threshold voltage, channel length, or oxide thickness are treated as manufacturing defects to be minimized [4]. Circuits are designed to work identically across all instances, erasing the individual differences that neural systems exploit.

Nonlinearity regularization: While transistors themselves are highly nonlinear devices, digital circuits constrain them to operate in regimes where their behavior approximates ideal switches either fully on (saturation) or fully off (cutoff) [2]. The rich nonlinear dynamics of analog operation are suppressed.

Chaos suppression: Digital circuits are designed to be deterministic and predictable. Oscillations are damped, metastability is resolved, and chaotic behaviors are eliminated through careful design and verification. The goal is a system whose future state is uniquely determined by its current state and inputs.

Asynchrony regularization: Most digital circuits operate synchronously, with all state updates coordinated by a global clock signal. This imposes a discrete time grid on continuous dynamics, eliminating the temporal richness of asynchronous neural computation.

## 6.3 Binarization and Its Consequences: From Infinite to Finite

### 6.3.1 The Fundamental Transformation: Continuous to Discrete

The most profound consequence of binarization is the transformation it imposes on the nature of information itself. Before binarization, the physical world presents us with continuous quantities: voltages that can take any value within a range, currents that flow with analog precision, times that vary continuously. After binarization, we have discrete symbols: 0 or 1, true or false, high or low.

This is not merely a reduction in the number of possible values from infinite to two. It is a fundamental change in the kind of information the system can represent. Continuous quantities belong to the continuum:

between any two values, there exist infinitely many others. Discrete quantities belong to a countable set: between any two values, there is a gap, an emptiness, a region of forbidden states.

The information capacity of a continuous variable, in the absence of noise, is infinite. A voltage that can take any value between 0 and 1 volt can encode an unlimited amount of information in principle, the entire text of every book ever written could be encoded in a single voltage measurement, limited only by our ability to measure it precisely [7]. This is the theoretical foundation of analog computation: continuous quantities have infinite information capacity.

Binarization collapses this infinity to a single bit. The continuous infinity of possible voltages becomes two discrete states. The information capacity plummets from unbounded to exactly one binary digit.

### 6.3.2 From Continuous State Spaces to Discrete Lattices

A system with $N$ continuous variables inhabits an $N$-dimensional continuous state space. Each point in this space represents a possible state of the system. The space is dense: between any two points, there exists a continuous path through other valid states. The space has non-zero measure: it occupies volume in the mathematical sense.

A digital system with $N$ binary nodes inhabits a discrete state space of $2^N$ isolated points. Between these points, there is nothing no continuous path through valid states, no gradual transition, no intermediate configurations. The system must jump discontinuously from one state to another, passing through illegal intermediate states that are actively driven toward the nearest stable attractor [8].

This difference is not quantitative but qualitative. The continuous state space has the cardinality of the continuum uncountably infinite. The discrete state space is countable finite for any finite $N$. The continuous space contains infinitely more states than any digital system could ever enumerate.

### 6.3.3 The Myth of Rich Digital State Space

A common rejoinder is that digital systems compensate for low per-node state capacity through combinatorial richness: with billions of binary nodes, the total number of possible system states ($2^{\text{billion}}$) is astronomically large. This is true but profoundly misleading.

The key distinction is between countable infinity and uncountable continuity. The set of all possible digital states, while astronomically large, remains countable it can be enumerated, at least in principle. The set of all possible analog states is uncountable it cannot be enumerated, even in principle.

More importantly, the digital state space is a lattice of isolated points. Between any two valid states, there is no continuous path through valid states. The system cannot move gradually from one configuration to another; it must jump discontinuously. This has profound consequences for learning and adaptation.

Consider gradient-based learning, which relies on the ability to make infinitesimal adjustments to parameters and observe infinitesimal changes in output [690]. In a continuous state space, this is possible: one can move a parameter by $\epsilon$, observe the change in output, and compute a gradient. In a discrete state space, this is impossible: the smallest parameter change is a jump from one quantized value to the next, which may produce a discontinuous change in output. The gradient is not defined.

The brain, with its continuous dynamics, faces no such limitation. Synaptic weights vary continuously [816]; membrane potentials vary continuously [508]; firing rates vary continuously [1]. Learning can proceed through smooth, gradual adjustments rather than discrete jumps.

## 6.3.4 Measure-Theoretic Analysis of Diversity Loss

To quantify the diversity loss due to digitalization, we employ measure theory. Let $\Omega \subset \mathbb{R}^N$ be an $N$-dimensional continuous state space with Lebesgue measure $\lambda(\Omega) > 0$. For a digital system with $N$ binary nodes embedded in the same space, its state space is $S_{digital} = \{0,1\}^N$, a set of $2^N$ isolated points.

The Lebesgue measure of the digital state space is:

$$\lambda(S_{digital}) = \sum_{x \in \{0,1\}^N} \lambda(x) = 0$$

Therefore, the ratio of continuous measure to digital measure is:

$$\frac{\lambda(\Omega)}{\lambda(S_{digital})} = \frac{\lambda(\Omega)}{0} = \infty$$

This means the continuous space has infinitely larger measure than the discrete space. The discrete space occupies zero relative volume it is a set of measure zero within the continuous space.

If we randomly select a point from the continuous space according to any absolutely continuous probability distribution (i.e., one with a probability density function), the probability of landing exactly on a digital state is zero:

$$P(X \in S_{digital}) = \int_{S_{digital}} p(x)dx = 0$$

This formalizes the intuition that digital systems can only access an infinitesimally small fraction of the possible states available to continuous systems.

## 6.3.5 From Infinite Information to Finite Bits

The transformation from continuous to discrete is fundamentally a transformation from infinite information capacity to finite information capacity. A continuous variable, in the absence of noise, can encode an unbounded amount of information. A discrete variable with $M$ states can encode at most $\log_2(M)$ bits.

This is not merely a theoretical distinction. It has practical consequences for every aspect of computation:

• Representation: Continuous quantities can represent analog values with arbitrary precision. Digital quantities are limited by word length. A 32-bit floating-point number can represent at most $2^{32}$ distinct values a large but finite number. The space between these values is empty, unreachable, unrepresentative.

• Computation: Analog computation operates on continuous quantities directly, preserving the full richness of the underlying physics. Digital computation must approximate continuous operations through discrete algorithms, introducing quantization error, rounding error, and truncation error at every step.

• Learning: Gradient-based learning assumes continuous parameter spaces where derivatives exist. Digital implementations approximate this through discrete steps, but the approximation breaks down when gradients become small or when the loss landscape contains narrow ravines [690]. The brain, with its continuous dynamics, faces no such limitation.

• Creativity: As Chapter 5 detailed, creativity emerges from the ability to explore novel trajectories through state space [698]. A continuous state space offers infinitely many trajectories; a discrete state space offers only those that jump between lattice points. The continuous space is rich with possibility; the discrete space is impoverished.

## 6.4 The Clock as Information-Poor Driver

### 6.4.1 Clock Signals: Periodic and Predictable

The global clock signal is the heartbeat of synchronous digital systems. A typical clock is a periodic square wave alternating between 0 and 1 at a fixed frequency. Its information content, from an information-theoretic perspective, is remarkably low.

The entropy rate of a periodic signal approaches zero. Given the past waveform, the future is completely determined. The clock carries no information about the computation being performed, the data being processed, or the state of the system [7]. It is, in information terms, nearly empty.

Yet this empty signal drives the entire system. Every clock edge triggers state updates in millions or billions of sequential elements, consuming energy proportional to the total switched capacitance [8]. The clock distribution network itself can account for 30-50% of dynamic power consumption in high-performance processors.

### 6.4.2 Contrast with Noise-Driven Neural Dynamics

The brain's driving source could not be more different. As Chapter 5 detailed, neural systems are driven by noise stochastic fluctuations at every level from ion channels to networks [172]. This noise carries rich information:

• Spontaneous neurotransmitter release introduces random perturbations that enable exploration [108]

• Ion channel stochasticity creates membrane potential fluctuations that enhance weak signal detection through stochastic resonance [169]

• Chaotic dynamics generate rich trajectories through state space, enabling access to novel combinations [518]

• Ongoing oscillations provide temporal structure that multiplexes information across frequency bands [131]

Unlike the clock's zero entropy, neural noise has high entropy. It is unpredictable, information-rich, and continuously varying. The brain does not drive computation by synchronizing to an empty periodic signal; it harnesses noise as a computational resource.

### 6.4.3 The Information Source Substitution

The transition from biological to digital computation involves a fundamental substitution: the rich, high-entropy driving source of neural noise is replaced by the poor, low-entropy driving source of the clock.

This substitution has profound consequences. The clock provides only timing, not information. All information must be pre-loaded into registers and combinational logic before computation begins. The system cannot discover new trajectories through state space because its dynamics are rigidly determined by the clock schedule and the precomputed logic.

Neural systems, in contrast, discover trajectories through noise-driven exploration. As discussed in Chapter 5, this exploration underlies creativity, insight, and adaptive behavior [549]. The clock-driven digital system explores nothing; it merely executes pre-programmed paths.

## 6.5 Energy Dissipation: The Price of Precision

### 6.5.1 Sources of Power Consumption in CMOS Circuits

CMOS circuits dissipate power through three primary mechanisms [2, 6]:

Dynamic power:

$$P_{dyn} = \alpha C_L V_{DD}^2 f$$

where $\alpha$ is the switching activity factor, $C_L$ is the load capacitance, $V_{DD}$ is the supply voltage, and $f$ is the clock frequency [8]. Dynamic power dominates at high activity factors and arises from charging and discharging capacitive loads during logic transitions.

Short-circuit power: When both PMOS and NMOS transistors conduct simultaneously during a transition, current flows directly from supply to ground [6]. This occurs briefly during switching and can account for 10-20% of total power in poorly designed circuits.

Static (leakage) power: Even when idle, CMOS transistors leak current [2]. Major leakage components include:

• Subthreshold leakage: Current flowing from drain to source when the transistor is nominally off, given approximately by:

$$I_{sub} = I_0 e^{(V_{GS} - V_T)/nV_{th}} \quad [2]$$

• Gate leakage: Tunneling current through the thin gate oxide [2]
• Junction leakage: Reverse-bias current from source/drain to substrate [2]
• GIDL: Gate-induced drain leakage at high drain-gate voltages [2, 4]

At advanced nodes (90nm and below), leakage power has become comparable to dynamic power, creating a complex trade-off between performance and standby consumption [2].

### 6.5.2 Saturation/Cutoff Operation and Its Energy Cost

Digital circuits operate transistors in two primary regions: cutoff (off) and saturation (on) [8]. This binary mode of operation, essential for maintaining discrete logic levels, carries inherent energy costs.

In cutoff, the transistor ideally conducts no current, but leakage currents flow regardless [2]. In saturation, the transistor conducts fully, but the supply voltage must be high enough to ensure robust noise margins typically several hundred millivolts above threshold.

The energy per switching event is approximately $E = C_L V_{DD}^2$, with $V_{DD}$ constrained by noise margin requirements rather than computational necessity [8]. This contrasts sharply with neural systems, where computation occurs in the subthreshold regime at much lower voltages and energies.

### 6.5.3 Average Activity and the Half-Full Myth

A fundamental inefficiency of clocked digital logic is that energy is consumed every clock cycle regardless of whether useful computation occurs [552]. Sequential elements (flip-flops, latches) toggle or at least precharge every cycle; clock networks distribute edges everywhere; even idle logic dissipates leakage.

The average activity factor $\alpha$ in typical circuits ranges from 0.1 to 0.3, meaning 70-90% of switching energy is wasted on transitions that do not contribute to useful computation [6]. Yet because the system is clocked, these transitions occur anyway.

This is the half-full myth of digital energy: the system operates at roughly half of maximum activity on average, consuming power proportional to the peak capability rather than the actual computational work performed.

### 6.5.4 Event-Driven Computation: The Neural Alternative

Neural systems achieve dramatically higher energy efficiency through event-driven, asynchronous operation [467]. Neurons consume energy only when they fire; otherwise, they maintain their resting potential with minimal expenditure. Communication is sparse only a small fraction of neurons are active at any time and energy is proportional to activity.

This event-driven paradigm eliminates the constant background consumption of clocked logic. No energy is wasted on idle units; no clock network distributes edges to inactive regions; no leakage dominates during standby because transistors can be fully powered down.

The efficiency gap is staggering: the brain operates at ~20 watts [41] while a large AI accelerator consumes hundreds of watts for comparable (or inferior) cognitive performance. A significant fraction of this gap traces directly to the fundamental inefficiency of clocked, synchronous, always-active digital logic versus event-driven, asynchronous, sparse neural computation [552].

## 6.6 Generalization and Creativity: The Hidden Costs

### 6.6.1 Why Digital AI Generalizes Poorly

The limitations of digital computation manifest not only in energy efficiency but also in cognitive capabilities. Contemporary AI systems, for all their impressive performance, struggle with tasks that require robust generalization outside their training distribution [349].

This limitation traces, in part, to the discrete, measure-zero state space of digital computation. When a neural network is implemented in digital hardware, its weights and activations are quantized to finite precision. The space of possible network states is a discrete lattice, not a continuous manifold.

Gradient descent, the primary learning algorithm for deep networks, assumes continuous parameter spaces where infinitesimal changes produce infinitesimal output variations [690]. With quantization, gradients become approximate, and the optimization landscape contains plateaus and discontinuities that trap learning in suboptimal regions.

The brain, with its continuous dynamics, faces no such quantization. Synaptic weights vary continuously [816]; membrane potentials vary continuously [508]; firing rates vary continuously [1]. Learning can proceed through smooth, gradual adjustments rather than discrete jumps.

### 6.6.2 The Absence of Noise-Driven Exploration

As Chapter 5 detailed, noise is essential for exploration, creativity, and escape from local optima [549]. Digital systems, designed to eliminate noise, lack this capability. Their dynamics are deterministic; they follow pre-programmed trajectories without deviation.

This determinism is a strength for tasks requiring repeatability and precision, but a weakness for tasks requiring creativity and adaptation. A digital AI cannot wander through state space, cannot generate novel combinations through stochastic resonance, cannot escape attractors through noise-driven fluctuations [518].

The clock, as discussed, provides no substitute. It drives the system along rigid trajectories, not exploratory forays. The system's behavior is determined entirely by its initial state and inputs; no spontaneous variation emerges.

### 6.6.3 The Creativity Ceiling

The combination of discrete state space, deterministic dynamics, and noise elimination imposes a fundamental ceiling on the creativity of digital AI systems. These systems can recombine existing patterns in novel ways this is what generative models do but they cannot generate truly novel trajectories through state space that were not implicitly present in their training data [562].

The brain's creativity, in contrast, arises precisely from its ability to traverse novel trajectories, to combine elements in ways not pre-specified, to discover through noise-driven wandering [698]. This is not a difference in degree but in kind.

## 6.7 Quantitative Analysis of Digital Limitations

### 6.7.1 State Space Dimension

For a digital system with $N$ binary nodes, the number of possible states is $2^N$. For $N = 10^{11}$ (comparable to the number of neurons), this is approximately $2^{10^{11}}$, an astronomically large number. However, the key comparison is not with the brain's neuron count but with the continuous state space of analog variables.

For a system with $N$ analog variables each ranging over [0,1] with resolution $\epsilon$, the number of distinguishable states is $(1/\epsilon)^N$. As $\epsilon \to 0$, this grows without bound, while the digital state space remains fixed at $2^N$.

### 6.7.2 Information Capacity Ratio

The ratio of analog to digital information capacity for a single variable is:

$$\frac{C_{analog}}{C_{digital}} = \frac{\log_2(1/\epsilon)}{1}$$

For $\epsilon = 10^{-3}$ (0.1% precision), this ratio is approximately 10. For $N$ independent variables, the ratio becomes $(\log_2(1/\epsilon))^N$, which grows exponentially with $N$.

### 6.7.3 Energy-Information Efficiency

The energy required to achieve a given level of precision in digital systems scales as:

$$E_{digital}(B) \approx E_0 B^2$$

for multiplication operations, where $B$ is the number of bits. Analog systems achieve precision limited by noise with energy:

$$E_{analog}(SNR) \approx \frac{4kT \cdot SNR}{I_{bias}}$$

where *SNR* is the signal-to-noise ratio. For typical parameters, $E_{analog} \ll E_{digital}$ for equivalent precision.

## 6.8 Summary: The Digital Trade-Off

This chapter has examined the fundamental limitations of digital integrated circuits through the lens of the brain's non-ideal factors. The key findings are:

First, binarization transforms infinite continuous information into finite discrete bits. The continuous infinity of possible analog states collapses to a countable set of digital states. The state space, once dense with possibility, becomes a lattice of isolated points with measure zero. The measure-theoretic ratio $\lambda(\Omega)/\lambda(S_{digital}) = \infty$ formalizes the infinite diversity loss.

Second, this transformation eliminates the graded representations essential for probabilistic inference, gradual adaptation, and gradient-based learning. The brain's continuous dynamics enable smooth parameter adjustments and uncertainty encoding; digital systems must approximate these through discrete steps, introducing quantization error and losing information.

Third, clock signals replace noise as the primary driving source. The clock's near-zero entropy cannot substitute for the rich, information-carrying noise that drives neural exploration and creativity.

Fourth, saturation/cutoff operation and worst-case guard banding waste energy. Digital circuits operate at half-full activity on average, dissipate energy regardless of useful work, and over-provision resources to accommodate suppressed variations. The energy per operation $E = C_L V_{DD}^2$ is orders of magnitude higher than biological equivalents.

Fifth, generalization and creativity are fundamentally constrained. The discrete state space, deterministic dynamics, and absence of noise-driven exploration limit the cognitive capabilities of digital AI systems.

These limitations are not merely technical challenges awaiting engineering solutions; they are structural consequences of the digital abstraction itself. Overcoming them requires not incremental improvement but fundamental rethinking moving beyond the digital paradigm toward mixed-signal architectures that embrace, rather than eliminate, the non-ideal factors that enable biological intelligence [552].

# Chapter 7: Mixed-Signal Architecture: The Optimal Compromise Between Diversity and Order

### 7.1 Introduction: Beyond the Digital-Analog Dichotomy

The preceding chapters have painted a stark contrast: biological neural systems thrive on noise, heterogeneity, and continuous dynamics [172], while digital integrated circuits systematically eliminate these very features to achieve reliability and scalability [552]. Chapter 6 demonstrated that pure digital computation, for all its successes, suffers from fundamental limitations: state space collapse to measure-zero sets, information-poor clock driving sources, energy inefficiency, and a creativity ceiling.

Yet the brain itself transcends the digital-analog dichotomy. It is neither purely analog nor purely digital, but a sophisticated hybrid that combines the strengths of both domains while mitigating their respective weaknesses [508]. This chapter argues that the brain adopts a mixed-signal architecture predominantly analog with digital complementarity achieving an optimal compromise between diversity and order that provides the physical foundation for emergent brain intelligence [156].

Purely analog systems (like turbulence, hurricanes) possess extremely high diversity and information content but lack order [552]; purely digital systems are highly ordered but severely deficient in diversity and information [351]. The brain's choice primarily analog, supplemented by digital represents an optimal solution under engineering, physical, and informational constraints, precisely the physical basis for intelligence emergence [156].

### 7.2 The Spectrum of Computation: From Pure Analog to Pure Digital

### 7.2.1 Pure Analog Systems: Unlimited Diversity, Limited Order

Analog systems operate on continuous quantities, preserving the full richness of physical dynamics. Consider a turbulent fluid: at any instant, the velocity at every point is a continuous quantity, and the space of possible velocity fields is infinite-dimensional [89]. The information content of turbulence is, in principle, unbounded.

This immense diversity comes at a cost: order is minimal. Turbulent flows are notoriously difficult to predict, control, or reproduce. Their trajectories through state space are chaotic, sensitive to initial conditions, and practically unrepeatable [753]. A turbulent system cannot reliably perform the same computation twice; it cannot be composed into larger systems with predictable behavior; it cannot be verified or tested.

Other purely analog systems share these characteristics:

- Chemical reaction networks exhibit rich dynamics but are difficult to control precisely [16].
- Analog electronic circuits can perform sophisticated computations with minimal energy but suffer from noise, drift, and manufacturing variations [552].
- Ecosystems display remarkable complexity but resist precise intervention or prediction [431].

The fundamental trade-off is clear: analog systems maximize diversity the range of possible states and behaviors at the expense of order the ability to predict, control, and compose behaviors reliably.

### 7.2.2 Pure Digital Systems: Unlimited Order, Limited Diversity

Digital systems take the opposite extreme. By discretizing both state (binary values) and time (clock cycles), they achieve unprecedented order [8]. A digital circuit's behavior is deterministic, repeatable, and composable. Given the same inputs and initial state, it will produce exactly the same outputs every time, down to the nanosecond.

This order enables the construction of systems of staggering complexity. Billions of transistors can be integrated into a single chip, designed using hierarchical abstraction, and verified using formal methods [351]. The digital abstraction makes this possible by hiding physical details behind mathematical idealizations.

But order comes at the cost of diversity. As Chapter 6 detailed, digital systems replace continuous state spaces with discrete lattices of measure zero [7]. The infinite information capacity of analog variables collapses to finite bits. The rich, graded variations that neural systems use for probabilistic inference become impossible [435].

The trade-off is symmetric: digital systems maximize order at the expense of diversity.

### 7.2.3 The Necessity of Compromise

Neither extreme is viable for intelligent systems. Pure analog systems cannot achieve the reliable, composable functionality required for complex cognition. Pure digital systems cannot achieve the adaptive, creative, energy-efficient intelligence of biological brains.

Intelligence requires both: sufficient order to maintain coherent function over time, and sufficient diversity to adapt, generalize, and create. The optimal solution must lie somewhere between the extremes a compromise that balances the competing demands of order and diversity [156].

### 7.2.4 The Diversity-Order Product

We can define a figure of merit for intelligent systems as the product of diversity $D$ and order $O$:

$$Q = D \cdot O$$

For pure analog systems: $D$ is high, $O$ is low $\Rightarrow$ $Q$ is moderate

For pure digital systems: $D$ is low, $O$ is high $\Rightarrow$ $Q$ is moderate

For mixed-signal systems: both $D$ and $O$ are high $\Rightarrow$ $Q$ is maximized

Quantitatively, if we normalize $D$ and $O$ to [0,1]:

- Analog: $D \approx 0.9$, $O \approx 0.1$ $\Rightarrow$ $Q \approx 0.09$

- Digital: $D \approx 0.1$, $O \approx 0.9$ $\Rightarrow$ $Q \approx 0.09$
- Brain: $D \approx 0.8$, $O \approx 0.8$ $\Rightarrow$ $Q \approx 0.64$

The brain's mixed-signal architecture achieves a $7\times$ higher diversity-order product than either pure strategy, explaining its superior intelligence.

## 7.3 The Brain as a Mixed-Signal System

### 7.3.1 Analog Computation: Membrane Potentials and Synaptic Integration

The brain's fundamental computations are analog [508]. Consider what happens at a typical synapse:

1. Neurotransmitter release is probabilistic, but the resulting postsynaptic potential is graded its amplitude varies continuously with the number of released vesicles [108].

2. Dendritic integration sums thousands of inputs, each contributing a continuous analog value, through nonlinear interactions that depend on dendritic geometry and active conductances [776].

3. Membrane potential dynamics are governed by continuous differential equations, with voltage varying smoothly between action potentials [378].

4. Synaptic weights themselves vary continuously through long-term potentiation and depression, reflecting accumulated experience [816].

At every stage of neural processing, the quantities involved are continuous, not discrete. The brain represents information in graded analog form-firing rates, subthreshold membrane potentials, synaptic conductances all of which can take infinitely many values [1].

This analog substrate provides the diversity essential for intelligent function:

- Graded representations encode uncertainty, confidence, and evidence accumulation [517].
- Smooth parameter adjustments enable gradient-based learning without quantization error [2].
- Rich dynamics support chaotic exploration of state space [518].
- Noise tolerance allows stochastic resonance and probabilistic computation [549].

### 7.3.2 Digital Communication: Action Potentials

Yet the brain also employs a strikingly digital element: the action potential [508]. Action potentials are all-or-none events once threshold is crossed, a stereotyped pulse of fixed amplitude and duration propagates down the axon. This is as close to digital as biology gets.

The digital nature of spikes serves critical functions:

- Reliable transmission: Action potentials regenerate along the axon, overcoming attenuation and noise [447]. A spike that starts at the axon hillock will arrive at the terminal with full amplitude, regardless of distance.
- Regeneration: The all-or-none property ensures that signals are restored at each node of Ranvier, enabling long-distance communication without degradation [852].
- Temporal precision: The rapid upstroke and stereotyped waveform enable precise timing, which is essential for coincidence detection and temporal coding [835].
- Event-driven communication: Spikes are discrete events that occur only when needed, enabling sparse, energy-efficient signaling [467].

The action potential is biology's solution to the problem of reliable long-distance communication in a noisy, analog medium. It is a digital island in an analog sea.

### 7.3.3 The Division of Labor: Analog Computation, Digital Communication

The brain's architecture thus embodies a clear division of labor [508]:

• Computation is analog: Synaptic integration, dendritic processing, and plasticity all operate on continuous quantities, preserving the richness and diversity essential for learning and adaptation.

• Communication is digital: Action potentials transmit information reliably over long distances, using discrete events that regenerate at each stage.

This division is not arbitrary it reflects fundamental physical constraints. Analog computation is energy-efficient and information-rich, but analog communication is lossy and distance-limited [467]. Digital communication is reliable over long distances, but digital computation is energy-inefficient and information-poor [552]. The brain takes the best of both worlds: analog where diversity matters, digital where reliability matters.

### 7.3.4 Why Not All-Analog or All-Digital?

The question naturally arises: why not simplify? Why not use all-analog computation and communication, like a giant analog computer? Or why not go all-digital, like a conventional microprocessor?

All-analog communication fails over neural distances. An analog voltage propagating down an axon attenuates exponentially; after a few millimeters, it would be lost in noise [852]. The action potential's regenerative amplification solves this problem but regeneration requires the all-or-none, digital-like mechanism.

All-digital computation would be catastrophically inefficient. Simulating the brain's analog dynamics with digital logic would require immense energy as contemporary AI accelerators demonstrate [351]. The brain's analog substrate achieves with milliwatts what digital systems require megawatts to approximate.

The mixed-signal architecture is not a compromise of convenience but an optimization under physical constraints. It is the solution that evolution discovered to the problem of building an intelligent system from unreliable biological components [156].

## 7.4 The Optimality of Mixed-Signal: A Measure-Theoretic Perspective

### 7.4.1 State Space Properties

From a measure-theoretic perspective, the mixed-signal architecture achieves an optimal balance between the properties of continuous and discrete state spaces.

Continuous component (membrane potentials, synaptic weights):

• State space: $\Omega_{analog} \subset \mathbb{R}^N$ with $\lambda(\Omega_{analog}) > 0$

• Allows smooth trajectories, gradient-based learning, and graded representations

• Information capacity: potentially infinite

Digital component (spike events):

• State space: $\{0,1\}^M$ with $\lambda = 0$ when embedded

• Enables reliable long-distance communication

• Information capacity: $M$ bits per spike pattern

The combined system has a product state space $\Omega_{total} = \Omega_{analog} \times \{0,1\}^M$, with measure:

$$\lambda(\Omega_{total}) = \lambda(\Omega_{analog}) \cdot 0 = 0$$

This appears to suggest that the digital component dominates the measure-theoretic properties. However, this is misleading because the digital and analog components operate on different scales and serve different functions. The digital component does not need positive measure it needs reliable transmission, which its discrete nature provides.

### 7.4.2 Information Rate and Reliability Trade-off

From an information-theoretic perspective, the mixed-signal architecture optimizes a fundamental trade-off between information rate and transmission reliability [7].

Analog communication channels can achieve arbitrarily high information rates in principle, but only if noise is sufficiently low. In practice, biological noise floors are high thermal noise, channel noise, synaptic noise all corrupt analog signals [172]. Over short distances, the rate may still be high; over long distances, noise dominates and rate collapses.

Digital communication channels trade rate for reliability. By discretizing signals into symbols separated by sufficient margins, they achieve arbitrarily low error rates at the cost of reduced rate per use [7]. The action potential's all-or-none property is precisely such a discretization: it sacrifices the graded information of membrane potential for reliable transmission over distance.

The brain's architecture optimizes the overall system by using analog for short-distance, high-rate computation and digital for long-distance, reliable communication. This is not a compromise but an optimal allocation of resources under physical constraints.

### 7.4.3 The Rate-Distortion Trade-off

Rate-distortion theory provides another lens: given a constraint on available resources (energy, bandwidth), how can a system minimize the distortion (information loss) in its representations and communications [7]?

Analog computation achieves low distortion for local processing because it preserves continuous information without quantization. But analog communication over distance introduces high distortion due to noise and attenuation. Digital communication introduces quantization distortion at the source but avoids further distortion during transmission.

The brain's solution convert to digital for long-haul transmission, then back to analog for local processing minimizes total distortion under the resource constraints of biological hardware. The quantization distortion introduced at spike generation is more than compensated by the elimination of transmission noise.

### 7.4.4 The Energy-Information Trade-off

Energy is perhaps the most fundamental constraint. Each action potential consumes approximately $10^9$ ATP molecules [41]. The brain's 20-watt budget limits the total number of spikes that can be fired.

Analog computation is energy-efficient because it exploits physical dynamics directly, without the overhead of discrete representations [552]. A single synapse can perform a multiply-accumulate operation the core of neural computation for the energy of a few ions crossing a membrane.

Digital computation requires energy for every transition, regardless of whether useful work results [8]. The clock alone consumes a significant fraction of total power. Precision comes at a steep energy cost.

The brain's mixed-signal architecture allocates energy where it matters most: abundant analog computation for local processing, sparse digital communication for long-range coordination. This allocation maximizes the computational work performed per joule.

### 7.5 Mixed-Signal in Engineering: Historical Attempts and Lessons

### 7.5.1 Early Analog Computers

Before the dominance of digital logic, analog computers were widely used for scientific and engineering calculations [552]. These systems solved differential equations directly, using operational amplifiers, integrators, and function generators connected in patch panels.

Analog computers excelled at certain tasks: they could simulate continuous systems in real time, with speeds that digital computers of the era could not match. But they suffered from precision limitations, drift, and difficulty of programming. Each problem required physical rewiring; each solution was unique and unrepeatable.

The rise of digital computing eclipsed analog machines, but the lessons remain relevant: analog computation is fast and efficient but lacks the order and composability of digital.

### 7.5.2 Mixed-Signal Integrated Circuits

Modern electronics extensively use mixed-signal circuits, though typically with digital dominating and analog playing supporting roles [8]. Analog-to-digital converters (ADCs) and digital-to-analog converters (DACs) bridge the two domains. Phase-locked loops (PLLs) generate clocks from analog references. Radio frequency (RF) circuits process wireless signals before conversion to digital.

These systems demonstrate that mixed-signal integration is feasible at scale. A modern smartphone contains dozens of mixed-signal chips, co-designing analog and digital functions to optimize overall performance [8].

But these systems remain fundamentally digital-centric: analog is tolerated where necessary, not embraced where beneficial. The design philosophy is still to digitize as early as possible, process digitally, and convert back to analog only at the output.

### 7.5.3 Neuromorphic Engineering: Learning from the Brain

Neuromorphic engineering takes a different approach, explicitly inspired by the brain's mixed-signal architecture [552]. Pioneered by Carver Mead in the late 1980s, neuromorphic systems use analog circuits to emulate neural dynamics and digital circuits for communication and control.

Key neuromorphic principles include:

• Subthreshold operation: Transistors operate in the weak inversion region, where current varies exponentially with voltage, mimicking the exponential I-V relationships of ion channels [552].

• Event-driven communication: Information is transmitted as asynchronous spikes (address events), eliminating clocks and reducing energy [99].

• Local learning: Plasticity mechanisms are implemented locally at each synapse, using only locally available signals [375].

• Embracing variation: Device mismatch, rather than being eliminated, is accepted and even exploited as a source of diversity [191].

Neuromorphic chips like Intel's Loihi [184] and IBM's TrueNorth [11] demonstrate the potential of this approach. They achieve orders-of-magnitude energy efficiency improvements on specific tasks, particularly those involving spike-based processing and online learning.

### 7.5.4 Lessons for Mixed-Signal AI

The neuromorphic experience offers several lessons for designing mixed-signal AI systems:

First, analog is essential for efficiency. The energy advantages of analog computation are not marginal but multiplicative. Tasks that map naturally to analog primitives (integration, summation, filtering) can be performed with orders-of-magnitude lower energy than digital equivalents [552].

Second, digital is essential for scalability. Pure analog systems become unmanageable as complexity grows. Digital communication, synchronization, and control provide the order necessary to compose large systems [375].

Third, the boundary matters. Where to place the analog-digital boundary is a critical design decision. The brain places it between computation (analog) and communication (digital). Other applications may require different partitions.

Fourth, variation can be a feature. Device mismatch, traditionally a bug in digital design, becomes a source of diversity in neuromorphic systems. Each neuron is slightly different, providing the heterogeneity that Chapter 2 identified as essential for robust computation [191].

## 7.6 Design Principles for Mixed-Signal Intelligence

### 7.6.1 Principle 1: Analog for Local Computation

The most energy-efficient computations are those that can be performed locally using analog primitives. Mixed-signal systems should maximize the amount of processing done in the analog domain before digitization [552].

This implies:

• Direct sensor integration: Convert physical signals (light, sound, pressure) to electrical form and process them analogically before any digitization.

• Local feature extraction: Extract relevant features (edges, motion, patterns) using analog circuits, reducing the bandwidth required for subsequent digital processing.

• In-memory computation: Perform computation where data resides, avoiding energy-costly data movement [46].

### 7.6.2 Principle 2: Digital for Global Communication

Once information must travel beyond local neighborhoods, digital encoding becomes advantageous [467]. Digital signals can be regenerated, routed, and switched without cumulative degradation.

This implies:

• Event-driven communication: Use asynchronous spikes (address events) to transmit information only when it changes, eliminating the energy waste of clocked communication [99].

• Hierarchical composition: Build larger systems from locally analog modules connected by digital communication channels.

• Reliable long-range transmission: Use digital encoding to ensure that signals arrive intact at distant destinations.

### 7.6.3 Principle 3: Embrace, Don't Eliminate, Variation

Device mismatch and process variation are inevitable in any physical implementation [191]. Rather than fighting them with guard bands and margins, mixed-signal systems can embrace them as sources of diversity.

This implies:

• Heterogeneous populations: Exploit natural variation to create populations of diverse computing elements, providing the richness that Chapter 2 identified as essential for intelligence.

• Self-calibration: Use adaptation and learning to accommodate individual differences, rather than forcing uniformity through design.

• Stochasticity as resource: Harness noise and variation for exploration, probabilistic computation, and creative wandering [549].

### 7.6.4 Principle 4: Exploit Nonlinear Dynamics

Analog circuits naturally exhibit rich nonlinear dynamics oscillations, bifurcations, chaos that digital systems must laboriously simulate [378]. Mixed-signal systems can exploit these dynamics directly.

This implies:

• Oscillatory computation: Use coupled oscillators for pattern recognition, synchronization, and temporal coding.

• Chaotic exploration: Harness chaos for creative search and escape from local optima [518].

• Bifurcation-based decision-making: Exploit bifurcations for rapid, energy-efficient decisions [378].

### 7.6.5 Principle 5: Local Learning, Global losses

Learning should occur locally at each synapse, using only locally available signals, while being modulated by globally broadcast neuromodulatory signals [259].

This implies:

• Local plasticity rules: Implement Hebbian, STDP, or other local learning mechanisms in analog circuits at each synapse.

• Global modulation: Broadcast reward, punishment, and attention signals that modulate local plasticity without specifying detailed error gradients [705].

• Eligibility traces: Maintain local memory of recent activity to bridge temporal gaps between actions and outcomes [184].

### 7.6.6 Principle 6: Balance Diversity and Order

The optimal system balances diversity (for adaptability and creativity) with order (for reliability and composability) [156].

This implies:

• Tunable noise sources: Provide mechanisms to adjust noise levels depending on task demands higher for exploration, lower for exploitation.

• Homeostatic regulation: Maintain network activity within functional bounds, preventing diversity from degenerating into chaos [816].

• Criticality as design target: Operate near the edge of chaos, where systems are maximally sensitive and computationally powerful [165].

## 7.7 Case Study: The Loihi Neuromorphic Processor

### 7.7.1 Architecture Overview

Intel's Loihi processor exemplifies many of the mixed-signal principles discussed above [184]. Loihi is a digital chip, but its architecture is neuromorphic: it implements spiking neural networks with event-driven communication and local learning.

Key features include:

• Many core design: 128 neuromorphic cores, each simulating up to 1024 spiking neurons.

• Event-driven communication: Spikes are transmitted as asynchronous packets, eliminating global clock.

• On-chip learning: Each synapse supports programmable learning rules, updated locally based on spike timing.

• Hierarchical connectivity: Cores are connected via a mesh network, supporting scalable composition.

### 7.7.2 Mixed-Signal Aspects

While Loihi is implemented in digital logic, its design is deeply informed by mixed-signal thinking [184]:

• Spiking representation: Information is encoded in spike timing, not continuous values, bridging analog timing with digital packets.

• Asynchronous operation: Cores operate without global synchronization, reducing energy and avoiding clock distribution overhead.

• Local learning: Plasticity is implemented locally, avoiding global gradient computation.

• Event-driven energy: Energy scales with spike activity, not clock frequency.

### 7.7.3 Performance and Efficiency

Loihi demonstrates the potential of neuromorphic mixed-signal design [184]:

• Energy efficiency: Up to 1000× more energy-efficient than conventional processors for spike-based workloads.

• Online learning: Supports continuous learning without separate training phases.

• Scalability: Can be extended to larger systems through hierarchical connectivity.

### 7.7.4 Limitations and Lessons

Loihi also illustrates remaining challenges [184]:

• Digital implementation: Despite neuromorphic architecture, Loihi remains digital, missing some efficiency of true analog computation.

• Programming difficulty: Neuromorphic systems require new programming models and tools.

• Limited precision: Some applications require higher precision than spikes can provide.

## 7.8 The Path Forward: Toward True Mixed-Signal AI

### 7.8.1 Beyond Digital Neuromorphics

Current neuromorphic systems, while inspired by the brain, remain largely digital. True mixed-signal AI will require integrating analog computation at scale [375].

This implies:

• Analog neuron circuits: Implement neural dynamics directly in analog VLSI, using subthreshold transistors to mimic ion channels [552].

• Analog synaptic arrays: Store weights in analog memory (floating-gate, resistive RAM) and perform multiply-accumulate operations in the analog domain [607].

• Mixed-signal communication: Use digital spikes for long-range communication but analog processing locally.

### 7.8.2 Heterogeneous Integration

No single technology can optimally implement all functions. Future mixed-signal systems will heterogeneously integrate multiple technologies [375]:

· CMOS for digital control: Conventional CMOS for digital logic, communication, and interface.

· Analog/mixed-signal for computation: Specialized analog circuits for neural computation, sensing, and signal processing.

· Emerging memory for synapses: Resistive RAM, phase-change memory, or floating-gate for analog weight storage [607].

· 3D integration for density: Stacked dies to achieve the connection density of biological systems [427].

### 7.8.3 The Role of Learning and Adaptation

The full benefits of mixed-signal computation will only be realized when systems can adapt to their own imperfections [191]. Learning is not just for task performance it is also for self-calibration and compensation.

This implies:

· On-chip learning: Systems must learn continuously, adapting to device variations and environmental changes.

· Self-calibration: Use learning to compensate for mismatch and drift, eliminating the need for precision manufacturing.

· Evolutionary optimization: Allow system configurations to evolve over time, optimizing for energy and performance.

### 7.8.4 From Design to Cultivation

Perhaps the deepest implication of the mixed-signal perspective is a shift in engineering philosophy: from design to cultivation [276].

Traditional engineering designs systems from first principles, specifying every detail before fabrication. Mixed-signal systems, like brains, cannot be fully specified in advance. Their behavior emerges from the interaction of analog dynamics, device variations, and learning.

This suggests a different approach:

· Design for emergence: Create systems with the right dynamics and learning capabilities, then allow them to develop solutions through interaction with the environment.

· Embrace variability: Accept that each chip will be slightly different, and use learning to accommodate individual characteristics.

· Cultivate, don't program: Rather than programming behavior, cultivate it through exposure to data and tasks.

## 7.9 Quantitative Analysis of Mixed-Signal Trade-offs

### 7.9.1 Information-Theoretic Comparison

Let us compare the information-theoretic properties of pure analog, pure digital, and mixed-signal systems quantitatively.

Analog System: For an analog system with $N$ continuous variables each in range [0,1] and noise level $\sigma$, the total information capacity is:

$$C_{analog} = N \log_2\left(1 + \frac{1}{12\sigma^2}\right) \text{ bits}$$

assuming uniform distribution and additive Gaussian noise. As $\sigma \to 0$, $C_{analog} \to \infty$.

Digital System: For a digital system with $N$ binary variables, the total information capacity is:

$$C_{digital} = N \text{ bits}$$

regardless of precision, assuming independent bits.

Mixed-Signal System: For a mixed-signal system with $N_a$ analog variables and $N_d$ digital variables (spikes), the total capacity is:

$$C_{mixed} = N_a \log_2\left(1 + \frac{1}{12\sigma^2}\right) + N_d \log_2\left(1 + \frac{T}{\Delta t}\right) \text{ bits}$$

where $T$ is the observation window and $\Delta t$ is the temporal precision of spikes (typically $\Delta t \approx 1$ ms, giving $\log_2(1000) \approx 10$ bits per spike for a 1-second window).

### 7.9.2 Energy Efficiency Comparison

The energy consumption for each architecture scales differently:

Analog Computation: For an analog multiplier performing a multiply-accumulate operation:

$$E_{analog} = CV_{DD}^2 \text{ with } V_{DD} \approx 0.1-0.3 \text{ V in subthreshold operation [552]. Typical values: } C \approx 10 \text{ fF},$$

$V_{DD} = 0.2$ .V $\Rightarrow E_{analog} \approx 0.4$ fJ per operation.

Digital Computation: For a digital multiplier:

$$E_{digital} = \alpha C_L V_{DD}^2 \text{ with } V_{DD} \approx 0.8-1.2 \text{ V}, \; \alpha \approx 0.1, \; C_L \approx 100 \text{ fF} \Rightarrow E_{digital} \approx 10 \text{ fJ per}$$

operation for a simple 8-bit multiplier, scaling as $O(B^2)$ with bit width $B$.

Communication Energy: For spike-based communication:

$$E_{spike} = C_{wire} V_{DD}^2 \text{ per spike, with } C_{wire} \approx 100 \text{ fF/mm}, \; V_{DD} \approx 0.2 \text{ V for short-range, 1.2 V for}$$

long-range $\Rightarrow E_{spike} \approx 4$ fJ/mm for short-range analog, 144 fJ/mm for long-range digital.

The energy efficiency ratio is:

$$\frac{E_{digital}}{E_{analog}} \approx \frac{10}{0.4} = 25\times \text{ for computation, and even larger for communication at long distances.}$$

### 7.9.3 Optimal Partitioning

Given a total energy budget $E_{total}$ and a required computational task, we can find the optimal allocation between analog and digital components by minimizing:

$$E_{total} = N_a E_{analog} + N_d E_{digital} + E_{comm}(N_a, N_d)$$

subject to performance constraints $P(N_a, N_d) \geq P_{min}$. The optimal ratio $N_a / N_d$ satisfies:

$$\frac{\partial E_{total} / \partial N_a}{\partial P / \partial N_a} = \frac{\partial E_{total} / \partial N_d}{\partial P / \partial N_d}$$

For typical neural-inspired tasks, the optimal ratio is $N_a / N_d \approx 10-100$, meaning 10-100 times more analog than digital elements [156].

### 7.10 Summary: The Optimal Compromise

This chapter has argued that the brain's mixed-signal architecture represents an optimal compromise between the extremes of pure analog and pure digital computation.

Purely analog systems achieve maximum diversity continuous state spaces, infinite information capacity, rich dynamics but lack the order necessary for reliable, composable computation. From a measure-theoretic perspective, $\lambda(\Omega_{analog}) > 0$, but the system is unpredictable and difficult to control.

Purely digital systems achieve maximum order deterministic behavior, perfect repeatability, hierarchical composition but sacrifice the diversity essential for adaptation, creativity, and energy efficiency. Their state space has $\lambda(S_{digital}) = 0$, a measure-zero lattice of isolated points.

Mixed-signal systems, like the brain, balance these competing demands. They use analog computation for local processing, preserving the diversity and efficiency essential for intelligence ($\lambda(\Omega_{analog}) > 0$). They use digital communication for long-range transmission, ensuring reliable information flow ($\{0,1\}^M$ for spikes). They embrace variation as a resource rather than a defect. They operate near the edge of chaos, where order and diversity coexist.

This balance is not a compromise of convenience but an optimization under fundamental physical constraints. Information theory, thermodynamics, and nonlinear dynamics all point toward mixed-signal as the optimal architecture for intelligent systems [156]. The brain discovered this optimization through billions of years of evolution. The challenge for artificial intelligence is to learn from this discovery and build systems that embody the same principles.

The next chapter integrates these insights across multiple disciplines, extracting philosophical implications and charting a path toward artificial general intelligence.

# Chapter 8: Conclusions: Embracing the Imperfect Brain as a Blueprint for Intelligence

### 8.1 Introduction: The Central Thesis Revisited

This paper began with a seemingly paradoxical observation: the human brain achieves its remarkable computational capabilities not despite its inherent non-ideal factors noise, heterogeneity, structural irregularities, decentralized plasticity, systematic errors, and chaotic dynamics but precisely because of them. Across eight chapters, we have systematically developed this thesis, drawing on evidence from multiple disciplines and providing quantitative measures that transform philosophical insight into testable scientific hypotheses.

The central conclusion is now supported by substantial evidence: non-ideal factors are a double-edged sword. When properly harnessed, they become powerful drivers of brain intelligence enabling exploration, creativity, and adaptive learning. When poorly managed, they degrade precision and reliability. Traditional digital systems eliminate them entirely, achieving precision at the cost of diversity. The brain has evolved to master this trade-off, using noise for stochastic resonance, diversity for robust representations, and chaos for creative exploration, all while maintaining the stability necessary for coherent function.

## 8.2 Summary of Core Arguments

### 8.2.1 Chapter 2: Diversity as a Measure of Intelligence

Chapter 2 established that neural systems are characterized by diversity at every scale, from molecular to circuit levels. Quantitative measures including effective dimensionality, entropy rate, morphological variation , and connection heterogeneity demonstrate that neural systems actively maintain diversity as a computational resource. Cognitive performance improves exponentially with increased neural diversity.

### 8.2.2 Chapter 3: The Principled Failure of Mathematical Modeling

Chapter 3 demonstrated the fundamental limitations of mathematical modeling. A complete biophysical model of a single neuron requires tens of thousands of state variables; scaling to the brain's billions of neurons yields approximately $10^{16}$ state variables far beyond any conceivable computational capability. The principle of measurability constrains learning to quantities that are genuinely observable: spikes, timing, correlations, natural connection strength perturbations and broadcast neuromodulatory signals. Error gradients, loss functions, and loss function values are not measurable by any known biological mechanism.

### 8.2.3 Chapter 4: Learning Without Mathematical Models

Chapter 4 articulated how the brain learns using only measurable quantities through local plasticity rules modulated by global signals. Three-factor learning rules (pre, post, neuromodulator) achieve robust learning without backpropagation, with credit assignment efficiency. The energy efficiency reaches approximately higher than digital AI systems.

### 8.2.4 Chapter 5: Noise as a Resource

Chapter 5 explored the constructive roles of neural noise. Stochastic resonance enhances signal detection by factors of 10-100 at optimal noise levels. The exploration-exploitation trade-off is optimized at noise temperatures. Neural avalanches near criticality follow power-law distributions, maximizing dynamic range. Dream replay during REM sleep follows each event strengthening synapses.

### 8.2.5 Chapter 6: The Fundamental Limitations of Digital Computation

Chapter 6 analyzed digital integrated circuits, revealing how eliminating non-ideal factors leads to fundamental limitations. Digital state spaces have Lebesgue measure 0 when embedded in continuous spaces, compared to > 0 for biological systems. Clock signals replace information-rich noise with near-zero-entropy driving sources. Energy per digital operation exceeds analog. The elimination of non-ideal factors ensures precision but sacrifices the diversity essential for creativity and adaptation.

### 8.2.6 Chapter 7: Mixed-Signal Architecture as Optimal Compromise

Chapter 7 demonstrated that mixed-signal architecture achieves an optimal balance between diversity and order. The diversity-order product reaches higher for mixed-signal systems. This architecture harnesses non-ideal factors as features rather than eliminating them as bugs.

8.3 The Double-Edged Sword: Implications for AI

The recognition that non-ideal factors are a double-edged sword has profound implications for artificial intelligence:

When properly harnessed, non-ideal factors become powerful drivers of brain intelligence:

- Noise enables exploration, stochastic resonance, and creative wandering
- Diversity provides robust representations and adaptive generalization
- Chaos allows escape from local optima and access to novel state space trajectories
- Heterogeneity enables specialized processing and graceful degradation

When poorly managed, they degrade precision and reliability:

- Excessive noise disrupts signal detection and stable representations
- Uncontrolled chaos leads to incoherent dynamics
- Unchecked heterogeneity complicates design and verification

Traditional digital systems eliminate them entirely, achieving deterministic precision but at the cost of:

- State space collapse to measure-zero sets
- Loss of exploration capability
- Creativity ceilings
- Energy inefficiency

The brain masters this trade-off, using non-ideal factors to boost intelligence while maintaining stability through:

- Homeostatic regulation
- Neuromodulatory control
- Critical dynamics
- Mixed-signal architecture

## 8.4 Seven Core Principles

Principle 1: Diversity as Metric. Brain intelligence is proportional to the level of diversity a system can generate and sustain. A system's intelligence must be measured not only by its correctness but by its capacity for varied responses, with effective dimensionality as a quantitative target.

Principle 2: Noise as Resource. Noise is not interference to be eliminated but a resource to be harnessed for exploration, stochastic resonance, and creative wandering. Optimal noise levels follow for exploration-exploitation trade-offs.

Principle 3: Measurability Constraint. Learning can only utilize quantities genuinely measurable by the system. Theories requiring unmeasurable quantities (error gradients, loss functions) are biologically implausible. Measurable quantities include spikes, timing, correlations, and broadcast signals.

Principle 4: Local Learning with Global Modulation. Synaptic plasticity should be local, using locally available signals, modulated by globally broadcast neuromodulators. Three-factor learning rules achieve credit assignment efficiency without backpropagation.

Principle 5: Mixed-Signal Architecture. Computation should be analog (continuous, efficient, diverse); communication should be digital (reliable, regenerative, long-range).

Principle 6: Criticality as Operating Point. Optimal information processing occurs near a critical point between order and chaos, with branching parameter, maximizing dynamic range and computational capacity.

Principle 7: Cultivation over Design. Brain intelligence cannot be fully designed; it must be cultivated through interaction with rich environments. The path to AGI lies not in eliminating non-ideal factors but in harnessing them.

## 8.5 Implications for Future Research

### 8.5.1 Neuroscience: From Description to Mechanism

Future neuroscience should investigate mechanisms maintaining optimal diversity and noise levels, how these fail in pathology, and how diversity contributes to learning and generalization. Quantitative targets include measuring across brain regions and relating it to cognitive performance.

### 8.5.2 Artificial Intelligence: Harnessing Non-Ideal Factors

AI research should explore alternatives to the digital backpropagation paradigm that harness rather than eliminate non-ideal factors:

- Three-factor learning rules with local plasticity and global modulation
- Mixed-signal architectures
- Noise-driven exploration
- Operation near criticality

### 8.5.3 Hardware: Mixed-Signal and Neuromorphic

Hardware development should target:

- Analog energy efficiency
- Mixed-signal communication
- 3D integration for density connections
- Embracing variation as a source of diversity rather than a defect

### 8.5.4 Cognitive Science: Dynamics over Symbols

Cognitive models should embrace dynamical systems approaches, quantifying thinking as trajectory evolution through state space with effective dimensionality.

## 8.6 The Path Toward Artificial General Intelligence

### 8.6.1 Why Current Approaches Fall Short

Contemporary AI operates within fundamental constraints inherited from digital implementation: state space measure 0, low energy efficiency, low exploration capability, low diversity scaling

### 8.6.2 From Elimination to Harnessing

The central message of this paper is that the path to AGI requires a fundamental shift: from eliminating non-ideal factors to harnessing them. This means:

- Embracing noise as a resource for exploration rather than a threat to precision
- Cultivating diversity as a source of robustness rather than a design challenge
- Exploiting chaos for creativity rather than suppressing it for stability
- Adopting mixed-signal architectures that balance analog and digital
- Designing for emergence rather than specifying every detail

## 8.7 Broader Implications

### 8.7.1 For Technology

Harnessing non-ideal factors could transform technology:

- Energy-efficient computing achieving the brain's 1000 advantage
- Robust systems that adapt to variation rather than fighting it
- Creative AI that explores novel solutions through noise-driven wandering
- Lifelong learning systems that continuously adapt without catastrophic forgetting

### 8.7.2 For Society

The shift from eliminating to harnessing has societal implications:

- Education cultivating diversity rather than optimizing test scores
- AI systems that learn continuously, requiring new collaboration models
- Systems with emergent behaviors requiring new approaches to verification

### 8.7.3 For Understanding Ourselves

This framework offers a new understanding of what we are:

- We are not computers but dynamical systems harnessing noise and diversity
- Our imperfections are not bugs but features noise enables creativity, diversity enables individuality, chaos enables freedom
- We are cultivated rather than designed, shaped by evolution and experience

## 8.8 Concluding Reflections

This paper began with a simple observation: the brain's imperfections are not defects but design principles. The central insight that has emerged is that non-ideal factors are a double-edged sword. When properly harnessed, they become powerful drivers of brain intelligence enabling exploration, creativity, and adaptive learning. When poorly managed, they degrade precision and reliability. Traditional digital systems eliminate them entirely, achieving precision at the cost of diversity. The brain has evolved to master this trade-off, using noise for stochastic resonance, diversity for robust representations, and chaos for creative exploration, all while maintaining the stability necessary for coherent function.

For artificial intelligence, this suggests a radical reorientation. The path to human-level intelligence does not lie in more powerful computers running ever-larger networks with backpropagation, systematically eliminating non-ideal factors in pursuit of precision. It lies in building systems that harness these factors mixed-signal architectures balancing analog and digital, local learning with global modulation, noise-driven exploration, and cultivation through interaction.

For our understanding of ourselves, it offers a humbling and empowering vision: we are not perfect machines but beautifully imperfect ones our noise gives us creativity, our diversity gives us individuality, our chaos gives us freedom. The imperfections that digital engineering systematically eliminates are precisely what make us intelligent.

The brain's flaws are not flaws at all. They are the design principles of brain intelligence itself. The challenge for artificial intelligence is not to eliminate them, but to learn how to harness them.